\documentclass[sigconf,review=false,anonymous=false, nonacm=true]{acmart}

\usepackage{xcolor}
\usepackage{tikz}
\usepackage{sidecap}
\usepackage{import}
\usepackage{cleveref}
\usepackage{enumitem}
\usepackage{subcaption}
\usepackage{adjustbox}
\usepackage{fancyhdr}

\begin{document}

\title{Near-Optimal Wafer-Scale Reduce}

\author{Piotr Luczynski}
\authornote{Both authors contributed equally.}
\affiliation{%
  \institution{ETH Zurich}
  \department{Department of Computer Science}
  \city{Zurich}
  \country{Switzerland}
  }
\email{piotr.luczynski@inf.ethz.ch}

\author{Lukas Gianinazzi}
\authornotemark[1]
\affiliation{%
  \institution{ETH Zurich}
  \department{Department of Computer Science}
  \city{Zurich}
  \country{Switzerland}
  }
\email{lukas.gianinazzi@inf.ethz.ch}

\author{Patrick Iff}
\affiliation{%
  \institution{ETH Zurich}
  \department{Department of Computer Science}
  \city{Zurich}
  \country{Switzerland}
  }

\author{Leighton Wilson}
\affiliation{%
  \institution{Cerebras Systems}
  \city{Sunnyvale, CA}
  \country{USA}}

\author{Daniele De Sensi}
\affiliation{%
  \institution{Sapienza University of Rome}
  \department{Department of Computer Science}
  \city{Rome}
  \country{Italy}}

\author{Torsten Hoefler}
\affiliation{%
  \institution{ETH Zurich}
  \department{Department of Computer Science}
  \city{Zurich}
  \country{Switzerland
  }
  }

\renewcommand{\shortauthors}{Luczynski \& Gianinazzi et al.}

\begin{abstract}
Efficient Reduce and AllReduce communication collectives are a critical cornerstone of high-performance computing (HPC) applications. We present the first systematic investigation of Reduce and AllReduce on the Cerebras Wafer-Scale Engine (WSE). This architecture has been shown to achieve unprecedented performance both for machine learning workloads and other computational problems like FFT. 
We introduce a performance model to estimate the execution time of algorithms on the WSE and validate our predictions experimentally for a wide range of input sizes. In addition to existing implementations, we design and implement several new algorithms specifically tailored to the architecture. 
Moreover, we establish a lower bound for the runtime of a Reduce operation on the WSE. Based on our model, we automatically generate code that achieves near-optimal performance across the whole range of input sizes. 
Experiments demonstrate that our new Reduce and AllReduce algorithms outperform the current vendor solution by up to 3.27$\times$. Additionally, our model predicts performance with less than 4$\%$ error.
The proposed communication collectives increase the range of HPC applications that can benefit from the high throughput of the WSE. Our model-driven methodology demonstrates a disciplined approach that can lead the way to further algorithmic advancements on wafer-scale architectures.
\vspace{-1.5em}
\end{abstract}





\usetikzlibrary{calc}
\usetikzlibrary{backgrounds}
\usetikzlibrary{arrows.meta}
\sidecaptionvpos{figure}{m}

\definecolor{color1}{HTML}{70DBFF}
\definecolor{color2}{HTML}{9E2B25}
\definecolor{color3}{HTML}{0072BB}
\definecolor{color4}{HTML}{ffafcc}

\definecolor{shade1}{HTML}{D6F5FF}
\definecolor{shade2}{HTML}{ADEBFF}
\definecolor{shade3}{HTML}{99E6FF}
\definecolor{shade4}{HTML}{85E0FF}
\definecolor{shade5}{HTML}{70DBFF}

\newcommand\heightswitch{1.5}
\newcommand\height{0.45}
\newcommand\length{0.25}
\newcommand\diff{0.20}

\newcommand{\Opt}[0]{\ensuremath{\mathcal{O} }}

\definecolor{grey-dark}{HTML}{343a40}
\definecolor{grey-light}{HTML}{adb5bd}

\usetikzlibrary{backgrounds}  
\usetikzlibrary{positioning}
\tikzstyle{proc}=[
    circle,
    minimum size =0.05cm,
    draw=black,
    fill=gray,
    text=white,
    font=\small,
    inner sep=2
]

\tikzstyle{node} = [
  circle,
  draw=black,
  fill=white,
  text=black
]

\tikzstyle{nodesmall}=[
  node,
  font=\tiny,
  inner sep=3,
  draw=grey-light,
  fill=grey-light,
]

\tikzstyle{child1} = [
  node,
  fill=color2
]

\tikzstyle{child2} = [
  node,
  fill=color3
]
\tikzstyle{child3} = [
  node,
  fill=color4
]

\tikzstyle{pruned} = [
    dashed,
    color=blue,
    line width=1
]

\tikzstyle{banded} = [
    color=color2,
    line width=1
]

\tikzstyle{outside} = [
    color=red,
    line width=1
]

\tikzstyle{vertex_name} = [
    font=\large
]

\tikzstyle{matrix_elem} = [
    font=\Huge
]

\tikzstyle{pe} = [
    draw=grey-light,
    fill=grey-light,
    inner sep =3,
    node distance=0.75cm,
    line width=0mm
]

\tikzstyle{packet} = [
    fill=grey-light,
    inner sep =2,
    node distance=0.4cm,
    line width=0mm,
    minimum size=2.5mm,
    font=\tiny
]


\tikzstyle{blue_line} = [
    color=black,
    line width=1.15
]

\tikzstyle{blue_arrow} = [
    -{Triangle[scale=0.6]},
    color=black,
    line width=1.15
]

\tikzstyle{switch} = [
  circle,
  draw=black,
  fill=red,
  text=white,
  font=\Large,
  inner sep = 1.5
]

\tikzstyle{red_arrow} = [
    -{Triangle[scale=0.6]},
    color=black,
    line width=1.15
]

\maketitle

\begingroup
\renewcommand\thefootnote{}\footnote{\copyright\ Piotr Luczynski \& Lukas Gianinazzi et. al. | ACM 2024. Author's version of the work intended for your personal use. Not for redistribution. The definitive Version of Record appeared at HPDC 2024, \url{https://dl.acm.org/doi/10.1145/3625549.3658693}.
}
\addtocounter{footnote}{-1}
\endgroup

\section{Introduction}

\subsection{Motivation}
Communication collectives are essential in numerous distributed applications~\cite{DBLP:journals/csur/Ben-NunH19, mpi40}. Consequently, their efficient implementation is crucial to achieve high communication performance. Among these, Reduce and AllReduce are the collectives most frequently utilized in typical HPC workloads~\cite{DBLP:conf/sc/ChunduriPBHK18, DBLP:conf/sc/LagunaMMRSS19}. Specifically, these operations are critical in GEMV and GEMM kernels for fields like deep learning~\cite{DBLP:conf/nips/VaswaniSPUJGKP17, gemv_nn, Cerebras-Neural-Net-Training, dist_gnn, DBLP:conf/sc/ChoJE21}, bioinformatics~\cite{shabalin2012matrix, qi2014krux}, and physics simulations~\cite{brandt1990multilevel, DBLP:conf/sc/LtaiefHWJRK23}. These heterogeneous applications demand a variety of input shapes.

The Cerebras WSE represents a groundbreaking architecture designed specifically to expedite machine learning workloads. Traditional architectures such as CPUs and GPUs use shared DRAM memories, which can lead to long access latencies. Instead, the WSE features hundreds of thousands of processing elements (PE), each equipped with a local fast static random-access memory (SRAM), thereby enabling single-cycle access latencies. These PEs communicate via an on-wafer 2D mesh network that supports multicasting, which influences communication efficiency and patterns, setting it apart from contemporary distributed systems that typically utilize low diameter networks~\cite{slingshot, DBLP:conf/sc/BestaH14, DBLP:conf/isca/KimDSA08, DBLP:conf/isca/KimDA07}. The architecture of the WSE delivers high throughput for machine learning training~\cite{Cerebras-Neural-Net-Training, DBLP:journals/corr/abs-2304-03208, DBLP:journals/micro/Lie23} and various other HPC applications~\cite{DBLP:journals/corr/cerebras_monte_carlo, DBLP:conf/ics/cerebras_fft, field_equation_modeling_cerebras, DBLP:conf/sc/LtaiefHWJRK23}. However, maximizing performance on this architecture necessitates tailoring communication patterns to its unique characteristics. This need motivates our investigation of Reduce and AllReduce on the WSE.

\subsection{Limitations of state-of-the-art}
Current wafer-scale Reduce and AllReduce implementations are primarily optimized for extreme vector sizes. This means they are suboptimal for the intermediate and variable vector lengths typical in HPC applications. 
Furthermore, certain implementations like ring, though efficient in conventional systems, underperform on specialized hardware like the WSE. Existing approaches for wafer-scale algorithms employ ad-hoc theoretical modeling on a per-problem basis~\cite{DBLP:conf/ics/cerebras_fft} or depend solely on experimental validation~\cite{massively_scalable_stencil_cerebras, DBLP:conf/sc/RockiESSMKPDS020, DBLP:conf/sc/LtaiefHWJRK23,DBLP:journals/corr/cerebras_monte_carlo}, leading to time-consuming trial-and-error or suboptimal performance. The results from traditional distributed memory computing models like the $\alpha-\beta$ model~\cite{DBLP:journals/concurrency/ChanHPG07} do not consider features such as pipelining and multicasting, which are essential in the wafer-scale setting. Therein we identify the gap for a model-driven approach to optimizing communication collectives.

\subsection{Key Insights and Contributions}

This work presents a robust methodology for designing, analyzing, and implementing algorithms tailored to architectures similar to the WSE. The specific contributions are:

\begin{figure*}[tb!]
\centering

    \begin{subfigure}[b]{0.2\textwidth}
    \includegraphics[width=\textwidth, trim={1cm 0cm 5.5cm 1cm}, clip]{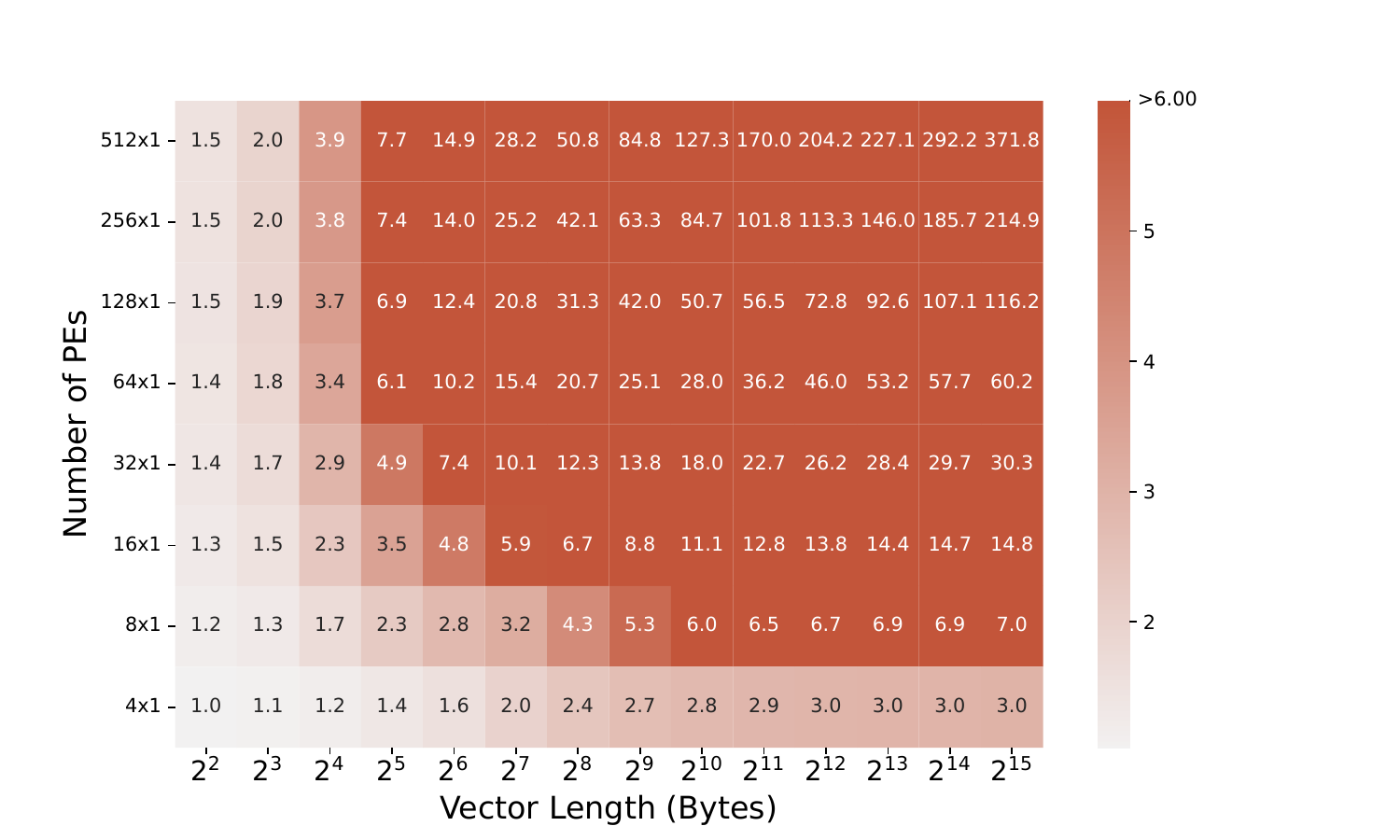}
        \caption{Star}
        \label{fig:lb_star}
    \end{subfigure}
    \begin{subfigure}[b]{0.176\textwidth}
        \includegraphics[width=\textwidth, trim={3.2cm 0cm 5.5cm 1cm}, clip]{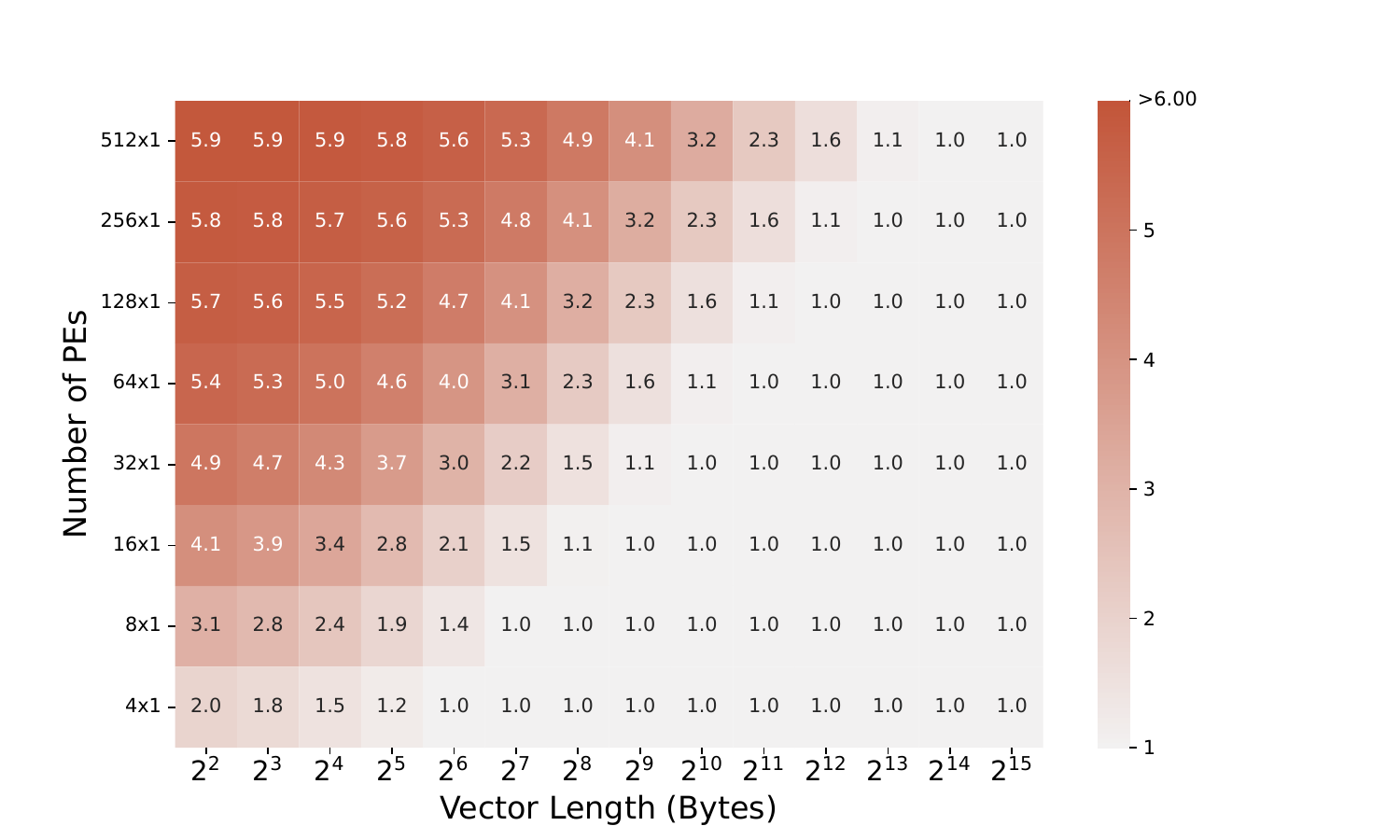}
        \caption{Chain}
        \label{fig:lb_chain}
    \end{subfigure}
    \begin{subfigure}[b]{0.176\textwidth}
        \includegraphics[width=\textwidth, trim={3.2cm 0cm 5.5cm 1cm}, clip]{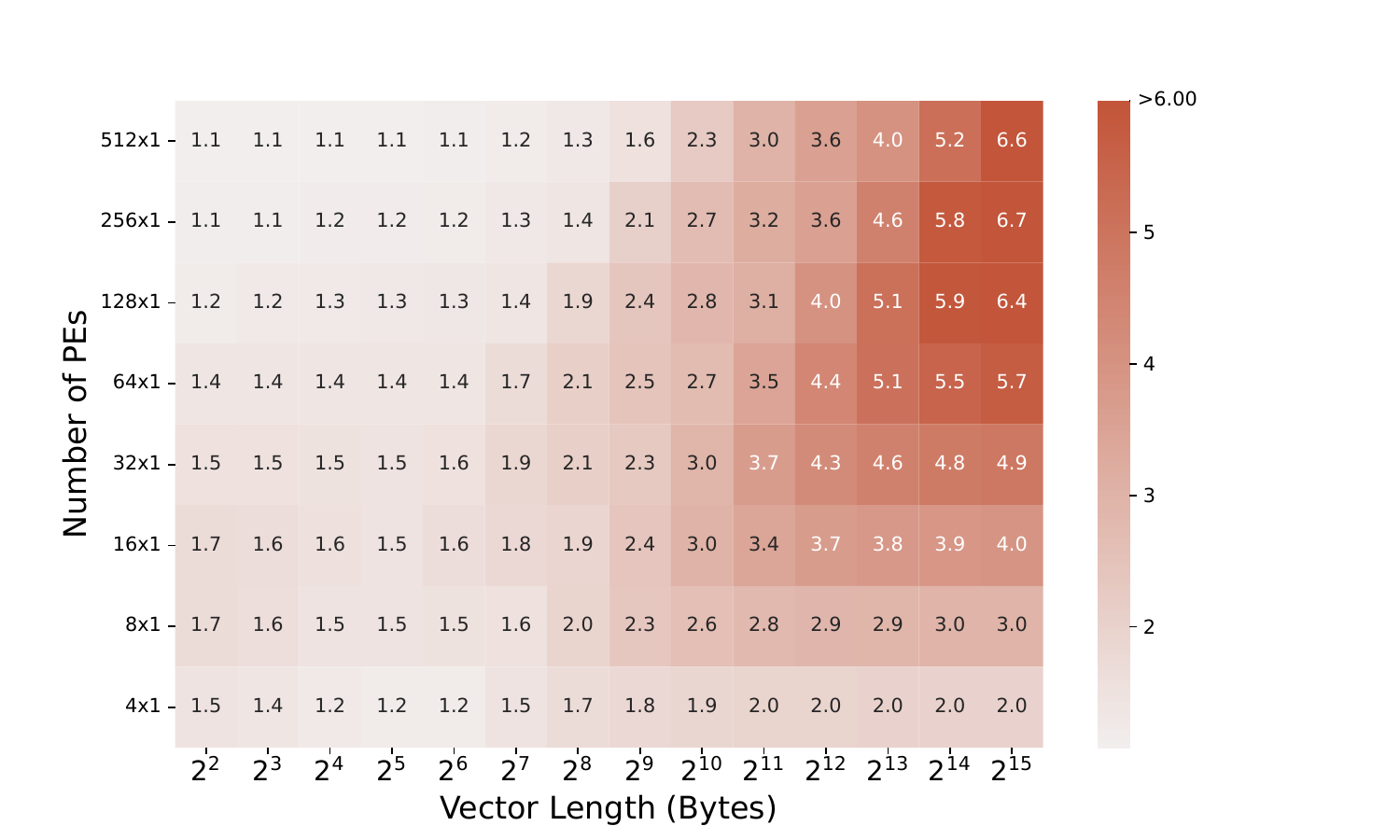}
        \caption{Tree}
        \label{fig:lb_tree}
    \end{subfigure}
    \begin{subfigure}[b]{0.176\textwidth}
        \includegraphics[width=\textwidth, trim={3.2cm 0cm 5.5cm 1cm}, clip]{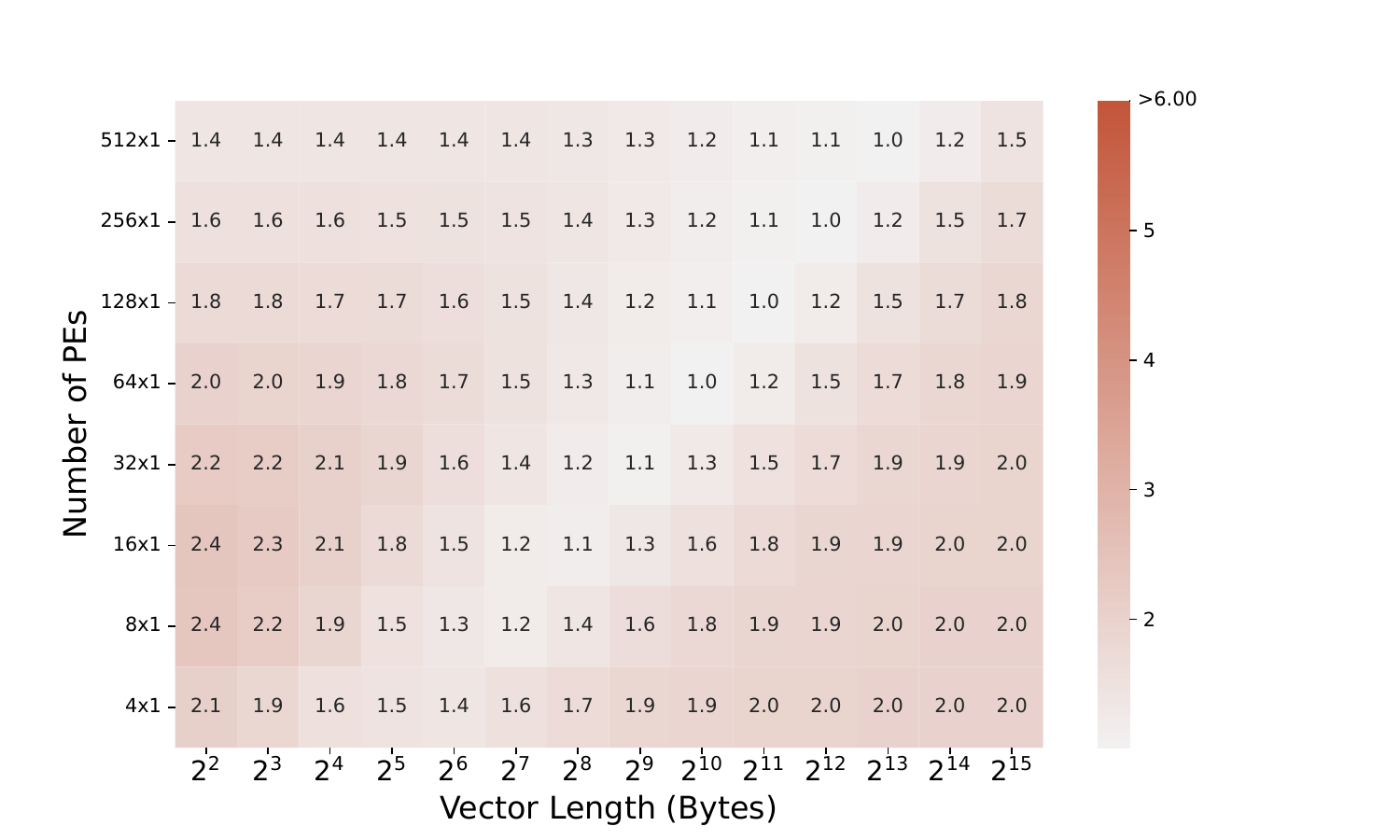}
        \caption{Two-Phase$^{\dag}$}
        \label{fig:lb_two_phase}
    \end{subfigure}
    \begin{subfigure}[b]{0.2\textwidth}
        \includegraphics[width=\textwidth, trim={3.2cm 0cm 3.3cm 1cm}, clip]{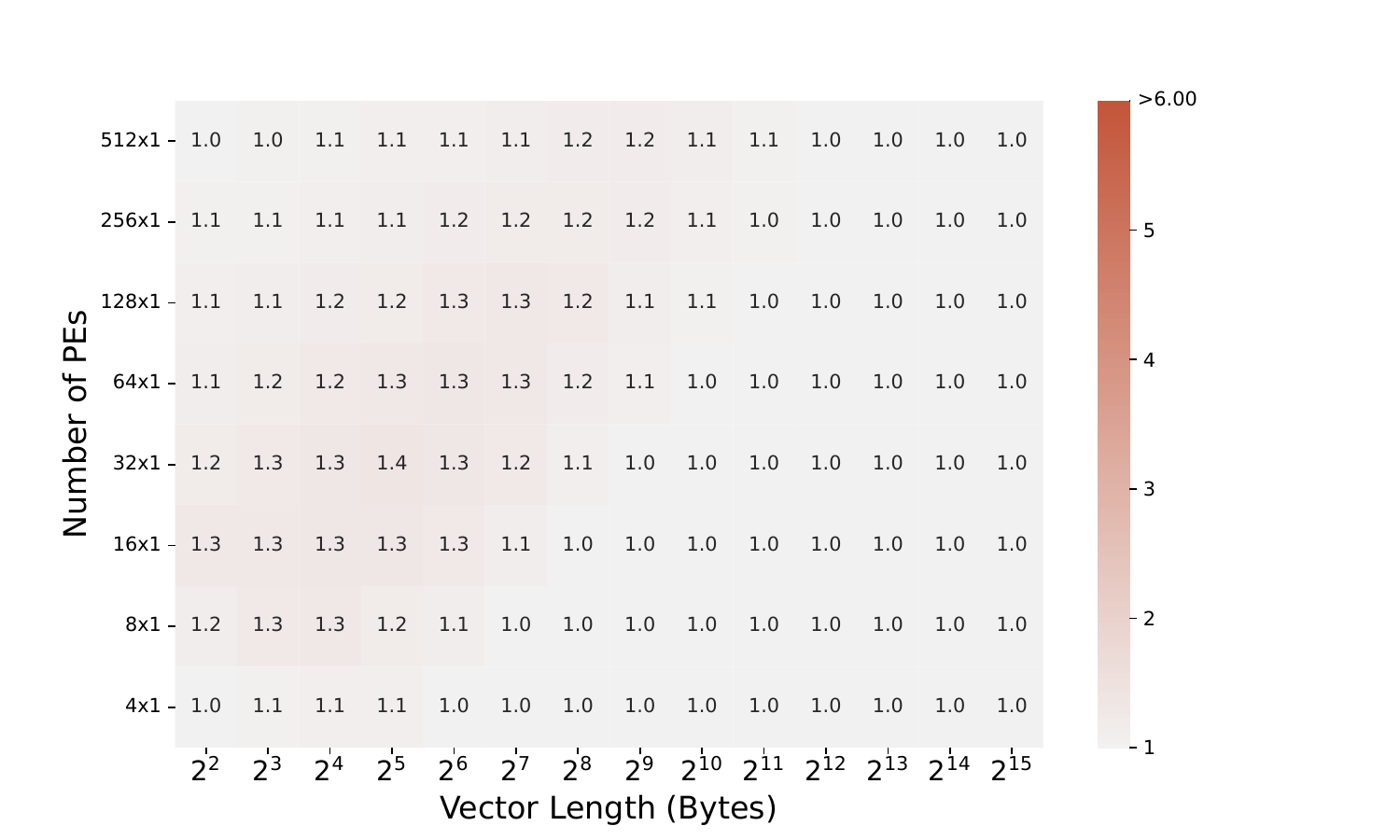}
        \caption{Auto-Gen$^{\dag}$}
        \label{fig:lb_pre}
    \end{subfigure}
    \vspace{-1em}
    \caption{
    Optimality ratios of 1D Reduce algorithms, where 1.0 is optimal. \ \
    $^{(\dag)}$ \emph{our contribution}.}
    \label{fig:heatmap_specific}
\end{figure*}

\begin{enumerate}[leftmargin=*]
    \item \textbf{Model.} We propose a novel model-driven approach. Our performance model accurately predicts execution times on the Cerebras WSE, providing a significant improvement over trial-and-error and ad-hoc methods. Our experiments confirm the model's accuracy in predicting execution times, accurately characterizing the relative strengths of each approach. 
    \item \textbf{Algorithms.} 
    We illustrate the effective use of this model by characterizing the performance trade-offs of \emph{Broadcast}, \emph{Reduce}, and \emph{AllReduce} algorithms. For Broadcast, our analysis shows that multicast support means that a simple flooding approach is optimal. We introduce a new (All)Reduce algorithm tailored to the architecture, which we call Two-Phase. Previous Reduce and AllReduce excel only within narrow ranges of input size. Two-Phase is the only approach that performs well for a wide range of vector lengths. In particular, on $512 \times 512$ PEs, Two-Phase is up to 3.32$\times$ and $2.56\times$ faster than the current vendor solution for Reduce and AllReduce, respectively. 
    We adapt classic ring AllReduce algorithm and asses its performance on the Cerebras WSE. Our model-driven approach enables a comparison of ring's performance against a direct Reduce-then-Broadcast method. Interestingly, the direct approach frequently outperforms the classical algorithm not designed with the Cerebras WSE’s particular hardware features, such as multicast, in mind.
    \item \textbf{Automatically Generated Collectives}
    Furthermore, we demonstrate how our model-driven approach facilitates the automatic generation of code for Reduce, significantly enhancing efficiency. This novel automatically-generated (\emph{Auto-Gen}) algorithm not only streamlines the optimization of complex kernels but also presents a less time-consuming alternative to manual tuning. Our experiments show that the Auto-Gen Reduce consistently matches or exceeds the performance of the best manual implementations across various input sizes. 
    \item \textbf{Lower Bound} 
    Additionally, our model is instrumental in establishing robust lower bounds for the runtime of Reduce. As summarized in \Cref{fig:heatmap_specific}, we prove that for the 1D case, our Auto-Gen Reduce is at most $1.4\times$ away from optimal across all input sizes. Two-Phase gives the best optimality ratio of the manual algorithms, being at most $2.4\times$ away from optimal. In contrast, previous algorithms are all up to $5.9\times$ away from optimal for some input size.
    \item \textbf{Implementation}
    We implement the proposed communication collectives for the second generation WSE, the Cerebras CS-2. 
\end{enumerate}

\subsection{Experimental Methodology}
We perform an extensive evaluation of our collectives on the CS-2 device. The benchmarks consider a row of PEs as well as a 2D grid of PEs. High precision measurements are of crucial importance when measuring short durations. Common problems with time measurements for distributed architectures~\cite{hoefler-collmea-sync, collective_benchmarking} need to be addressed. We propose a solution to synchronizing the clock between PEs and establishing a common start time. Each benchmark is evaluated 5 times with negligible standard deviation $(< 4\%)$. The small number of evaluations suffice because the CS-2 exhibits small runtime variance. Because execution of a thread cannot be preempted, the PE programs exhibit deterministic, state-machine like behavior which can be modeled with a cycle-accurate fabric simulator. The only notable deviation between the fabric simulator and the physical chip is overheating, which can cause a PE to insert no-ops to prevent wafer cracking. The source code is available on GitHub\footnote{\href{https://github.com/spcl/spatial-collectives}{https://github.com/spcl/spatial-collectives}}. 

\subsection{Limitations of the Proposed Approach}
%

Although we demonstrate that our lower bounds are near-optimal in a 1D row or column of PEs, for a 2D grid of PEs the optimality gap remains large. This is in part due to the lack of a strong lower bound for the 2D case. In turn, our model suggests further improvements are possible for the general 2D case.

\section{Background}\label{sec:background}

\subsection{Communication Collectives}
The \emph{Message Passing Interface} (MPI) standard \cite{DBLP:journals/pc/GroppLDS96, mpi40} defines semantics for collective operations in systems with distributed memory. MPI collectives have been extensively studied and optimized for a variety of network topologies~\cite{hoefler-moor-collectives,DBLP:journals/concurrency/ChanHPG07,DBLP:journals/ijhpca/ThakurRG05,DBLP:conf/sc/VadhiyarFD00,DBLP:journals/tc/JohnssonH89,DBLP:journals/pc/SaadS89, DBLP:conf/ipps/KaronisSFGLB00}. 
%
%
In a \emph{Reduce}, initially each PE holds a vector of equal length. The goal is to compute the sum of the vector and store it at a designated \emph{root} PE. We consider the sum over the vector, although any associative operation may be used interchangeably. In an \emph{AllReduce}, the result must be stored in every PE. 
Many patterns and techniques have been developed for  
\emph{Reduce} and \emph{AllReduce} \cite{DBLP:conf/pvm/RabenseifnerT04, DBLP:journals/jpdc/PatarasukY09, DBLP:conf/iccS/Rabenseifner04, DBLP:conf/ics/JainS10} offering different tradeoffs. 

The ring algorithm is a bandwidth optimal AllReduce~\cite{DBLP:conf/sc/HoeflerBSGLHBGCS22, DBLP:journals/jpdc/BarnettLPG95}, but it is mostly used to reduce large vectors or when running on a few nodes, since it performs a number of steps equal to the number of nodes minus one. 
Another notable example is the butterfly pattern \cite{DBLP:conf/iccS/Rabenseifner04}, which relies on recursive halving and doubling to reduce the number of steps compared to the ring algorithm. 

Although some algorithms have been optimized for torus~\cite{swing,DBLP:journals/topc/SackG15, DBLP:conf/ics/JainS10} or mesh networks~\cite{DBLP:journals/corr/abs-2011-03605}, the specific features of Cerebras CS-2, like hardware support for multicast and pipelining require the design of novel algorithms that can fully exploit those capabilities.

Some AllReduce implementations exploit the network hardware support for multicast~\cite{allreduce_multicast} or in-network compute capabilities to perform vector aggregation in the network switches~\cite{flare, Graham2020ScalableHA}. 

\subsection{Wafer-Scale Engine}
\begin{figure}
    \centering
    \includegraphics[width=0.83\columnwidth, trim={0.3cm 0 1.2cm 0}, clip]{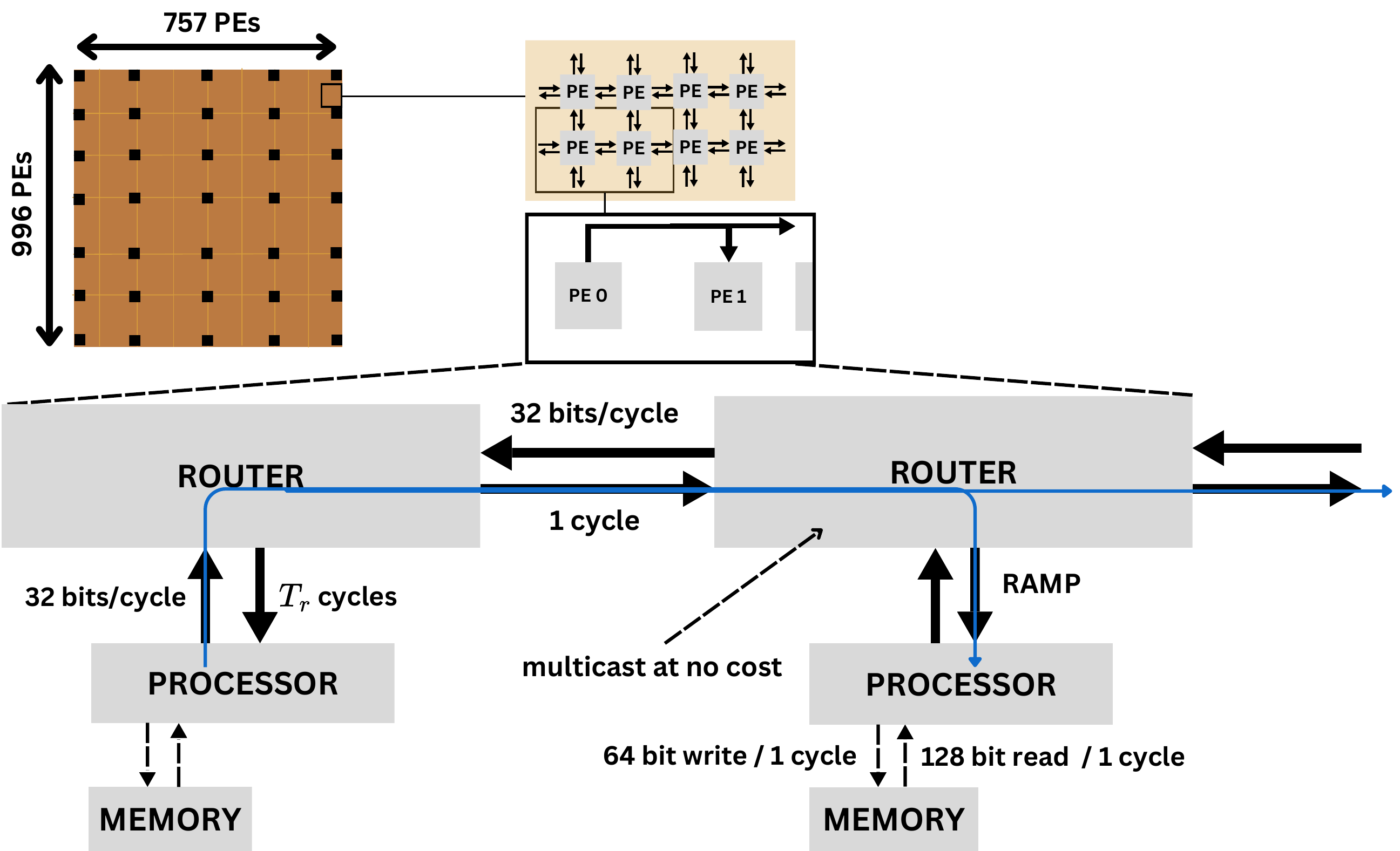}
    \vspace{-0.5em}
    \caption{PE 0 sends a wavelet to the neighbouring PE 1 on the blue color. The router connected to PE 1 forwards the wavelet to the right \emph{and} sends it up the ramp towards PE 1. This demonstrates the \emph{multicasting} capability of the network.}
    \label{fig:send_recv_2_pes}
    \vspace{-2em}
\end{figure}

The Cerebras CS-2~\cite{Cerebras, massively_scalable_stencil_cerebras, DBLP:conf/hotchips/Lie21} consists of around 750,000 processing elements (PEs) structured in a 2D grid. See Figure~\ref{fig:send_recv_2_pes} for an overview of the hardware architecture. Each PE consists of a router connected to a processor with 48KB of dedicated SRAM memory. In each cycle, a PE can read up to 128 bits from memory and write up to 64 bits. It can also execute up to 8 16-bit operations per cycle. The router manages 5 bidirectional links: 4 between the neighboring routers and a \emph{ramp} link connected to the processor. A link has a bandwidth of 32 bits/cycle in each direction. Data is sent in 32-bit packets called \emph{wavelets}. A wavelet can travel any link in a single cycle, but it takes $T_R$ cycles between when it enters the router and when an instruction can be issued by the processor using that wavelet and similarly between after a send instruction completes and when the resulting wavelet enters the router. $T_R$ is a small value, which Tramm et al.~\cite{DBLP:journals/corr/cerebras_monte_carlo} say to be around $7$. We found it to equal $2$ by inspection of the cycle-accurate simulator, which models ideal conditions.

\begin{figure}[t]
    \centering
    \resizebox{1.0\linewidth}{!}{  \centering
  \scalebox{1}{
    \begin{tikzpicture}[]
        \node[pe] (0) {0};
        \node[pe] (1) [right of=0]{1};
        \node[pe] (2) [right of=1]{2};
        \node[pe] (3) [right of=2]{3};

        \draw[blue_arrow] ($(1) + (0, \diff)$) -- ++(0, 2*\length) -- ($(0) + (0, 2*\height - \diff)$) -- ++(0,-2*\length);

        \draw[blue_arrow]($(3) + (0, \diff)$) -- ++(0, 2*\length) -- ($(2) + (-0.5, 2*\height - \diff)$);
         \node at (current bounding box.south) [below, yshift=-1mm] {$t$};
        \node[packet, fill=color2] at ($(2) + (0, 2*\height - \diff+0.1)$){};
        \node[packet, fill=color2] at ($(3) + (0, 2*\height-\diff+0.1)$){};
        \node[packet, fill=color2] at ($(3) + (0, \height+0.05)$){};
                \node[packet, fill=color3] at ($(1) + (0, \height+0.05)$){};
    \end{tikzpicture}
    \hspace{2mm}
    \begin{tikzpicture}[]
        \node[pe] (0) {0};
        \node[pe] (1) [right of=0]{1};
        \node[pe] (2) [right of=1]{2};
        \node[pe] (3) [right of=2]{3};

        \draw[blue_arrow] ($(1) + (0, \diff)$) -- ++(0, 2*\length) -- ($(0) + (0, 2*\height - \diff)$) -- ++(0,-2*\length);

        \draw[blue_arrow]($(3) + (0, \diff)$) -- ++(0, 2*\length) -- ($(2) + (-0.5, 2*\height - \diff)$);
         \node at (current bounding box.south) [below, yshift=-1mm] {$t'$};
        \node[packet, fill=color2] at ($(2) + (0, 2*\height - \diff+0.1)$){};
        \node[packet, fill=color2] at ($(3) + (0, 2*\height-\diff+0.1)$){};
        \node[packet, fill=color2] at ($(3) + (0, \height+0.05)$){};
        \node[packet, fill=color3] at ($(1) + (0, 2*\height - \diff+0.1)$){};
        \node[packet, fill=color3] at ($(0) + (0, 2*\height-\diff+0.1)$){};
        \node[packet, fill=color3] at ($(0) + (0, \height+0.05)$){};
    \end{tikzpicture}
        \hspace{2mm}
    \begin{tikzpicture}[]
        \node[pe] (0) {0};
        \node[pe] (1) [right of=0]{1};
        \node[pe] (2) [right of=1]{2};
        \node[pe] (3) [right of=2]{3};

        \draw[blue_arrow] ($(3) + (0, \diff)$) -- ++(0, 2*\length) -- ($(0) + (0, 2*\height - \diff)$) -- ++(0,-2*\length);

         \node at (current bounding box.south) [below, yshift=-1mm] {$t' + 1$};
        \node[packet, fill=color2] at ($(2) + (0, 2*\height - \diff+0.1)$){};
        \node[packet, fill=color2] at ($(3) + (0, 2*\height-\diff+0.1)$){};
        \node[packet, fill=color2] at ($(3) + (0, \height+0.05)$){};
        \node[packet, fill=color2] at ($(1) + (0, 2*\height - \diff+0.1)$){};
        \node[packet, fill=color3] at ($(0) + (0, 2*\height-\diff+0.1)$){};
        \node[packet, fill=color3] at ($(0) + (0, \height+0.05)$){};
    \end{tikzpicture}
  }}
    \vspace{-2em}
    \caption{Synchronization on the WSE occurs through routing configurations. In cycle $t$, router 1 is configured to forward the blue wavelets it gets from PE 1 towards PE 0. As a result, the red wavelets from PE 3 stall at router 2. At cycle $t'$, the last element of the vector from PE $1$ arrives at the router $1$. This triggers a change in routing configuration, such that in cycle $t'+1$ the red wavelets are propagated towards PE 0.    
    }
    \vspace{-0.5em}
    \label{fig:receive_two}
\end{figure}
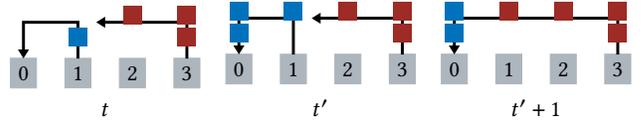

\textbf{Routing.}
Each wavelet is assigned a \emph{color}, which determines its routing.
When a router receives a wavelet, it selects in which directions to forward the wavelet based on the current configuration for the color the wavelet arrived from. Routers support \emph{multicast}, which allows them to duplicate a wavelet and send it in multiple directions at no additional cost. 
If a wavelet arrives at a router from some direction from which the router is currently not accepting wavelets, it stalls until the routing configuration changes accordingly. If two wavelets arrive at a router on the same color in the same cycle, the behaviour is undefined.  
For each color, a router stores up to four routing configurations. Initially, one of those is active. Control wavelets allow cycling through those configurations. For more flexibility, a \emph{teardown} wavelet on a color allows re-configuration from scratch. A normal wavelet can also advance the routing configuration to receive from a different direction. See Figure~\ref{fig:receive_two} for an example on how routing configuration changes in order to receive vectors from two different PEs.


\textbf{Dataflow.}
 The Cerebras chip has a dataflow architecture \cite{dataflow_architectures}. Tasks can be activated by wavelets of a specific color that arrive at a PE. This means that the order of tasks may differ depending on the order in which the wavelets arrive. Tasks may also be activated at compile time or by other tasks. Most of the operations are described using Data Structure Descriptors (DSDs), a way to describe certain vectorized operations. DSDs represent some part of memory or a sequence of incoming or outgoing wavelets. By using DSDs, we can simplify repeated operations down to a single hardware instruction. 

\section{Performance Model}\label{sec:model}

To effectively design algorithms, it is paramount to be guided by a performance model. An ideal performance model should be accurate and straightforward to evaluate, facilitating quick design iterations without extensive implementation and measurement.

We base our performance model on the \emph{spatial} \emph{computer} model, which provides a general framework for assessing algorithm performance in a disaggregated on-chip setting~\cite{22SpatialGianinazzi}. For cycle-level predictions, we parameterize the model to the properties of the WSE.
The model isolates the individual contributions to the application performance. Then, the individual contributions are aggregated into a runtime estimate. This approach allows us to identify and name the bottlenecks of a given algorithm and simplifies the analysis. 
\Cref{table:vars} explains the the individual cost terms that contribute to the model, namely \emph{depth}, \emph{distance},  \emph{contention}, and \emph{energy}.

\begin{table}[!bt]
    \centering
    \renewcommand{\arraystretch}{1.3}
    \caption{Summary of our Performance Model.}
    \vspace{-1em}
    \begin{tabular}{@{}lp{6.5cm}@{}}
    \toprule
        \textbf{Symbol} & \textbf{Description} \\ \midrule
        $E$ &  \textbf{Energy} is the \emph{total} number of hops the network needs to route wavelets for. I.e., every wavelet $m$ that travels $E_m$ hops increments the energy by $E_m$. \\ 
        $L$ & \textbf{Distance} is the largest number of hops a wavelet has to travel. \\ 
        $D$ & \textbf{Depth} is the longest sequence of PEs that perform operations that depend on each other's output. \\
        $C$ & \textbf{Contention} is the largest number of wavelets a PE sends/receives.\\ 
        $N$ & \textbf{Number of links} being used overall \\ 
        $T_R$ & \textbf{Ramp latency} from a processor to its router\\ 
        $P$ & \textbf{Number of PEs} \\ 
        $B$ & \textbf{Vector length} in wavelets \\ 
    \bottomrule
    \end{tabular}
    \label{table:vars}
\end{table}

Intuitively, a high depth means that the computation is highly sequential. A large distance means a higher communication latency. A high contention means that wavelets might stall at the contended PE. Lastly, a high energy indicates that the network might become congested. By reducing all these contributing costs, we can obtain a high-performance algorithm.

We synthesize the spatial cost metrics into an estimate for the cycle count for the WSE. Receiving and sending a wavelet costs $2 T_R$ cycles to go down and up the ramp. Additionally, it costs $1$ cycle to store the received element. Hence, the number of cycles increases by $(2 T_R + 1)$ times the depth $D$. Note that on the WSE-2, $T_R$ is about $2$ cycles on average.
%
Moreover, a wavelet needs at least $L$ cycles to travel the distance $L$. 
Next, observe that with a bandwidth of $1$ element per link per cycle, using $N$ links it takes at least $E/N$ cycles to route a communication pattern with energy $E$.
Finally, when a PE experiences contention $C$, it takes at least $C$ cycles to receive those elements. We observe that when PEs experience high contention relative to the congestion in the network, the system approaches the behavior of a pipeline, where in each of the $C$ cycles one element arrives at each PE. In this case, the congestion and latency in the network becomes negligible, as the system will stall at the contended PE.

Synthesizing these observations, we propose the following estimate $T$ for the number of cycles on a grid using $N$ links:
\begin{align}
T = \max\left(C, \frac{E}{N} + L \right) + (2 T_R + 1) D  \enspace .
\end{align}
Note that the number of links being used should be determined based on the algorithm at hand. For example, if only the links in one direction are used, those should be used for estimating the contribution of energy to the number of cycles.

Equipped with a performance model, we now dive into the design and analysis communication collectives. We begin with the 1D case where we are operating on a part of a row or column of the device. This case is important in its own right for applications such as GEMV~\cite{Cerebras-Neural-Net-Training}. 
In \Cref{sec:2d}, we consider the general 2D case.
\section{1D Broadcast}\label{sec:broadcast}
A broadcast is a simple, yet ubiquitous communication collective which can be used to build an AllReduce from a Reduce. A crucial feature of the WSE is the multicast support. It greatly simplifies operations where all PEs end up with the same data. We show with our model that this allows us to perform a broadcast with the same runtime as sending a message.

\subsection{Message}
The simplest communication primitive is sending a vector of length $B$ to some PE. Consider a setting where we are sending a vector from the rightmost to the leftmost PE in the row. 

The depth when sending a message is $1$, the distance is $P - 1$, the energy $B(P - 1)$, and the contention is $B$. The number of links is $P - 1$. Hence, we conclude:

$$T_{\textsc{Message}} = B + P  + 2T_R$$

Note that this cost is optimal for sending such a message as there is no benefit of using the links that point towards the sender, there is no benefit in using other PEs, and so there is a single path.

\subsection{Flooding Broadcast}
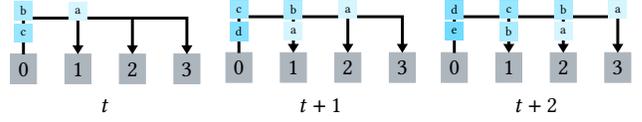
\begin{figure}[t]
    \centering
    \resizebox{1.0\linewidth}{!}{    \centering
  \scalebox{1}{
    \begin{tikzpicture}[]

        \node[pe] (0) {0};
        \node[pe] (1) [right of=0]{1};
        \node[pe] (2) [right of=1]{2};
        \node[pe] (3) [right of=2]{3};

        \draw[blue_arrow] ($(0) + (0, \diff)$) -- ++(0, 2*\length) -- ($(3) + (0, 2*\height - \diff)$) -- ++(0,-2*\length);

        \draw[blue_arrow]($(1) + (0, 2*\height- \diff)$) -- ++(0,-2*\length);
        \draw[blue_arrow]($(2) + (0, 2*\height- \diff)$) -- ++(0,-2*\length);
         \node at (current bounding box.south) [below, yshift=-1mm] {$t$};
        \node[packet, fill=shade1] at ($(1) + (0, 2*\height - \diff+0.1)$){a};
        \node[packet, fill=shade2] at ($(0) + (0, 2*\height-\diff+0.1)$){b};
        \node[packet, fill=shade3] at ($(0) + (0, \height+0.05)$){c};
    \end{tikzpicture}
    \hspace{2mm}
        \begin{tikzpicture}[]

        \node[pe] (0) {0};
        \node[pe] (1) [right of=0]{1};
        \node[pe] (2) [right of=1]{2};
        \node[pe] (3) [right of=2]{3};

        \draw[blue_arrow] ($(0) + (0, \diff)$) -- ++(0, 2*\length) -- ($(3) + (0, 2*\height - \diff)$) -- ++(0,-2*\length);

        \draw[blue_arrow]($(1) + (0, 2*\height- \diff)$) -- ++(0,-2*\length);
        \draw[blue_arrow]($(2) + (0, 2*\height- \diff)$) -- ++(0,-2*\length);
         \node at (current bounding box.south) [below, yshift=-1mm] {$t + 1$};
        \node[packet, fill=shade1] at ($(2) + (0, 2*\height - \diff+0.1)$){a};
        \node[packet, fill=shade2] at ($(1) + (0, 2*\height-\diff+0.1)$){b};
        \node[packet, fill=shade3] at ($(0) + (0, 2*\height-\diff+0.1)$){c};
        \node[packet, fill=shade4] at ($(0) + (0, \height+0.05)$){d};

        \node[packet, fill=shade1] at ($(1) + (0, \height+0.05)$){a};
    \end{tikzpicture}
    \hspace{2mm}
        \begin{tikzpicture}[]

        \node[pe] (0) {0};
        \node[pe] (1) [right of=0]{1};
        \node[pe] (2) [right of=1]{2};
        \node[pe] (3) [right of=2]{3};

        \draw[blue_arrow] ($(0) + (0, \diff)$) -- ++(0, 2*\length) -- ($(3) + (0, 2*\height - \diff)$) -- ++(0,-2*\length);

        \draw[blue_arrow]($(1) + (0, 2*\height- \diff)$) -- ++(0,-2*\length);
        \draw[blue_arrow]($(2) + (0, 2*\height- \diff)$) -- ++(0,-2*\length);
         \node at (current bounding box.south) [below, yshift=-1mm] {$t + 2$};
        \node[packet, fill=shade1] at ($(3) + (0, 2*\height - \diff+0.1)$){a};
        \node[packet, fill=shade2] at ($(2) + (0, 2*\height-\diff+0.1)$){b};
        \node[packet, fill=shade3] at ($(1) + (0, 2*\height-\diff+0.1)$){c};
        \node[packet, fill=shade4] at ($(0) + (0, 2*\height-\diff+0.1)$){d};
        \node[packet, fill=shade5] at ($(0) + (0, \height+0.05)$){e};
        \node[packet, fill=shade1] at ($(2) + (0, \height+0.05)$){a};

        \node[packet, fill=shade2] at ($(1) + (0, \height+0.05)$){b};
    \end{tikzpicture}
  }}
    \vspace{-2em}
    \caption{Broadcast for 3 consecutive cycles. This example showcases both pipelining and multicasting.}
    \vspace{0em}
    \label{fig:bcast}
\end{figure}

We consider a broadcast where the root is the rightmost PE. It is implemented by sending a message from the rightmost to the leftmost PE. Each router is configured such that it duplicates the message and sends it to the corresponding processor as well as in the direction of the leftmost PE. See Figure~\ref{fig:bcast} for an illustration. We model the runtime as:

\begin{lemma}
$T_{\textsc{Bcast}} = B + P + 2T_R = T_{\textsc{Message}}$
\end{lemma}
\begin{proof}
The depth of the broadcast is $1$, the distance is $P - 1$, the energy $B(P - 1)$, and the contention is $B$. Since the messages are sent only towards the root, the number of links is $P - 1$. 
\end{proof}

Because broadcasting sends the message to each PE, the optimality of flooding follows from the lower bound on messaging.

\section{1D Reduce}\label{sec:reduce}

Guided by our performance model, we analyze the tradeoffs of different Reduce patterns. In this section, we will focus on reduction to the leftmost PE in a single row of PEs. We show how to generalize these ideas to the full 2D grid in \Cref{sec:2d}.
%
We first discuss two patterns which have already been introduced in previous works. We then introduce two of our own patterns. 



\subsection{Star Reduce}
If we minimize the depth, we get the following algorithm: every PE sends its vector directly to the root. See Figure \ref{fig:reduce_scalar} for an illustration. This pattern has been used as part of a stencil computation algorithm for the CS-1~\cite{DBLP:conf/sc/RockiESSMKPDS020}. We can model its performance as follows:
\begin{lemma}
$T_{\textsc{Star}} \leq \max\left(B(P - 1), \frac{P}{2}B + P - 1\right) + 2 T_R + 1$
\end{lemma}
\begin{proof}
The depth is $1$, because messages go from each PE to the root directly. The distance is $P-1$, because the message from PE $P$ to PE $1$ needs $P - 1$ hops. Each PE sends $B$ messages to PE 1. Each message from PE $i$ will need $i - 1$ hops, which leads to $B \frac{P(P-1)}{2}$ energy. The contention is $B(P-1)$ at PE 1. 
\end{proof}

We can actually find here a better performance prediction than our model would suggest. For the case $B=1$, the direct upper bound predicts that the energy plus distance term would dominate. However, a closer look reveals that actually there is no congestion in the network in this case. Instead, the communication forms a perfect pipeline. Hence, the runtime is still $P-1$, rather than $\frac{3P}{2} - 1$.  We conclude that
$$T_{\textsc{Star}} = B(P-1) + 2 T_R + 1 \enspace .$$

From the model, we expect this pattern to perform well when reducing a scalar, i.e., $B = 1$. In this case, the runtime approaches the distance lower bound $P-1$.

\subsection{Chain Reduce}
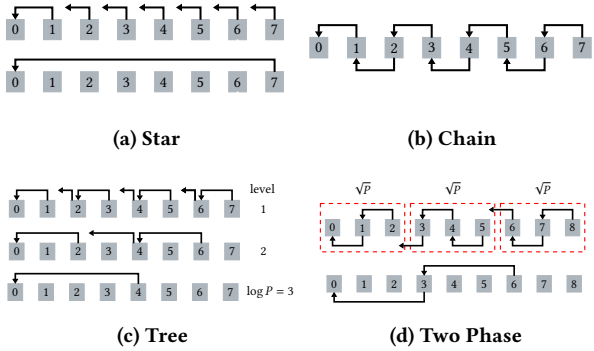
\begin{figure}[t]
    \begin{subfigure}[b]{0.48\columnwidth}
        \resizebox{\linewidth}{!}{  \centering
  \scalebox{1}{
  \begin{tabular}{c}
    \begin{tikzpicture}[]

        \node[pe] (0) {0};
        \node[pe] (1) [right of=0]{1};
        \node[pe] (2) [right of=1]{2};
        \node[pe] (3) [right of=2]{3};
        \node[pe] (4) [right of=3]{4};
        \node[pe] (5) [right of=4]{5};
        \node[pe] (6) [right of=5]{6};
        \node[pe] (7) [right of=6]{7};

        \draw[blue_arrow] ($(1) + (0, \diff)$) -- ++(0, \length) -- ($(0) + (0, \height)$) -- ++(0,-\length);

        \draw[blue_arrow] ($(2) + (0, \diff)$) -- ++(0, \length) -- ($(1) + (+0.25, \height)$);
       \draw[blue_arrow] ($(3) + (0, \diff)$) -- ++(0, \length) -- ($(2) + (+0.25, \height)$);
       \draw[blue_arrow] ($(4) + (0, \diff)$) -- ++(0, \length) -- ($(3) + (+0.25, \height)$);
       \draw[blue_arrow] ($(5) + (0, \diff)$) -- ++(0, \length) -- ($(4) + (+0.25, \height)$);
       \draw[blue_arrow] ($(6) + (0, \diff)$) -- ++(0, \length) -- ($(5) + (+0.25, \height)$);
       \draw[blue_arrow] ($(7) + (0, \diff)$) -- ++(0, \length) -- ($(6) + (+0.25, \height)$);
        \draw[blue_line, color=white]($(6) + (0, -\height)$) -- ++(0,\length);

    \end{tikzpicture}
\cr
        \begin{tikzpicture}[]

        \node[pe] (0) {0};
        \node[pe] (1) [right of=0]{1};
        \node[pe] (2) [right of=1]{2};
        \node[pe] (3) [right of=2]{3};
        \node[pe] (4) [right of=3]{4};
        \node[pe] (5) [right of=4]{5};
        \node[pe] (6) [right of=5]{6};
        \node[pe] (7) [right of=6]{7};

        \draw[blue_arrow] ($(7) + (0, \diff)$) -- ++(0, \length) -- ($(0) + (0, \height)$) -- ++(0,-\length);

        \draw[blue_line, color=white]($(6) + (0, -\height)$) -- ++(0,\length);

    \end{tikzpicture}
    \end{tabular}
  }}
        \caption{Star}
        \label{fig:reduce_scalar}
    \end{subfigure}
    \begin{subfigure}[b]{0.46\columnwidth}
        \resizebox{\linewidth}{!}{  \centering
  \scalebox{1}{
    \begin{tikzpicture}[]
        \node[pe] (0) {0};
        \node[pe] (1) [right of=0]{1};
        \node[pe] (2) [right of=1]{2};
        \node[pe] (3) [right of=2]{3};
        \node[pe] (4) [right of=3]{4};
        \node[pe] (5) [right of=4]{5};
        \node[pe] (6) [right of=5]{6};
        \node[pe] (7) [right of=6]{7};

        \draw[blue_arrow] ($(1) + (0, \height - \length)$) -- ++(0,\length) -- ($(0) + (0, \height)$) -- ++(0,-\length);

        \draw[blue_arrow] ($(3) + (0, \height - \length)$) -- ++(0,\length) -- ($(2) + (0, \height)$) -- ++(0,-\length);

        \draw[blue_arrow] ($(5) + (0, \height - \length)$) -- ++(0,\length) -- ($(4) + (0, \height)$) -- ++(0,-\length);

         \draw[blue_arrow] ($(7) + (0, \height - \length)$) -- ++(0,\length) -- ($(6) + (0, \height)$) -- ++(0,-\length);

        \draw[red_arrow] ($(4) + (0, \length - \height)$) -- ++(0,-\length) -- ($(3) + (0, -\height)$) -- ++(0,\length);

        \draw[red_arrow] ($(6) + (0, \length - \height)$) -- ++(0,-\length) -- ($(5) + (0, -\height)$) -- ++(0,\length);

        \draw[red_arrow] ($(2) + (0, \length - \height)$) -- ++(0,-\length) -- ($(1) + (0, -\height)$) -- ++(0,\length);
        \draw[white] (0,-1.25);
    \end{tikzpicture}
  }}
        \caption{Chain}
        \label{fig:reduce_sequential}
    \end{subfigure}
    \begin{subfigure}[b]{0.53\columnwidth}
        \vspace{1em}
        \resizebox{\linewidth}{!}{  \centering
  \scalebox{0.95
  }{
    \setlength{\tabcolsep}{12pt}
    \begin{tabular}{c}
        \begin{tikzpicture}[]
            \node[pe] (0) {0};
            \node[pe] (1) [right of=0]{1};
            \node[pe] (2) [right of=1]{2};
            \node[pe] (3) [right of=2]{3};
            \node[pe] (4) [right of=3]{4};
            \node[pe] (5) [right of=4]{5};
            \node[pe] (6) [right of=5]{6};
            \node[pe] (7) [right of=6]{7};
    
            \draw[red_arrow] ($(5) + (0, \diff)$) -- ++(0,\length) -- ($(4) + (0, \height)$) -- ++(0,-\length);
            \draw[blue_arrow] ($(7) + (0, \diff)$) -- ++(0,\length) -- ($(6) + (0, \height)$) -- ++(0,-\length);
            \draw[red_arrow] ($(3) + (0, \diff)$) -- ++(0,\length) -- ($(2) + (0, \height)$) -- ++(0,-\length);
            \draw[blue_arrow] ($(1) + (0, \diff)$) -- ++(0,\length) -- ($(0) + (0, \height)$) -- ++(0,-\length);
            \draw[blue_arrow] ($(2) + (-0.15, \diff)$) -- ++(0, \length) -- ($(1) + (+0.25, \height)$);
            \draw[blue_arrow] ($(4) + (-0.15, \diff)$) -- ++(0, \length) -- ($(3) + (+0.25, \height)$);
            \draw[blue_arrow] ($(6) + (-0.15, \diff)$) -- ++(0, \length) -- ($(5) + (+0.25, \height)$);
    
            
            \node [right=0.15cm of 7, yshift=5mm] {level};
            \node [right=0.4cm of 7] {$1$};
            \node [right=1.1cm of 7] {};
                            \draw[blue_arrow] ($(6) + (-0.15, \diff)$) -- ++(0, \length) -- ($(5) + (+0.25, \height)$);
            
        \end{tikzpicture}
        \vspace{2mm}
        \cr
            \begin{tikzpicture}[]
            \node[pe] (0) {0};
            \node[pe] (1) [right of=0]{1};
            \node[pe] (2) [right of=1]{2};
            \node[pe] (3) [right of=2]{3};
            \node[pe] (4) [right of=3]{4};
            \node[pe] (5) [right of=4]{5};
            \node[pe] (6) [right of=5]{6};
            \node[pe] (7) [right of=6]{7};

            \draw[red_arrow] ($(6) + (0, \diff)$) -- ++(0,\length) -- ($(4) + (0, \height)$) -- ++(0,-\length);
            \draw[blue_arrow] ($(2) + (0, \diff)$) -- ++(0,\length) -- ($(0) + (0, \height)$) -- ++(0,-\length);
                \draw[blue_arrow] ($(4) + (-0.15, \diff)$) -- ++(0, \length) -- ($(2) + (+0.25, \height)$);
            \node [right=0.4cm of 7] {$2$};
            \node [right=1.1cm of 7] {};
    
        \end{tikzpicture}
        \vspace{2mm}
        \cr
        \begin{tikzpicture}[]
            \node[pe] (0) {0};
            \node[pe] (1) [right of=0]{1};
            \node[pe] (2) [right of=1]{2};
            \node[pe] (3) [right of=2]{3};
            \node[pe] (4) [right of=3]{4};
            \node[pe] (5) [right of=4]{5};
            \node[pe] (6) [right of=5]{6};
            \node[pe] (7) [right of=6]{7};

            \node [right=0.1cm of 7] {$\log P = 3$};

            \draw[blue_arrow] ($(4) + (0, \diff)$) -- ++(0,\length) -- ($(0) + (0, \height)$) -- ++(0,-\length);
            
        \end{tikzpicture}
    \end{tabular}
  }}
        \caption{Tree}
        \label{fig:reduce-1color-low-depth}
    \end{subfigure}
    \hskip -1.8ex
    \begin{subfigure}[b]{0.46\columnwidth}
        \resizebox{\linewidth}{!}{  \centering
  \scalebox{0.75}{
  \begin{tabular}{c}
    \begin{tikzpicture}[]
        \node[pe] (0) {0};
        \node[pe] (1) [right of=0]{1};
        \node[pe] (2) [right of=1]{2};
        \node[pe] (3) [right of=2]{3};
        \node[pe] (4) [right of=3]{4};
        \node[pe] (5) [right of=4]{5};
        \node[pe] (6) [right of=5]{6};
        \node[pe] (7) [right of=6]{7};
        \node[pe] (8) [right of=7]{8};

        \draw [thick, dashed, red]([xshift=-1.25mm,yshift=0.4cm]6.north west) -- ([xshift=0.125cm,yshift=0.4cm]8.north east) -- ([xshift=0.125cm,yshift=-0.4cm]8.south east) -- ([xshift=-0.125cm,yshift=-0.4cm]6.south west) -- cycle;

        \node [above=0.5cm of 7] {$\sqrt{P}$};
        \node [above=0.5cm of 4] {$\sqrt{P}$};
        \node [above=0.5cm of 1] {$\sqrt{P}$};

                \draw [thick, dashed, red]([xshift=-1.25mm,yshift=0.4cm]3.north west) -- ([xshift=0.125cm,yshift=0.4cm]5.north east) -- ([xshift=0.125cm,yshift=-0.4cm]5.south east) -- ([xshift=-0.125cm,yshift=-0.4cm]3.south west) -- cycle;

                        \draw [thick, dashed, red]([xshift=-1.25mm,yshift=0.4cm]0.north west) -- ([xshift=0.125cm,yshift=0.4cm]2.north east) -- ([xshift=0.125cm,yshift=-0.4cm]2.south east) -- ([xshift=-0.125cm,yshift=-0.4cm]0.south west) -- cycle;

        \draw[blue_arrow] ($(8) + (0, \diff)$) -- ++(0,\length) -- ($(7) + (0, \height)$) -- ++(0,-\length);

        \draw[blue_arrow] ($(4) + (0, \diff)$) -- ++(0,\length) -- ($(3) + (0, \height)$) -- ++(0,-\length);

        \draw[blue_arrow] ($(2) + (0, \diff)$) -- ++(0,\length) -- ($(1) + (0, \height)$) -- ++(0,-\length);

        \draw[red_arrow] ($(1) + (0, \length - \height)$) -- ++(0,-\length) -- ($(0) + (0, -\height)$) -- ++(0,\length);
        \draw[red_arrow] ($(5) + (0, \length - \height)$) -- ++(0,-\length) -- ($(4) + (0, -\height)$) -- ++(0,\length);
        \draw[red_arrow] ($(7) + (0, \length - \height)$) -- ++(0,-\length) -- ($(6) + (0, -\height)$) -- ++(0,\length);
                                    \draw[blue_arrow] ($(6) + (-0, \diff)$) -- ++(0, \length) -- ($(5) + (+0.15, \height)$);

\draw[blue_arrow] ($(3) + (0, -\diff)$) -- ++(0, -\length) -- ($(2) + (+0.15, -\height)$);
    \end{tikzpicture}
    \vspace{2mm}
    \cr
    \begin{tikzpicture}[]
        \node[pe] (0) {0};
        \node[pe] (1) [right of=0]{1};
        \node[pe] (2) [right of=1]{2};
        \node[pe] (3) [right of=2]{3};
        \node[pe] (4) [right of=3]{4};
        \node[pe] (5) [right of=4]{5};
        \node[pe] (6) [right of=5]{6};
        \node[pe] (7) [right of=6]{7};
        \node[pe] (8) [right of=7]{8};


        \draw[blue_arrow] ($(6) + (0, \diff)$) -- ++(0,\length) -- ($(3) + (0, \height)$) -- ++(0,-\length);



        \draw[red_arrow] ($(3) + (0, \length - \height)$) -- ++(0,-\length) -- ($(0) + (0, -\height)$) -- ++(0,\length);
    \end{tikzpicture}
  \end{tabular}
  }}
        \caption{Two Phase}
        \label{fig:reduce-2color-2-phase}
    \end{subfigure}
    \vspace{-0em}
    \caption{Routing configurations for 1D Reduce schemes. Each row shows a configuration. When a PE has sent all its data, it switches to the next configuration. Observe that every path is set up to process a vector of elements in a \emph{pipeline}. 
    However, if a router is not ready yet to forward data because its PE is still in a previous configuration, this will stall the preceding PE. In this way, the operation is loosely synchronized between configurations. 
    }
    \label{fig:reduce_patterns}
\end{figure}

A lot of distributed applications require reduction of longer vectors, where Star-Reduce would be very inefficient. We could use the chain pattern for that, which is currently implemented as part of the existing collectives library \cite{Cerebras_Collectives} and used in Cerebras' matrix multiplication algorithm \cite{Cerebras-Neural-Net-Training}.
Every PE sends its vector to its left neighbor, forming a chain as shown in \Cref{fig:reduce_sequential}. The operation is pipelined, i.e., when a PE is receiving wavelets it is also sending out the already processed ones.
We will later show that this pattern is optimal for very large vectors. 

The pattern uses two colors. Every PE receives on the red color and sends them out on the blue color. Routing decisions cannot depend on where a wavelet came from. If we had only one color, we would need to treat the wavelets coming from the RAMP differently than the ones coming from the EAST (see also \Cref{fig:send_recv_2_pes}).

\begin{lemma}
$
T_{\textsc{Chain}} = B + (2 T_R + 2) (P-1)
$
\end{lemma}
\begin{proof}
The depth is $P-1$, because PE $i$ can only start sending messages after it receives them from PE $i + 1$. The distance is $P-1$ as it is the number of hops from PE $P$ to PE $1$. Each PE $(i + 1)$ sends $B$ messages to PE $i$. This requires $1$ hop per message, which leads to $(P-1)B$ energy. Every PE  receives $B$ messages from its right neighbouring PE, which results in $B$ contention. 
\end{proof}
The Chain-Reduce shines for vector lengths $B \gg T_R P$, when its runtime approaches the contention lower bound $B$.

\subsection{Tree Reduce}
The main issue with the Chain Reduce is that the runtime increases linearly with the number of PEs. Therefore, we propose a binary tree reduction pattern. We assume in our description that the number of PEs is a power of two. This assumption can be easily removed. 
The reduction proceeds in $\log P$ rounds. Initially, all PEs are \emph{active}. In every round, every second active PE sends a message containing its partial result to the previous active PE and then becomes inactive. This way, we halve the number of active PEs in every round until the root holds the result. 
See Figure \ref{fig:reduce-1color-low-depth} for an illustration. 

Note that the router configuration changes during the execution, is achieved using control wavelets. An active PE that is sending wavelets has a router configuration to receive from the RAMP and propagate to the WEST. Then, when it has sent everything out and becomes inactive, it switches the configuration to receive from the EAST and propagate to the WEST. 


\begin{lemma}
    \begin{align*}
T_{\textsc{Tree}} = \max \left (B \log_2 P, \frac{B\cdot P}{2(P - 1)} \log_2 P + P-1 \right) + (2 T_R + 1) \log_2 P \enspace 
    \end{align*}
\end{lemma}

\begin{proof}
   For a Tree-Reduce on $P$ processors, where $P$ is a power of two, the depth is $\log_2 P$ because we halve the number of active PEs each round. The distance is $P-1$. In the $i$-th round, we have $\frac{P}{2^{i-1}}$ active PEs. Half of those send $B$ wavelets that travel a distance of $2^{i - 1}$. The energy round $i$ is therefore $\frac{PB}{2^{i}}2^{i - 1} = \frac{PB}{2}$. Because we have a total of $\log_2 P$ rounds, the energy is $\frac{B}{2} P\log_2 P$. The root will receive $B$ messages in each round which leads to $B\log_2 P$ contention. 
\end{proof}
The tree pattern overcomes the large depth of the Chain Reduce. However, it comes at the cost of a significantly increased contention. This becomes an issue for large vector lengths.

%


\subsection{Two Phase Reduce}
We introduce an approach that combines the beneficial aspects of the Tree and the Chain pattern, namely low depth and low contention.
The algorithm is parameterized by the group size $S$ and has two phases.
In the first phase, we perform Chain-Reduce in groups of $S$ consecutive PEs. Only the leftmost PE in each group participate in the second phase. In that phase, we perform Chain-Reduce on the remaining $\left\lceil\frac{P}{S}\right\rceil$ PEs. It is important that we assign the groups from the end, i.e., starting from $p_{P - 1}$. See Figure \ref{fig:reduce-2color-2-phase} for an illustration of the approach. A choice of $S=\sqrt{P}$ reduces the depth and energy costs. Hence, we use this choice of $S$ throughout.


\begin{lemma}
When $P=S^2$, we have:
\begin{align*}
    T_{\textsc{TwoPhase}}  \leq \max\left(2B, 2B-2\frac{B}{\sqrt{P}} + P\right) + (2\sqrt{P} - 2)\cdot \left(2T_R+1\right)
\end{align*}
\end{lemma}
\begin{proof}
    The depth is $S - 1 + \left\lceil \frac{P}{S} \right\rceil - 1$. Executing chain reduce on $S$ PEs has depth $S - 1$. In the second phase, we have $\left\lceil \frac{P}{S} \right\rceil$ PEs left, which again requires $\left\lceil \frac{P}{S} \right\rceil - 1$ depth with chain reduce.
    The energy of the first phase equals $\lceil\frac{P}{S}\rceil$ times that of a chain pattern on $S$ PEs, that is, $(S - 1)B \lceil\frac{P}{S}\rceil$ energy.
    %
    In the second phase, we have $\lceil \frac{P}{S} \rceil - 1$ vectors of length $B$ that travel $S$ hops. This totals $SB (\lceil\frac{P}{S}\rceil -1)$ energy.  
    When $S=P^2$, the energy of each phase simplifies to $PB - B\sqrt{P}$. The result follows since there are at most $P$
     links active. 
     
%
%
\end{proof}
We observe that the two phase pattern has a contention that is only a factor $2$ worse than the chain reduce, while vastly reducing the depth from $P-1$ to $2\sqrt{P}$. Hence, we expect it to perform well for intermediate ranges of vector sizes.

\subsection{Auto-Gen Reduce}
Our model reveals that none of the existing algorithms provide a consistent performance throughput the whole range of vector lengths $B$ and number of PEs $P$. In this section, we provide a method that achieves good performance across the board.
%
%

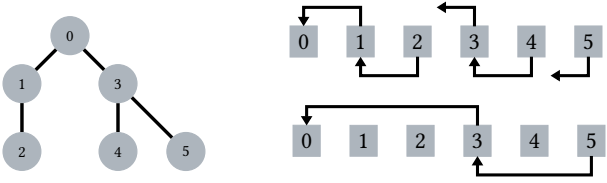
\begin{figure}
    \centering
    \begin{minipage}{.35\columnwidth}
        \resizebox{\linewidth}{!}{\centering
\scalebox{1}{
\begin{tikzpicture}[node distance=8mm]

    \node[nodesmall] (0) {$0$};
    \node[nodesmall] (1) [below left of=0] {$1$};
    \node[nodesmall] (2) [below of=1] {$2$};
    \node[nodesmall] (3) [below right of=0] {$3$};
    \node[nodesmall] (4) [below of=3] {$4$};
    \node[nodesmall] (5) [right of=4] {$5$};
    
    \draw[blue_line] (0) -- (1);
    \draw[blue_line] (1) -- (2);
    \draw[blue_line] (0) -- (3);
    \draw[blue_line] (3) -- (4);
    \draw[blue_line] (3) -- (5);
\end{tikzpicture}
}}
    \end{minipage}%
    \hfill
    \begin{minipage}{.55\columnwidth}
        \resizebox{\linewidth}{!}{\begin{tabular}{c}
\begin{tikzpicture}[]
    \node[pe] (0) {0};
    \node[pe] (1) [right of=0]{1};
    \node[pe] (2) [right of=1]{2};
    \node[pe] (3) [right of=2]{3};
    \node[pe] (4) [right of=3]{4};
    \node[pe] (5) [right of=4]{5};

    \draw[blue_arrow] ($(1) + (0, \diff)$) -- ++(0,\length) -- ($(0) + (0, \height)$) -- ++(0,-\length);
       \draw[blue_arrow] ($(3) + (0, \diff)$) -- ++(0, \length) -- ($(2) + (+0.25, \height)$);
              \draw[blue_arrow] ($(5) + (0, -\diff)$) -- ++(0, -\length) -- ($(4) + (+0.25, -\height)$);

    \draw[red_arrow] ($(2) + (0, \length - \height)$) -- ++(0,-\length) -- ($(1) + (0, -\height)$) -- ++(0,\length);
    \draw[red_arrow] ($(4) + (0, \length - \height)$) -- ++(0,-\length) -- ($(3) + (0, -\height)$) -- ++(0,\length);
\end{tikzpicture}
\vspace{2mm}
\cr
\begin{tikzpicture}[]
    \node[pe] (0) {0};
    \node[pe] (1) [right of=0]{1};
    \node[pe] (2) [right of=1]{2};
    \node[pe] (3) [right of=2]{3};
    \node[pe] (4) [right of=3]{4};
    \node[pe] (5) [right of=4]{5};

    \draw[blue_arrow] ($(3) + (0, \diff)$) -- ++(0,\length) -- ($(0) + (0, \height)$) -- ++(0,-\length);



    \draw[red_arrow] ($(5) + (0, \length - \height)$) -- ++(0,-\length) -- ($(3) + (0, -\height)$) -- ++(0,\length);
\end{tikzpicture}
\end{tabular}}
    \end{minipage}
    \vspace{-1em}
    \caption{Reduction tree labelled in pre-order and the corresponding Auto-Gen Reduce routing configuration. }
    \label{fig:pre_order}
\end{figure}

This Reduce algorithm generates a different reduction tree for a given set of input sizes and PE counts. We call this algorithm the Auto-Gen algorithm, because it traverses an automatically generated reduction tree in pre-order. Each tree stored in pre-order represents some reduce execution. In such an execution, each vertex, representing a unique PE, receives messages from its children in-order. During the execution, each PE can send messages only to one other PE. Moreover, we do not allow for overlapping communication edges. This means that if PE 3 is sending messages to PE 0, then PE 4 can send messages to neither PE 1 or PE 2. See \Cref{fig:pre_order} for an illustration.  
Note that this general approach generalizes every algorithm we have presented so far. For example, a Star reduce is represented by a star graph and a Chain reduce by a path. Hence, by finding the optimal tree, we can guarantee to match or outperform those fixed algorithms.



Let $E_{\textsc{Auto-Gen}}(P, B, D, C)$ be the minimum energy of Auto-Gen reduce with $P$ PEs, $B$ vector length, $D$ depth and contention $C$. We can calculate it recursively as:

$$\min_i E_{\textsc{Auto-Gen}}(i, B, D, C - 1) + E_{\textsc{Auto-Gen}}(P - i, B, D - 1, C) + i$$

\noindent
The root PE needs to receive at least one message. Let the last message it receives be from PE $i + 1$. When it receives that message, it needs to already have the sum of $i$ PEs. This needs to be done with contention at most $C - 1$, since it will receive one more message. The message sent from PE $i + 1$ needs to have the value of reduce on $P - i$ rightmost PEs. This can be done with depth at most $D - 1$, because it is followed by sending a message. We can now calculate $T_{\textsc{Auto-Gen}}(P, B)$ as:
$$\min_{(D, C)} \max\left(C, \frac{E_{\textsc{Auto-Gen}}(P, B, D, C)}{P - 1} + P - 1\right) + D(2T_R + 1)$$
\noindent
We add $P - 1$, because the message from the rightmost PE needs $P - 1$ hops. We divide the energy by $P - 1$, because we assume that messages are sent towards the root. 


To generate the code for the Auto-Gen reduce, we first compute the best pre-order tree via backtracking. We can compute this optimal tree in $O(P^4)$ by finding the lowest energy tree according to the  formula for $T_{\textsc{Auto-Gen}}$. Based on the tree we compute the routing configuration for each PE. This includes the colors on which the wavelets are sent and received. We need to compute whether a PE should be sending a control wavelet in order to update the routing configuration of the receiver. We implement Auto-Gen by providing a python program which computes the optimal tree and generates the code with the routing and PE code.

\subsection{Lower Bound}\label{sec:lower_bound}
We present a lower bound on the cost of 1D Reduce for a broad class of algorithms in our model. 
The idea is to bound the energy required for a given depth recursively. We can then find the best algorithm by considering every possible depth and summing all cost terms.
We assume that PEs send messages in the direction of the root. However, we allow for a PE to send one wavelet to PE $i$ and another to PE $j$. 

Let $T^{\star}(P, B)$ be the minimum time it takes to Reduce a vector of length $B$ between $P$ consecutive PEs. Let $E^{\star}(P, B, D)$ be the minimum energy needed for this reduction using depth at most $D$.

\begin{lemma}\label{lem:opt}
    $$E^{\star}(P, 1, D) \geq \min_{0 < i < P} E^{\star}(i, 1, D) + E^{\star}(P - i, 1, D - 1) + \min(i, P - i + 1)$$
\end{lemma}

\begin{proof}
In order to decompose the energy term, we consider a generalized problem, where the distance between the $j$-th and $j+1$-th PE is some integer $s_j\geq 1$. Then, $E^{\star}(i, 1, D, S)$ denotes the energy to perform an optimal reduce on $i$ PEs with depth at most $D$ where $S$ denote the sum of the distances $s_j$.
%
    Note that $S = i - 1$ if and only if all PEs are neighbouring. Let now $S = i - 1 + k$, we want to show that $E^{\star}(i, 1, D, S) \geq E^{\star}(i, 1, D, i - 1) + k = E^{\star}(i, 1, D) + k$. To see that the first inequality holds, consider the pattern used in $E^{\star}(i, 1, D, S)$. If we made all PEs neighbouring, i.e., decreased the total distance by $k$, the energy would need to decrease by at least $k$. This is because each link in the reduce (in the direction of the root) needs to be used at least once. By decreasing the sum of distances by $k$ we are shortening the links by a total of $k$. Since each of such link was used at least once, they contribute at least $k$ to the energy. The second equality holds because when all PEs are neighbouring we are in the usual cost setting. 

    Now, we are ready to derive the main recursion of the lemma. 
    Let the last message received by the root contain a partial sum of $P - i$ PEs for some $i$. Since this is the last message, the root must already have a partial sum of $i$ PEs. To reduce $i$ PEs with depth $D$, we need at least $E^{\star}(i, 1, D)$ energy. The reduction of $P - i$ PEs needs to be done with depth at most $D - 1$, which needs $E^{\star}(P - i, 1, D - 1)$ energy.    Let $S_3$ be the energy 
    of the last message. Then, for some total distances $S_1$ and $S_2$ we have:
    $$E^{\star}(P, 1, D) = E^{\star}(i, 1, D, S_1) + E^{\star}(P - i, 1, D - 1, S_2) + S_3 \enspace  .$$
Using our previous observation, we get
$$E^{\star}(P, 1, D) \geq E^{\star}(i, 1, D) + E^{\star}(P - i, 1, D - 1) + S_1 + S_2 + S_3 - (P - 1) \enspace . $$
    It remains to bound $S_1 + S_2 + S_3$. 
    Let $x_{1,1}$ and $x_{1,2}$ be the leftmost and rightmost PEs in the first reduction, respectively. Define $x_{2,1}$ and $x_{2,2}$ similarly for the second reduction. Observe that:
    $$S_1 + S_2 + S_3 = (x_{1,2} - 1) + (x_{2,2} - x_{2,1}) + x_{2,1}$$

    We know that $x_{1,2} \geq i$ and $x_{2, 2} \geq P - i + 1$. We know that either $x_{1,2} = P$ or $x_{2,2} = P$. Hence, we conclude that
    $$S_1 + S_2 + S_3 = P - 1 + \min(i, P - i + 1) \enspace .$$
\end{proof}

 We can compute the energy for reducing a scalar on $P$ PEs with depth at most $D$ in $O(P^2)$ with a dynamic programming approach. We use this result to bound the optimal runtime $T^{\star}$:

\begin{equation*}
\begin{split}
T^{\star}(P, B) &\geq \min_D \frac{E^{\star}(P, B, D)}{P - 1} + P - 1 + D(2T_R + 1)\\
&\geq \min_D B\frac{E^{\star}(P, 1, D)}{P - 1} + P - 1 + D(2T_R + 1)
\end{split}
\end{equation*}

The first inequality follows because we can omit contention when calculating the lower bound. The second inequality follows because the energy to Reduce a vector of length $B$ needs to be at least $B$ times the energy to Reduce a scalar. Solving the dynamic program takes $O(P^3)$ time.

\subsection{Comparison}

We run the predictions of the Auto-Gen reduce and compare them against the lower bound and the fixed patterns.
%
%
%
 Figure~\ref{fig:heatmap_specific} compares each pattern against the lower bound.  Star-Reduce is effective at $B=1$, Tree-Reduce is effective for slightly larger $B$, and Chain Reduce excels for large $B$. Finally, the Two-phase pattern is effective for intermediate vector sizes, that is when $P \approx B$. We can see that each pattern outside of its ideal range is often at least 3x worse than the best one. The Two-Phase pattern performs quite well throughout the whole range, although it is up to 2.4x away from the lower bound. Our Auto-Gen reduce strictly dominates all other patterns and is at most 1.4x away from our lower bound.


\section{1D Allreduce}\label{sec:allreduce}
We now consider different AllReduce patterns and analyze them using our performance model. We focus on AllReduce in a single row or column, but show in Section~\ref{sec:2d} how to generalize those ideas to the whole 2D grid.

\subsection{Reduce-then-Broadcast}
We first consider a Reduce-then-Broadcast implementation of AllReduce. Let us assume that we use a reduction pattern \textsc{Reduce}. The total predicted runtime is simply
$$T_{\textsc{Naive}} = T_{\textsc{Reduce}} + T_{\textsc{Bcast}} \enspace .$$

Note that this naive implementation could be further optimized by choosing an optimal root to reduce to. We could choose it based on our performance model. This is done in optimized stencil implementations~\cite{massively_scalable_stencil_cerebras}, in which they first reduce to the middle PE and broadcast from there.

\subsection{Ring AllReduce}
The main issue with a Reduce-then-Broadcast approach is that the root receives the whole vector and then sends it out, which has a runtime of at least $2B$ because of contention. This is suboptimal for larger vector lengths.

\begin{figure}[t]
    \centering
    \begin{subfigure}[b]{0.45\linewidth}
        \resizebox{\linewidth}{!}{  \centering
  \scalebox{1}{
    \begin{tikzpicture}[]
        \node[pe] (0) {0};
        \node[pe] (1) [right of=0]{1};
        \node[pe] (2) [right of=1]{2};
        \node[pe] (3) [right of=2]{3};
        \node[pe] (4) [right of=3]{4};
        \node[pe] (5) [right of=4]{5};






        \draw[red_arrow] ($(2) + (0.1, \length - \height)$) -- ++(0,-\length) -- ($(3) + (-0.1, -\height)$) -- ++(0,\length);

            \draw[red_arrow] ($(0) + (0, \length - \height)$) -- ++(0,-\length) -- ($(1) + (-0.1, -\height)$) -- ++(0,\length);

                        \draw[red_arrow] ($(4) + (0.1, \length - \height)$) -- ++(0,-\length) -- ($(5) + (0, -\height)$) -- ++(0,\length);

        \draw[blue_arrow] ($(1) + (0.1, \length - \height)$) -- ++(0,-\length) -- ($(2) + (-0.1, -\height)$) -- ++(0,\length);

                \draw[blue_arrow] ($(3) + (0.1, \length - \height)$) -- ++(0,-\length) -- ($(4) + (-0.1, -\height)$) -- ++(0,\length);

        \draw[blue_arrow] ($(5) + (0, \height - \length)$) -- ++(0,\length + 0.1) -- ($(0) + (0, \height + 0.1)$) -- ++(0,-\length - 0.1);

    \end{tikzpicture}
  }}
        \caption{simple}
        \label{fig:ring_simple}
    \end{subfigure}
    \hfill
    \begin{subfigure}[b]{0.45\linewidth}
        \resizebox{\linewidth}{!}{  \centering
  \scalebox{1}{
    \begin{tikzpicture}[]
        \node[pe] (0) {0};
        \node[pe] (1) [right of=0]{1};
        \node[pe] (2) [right of=1]{2};
        \node[pe] (3) [right of=2]{3};
        \node[pe] (4) [right of=3]{4};
        \node[pe] (5) [right of=4]{5};






        \draw[blue_arrow] ($(0) + (0, \height - \length)$) -- ++(0,\length) -- ($(2) + (0, \height)$) -- ++(0,-\length);

        \draw[red_arrow] ($(2) + (0, \length - \height)$) -- ++(0,-\length) -- ($(4) + (0, -\height)$) -- ++(0,\length);
        \draw[red_arrow] ($(1) + (0, \length - \height)$) -- ++(0,-\length) -- ($(0) + (0, -\height)$) -- ++(0,\length);

        \draw[blue_arrow] ($(4) + (0, \height - \length)$) -- ++(0,\length) -- ($(5) + (0, \height)$) -- ++(0,-\length);

        \draw[blue_arrow] ($(5) + (0.1, \height - \length)$) -- ++(0,\length + 0.15) -- ($(3) + (0, \height + 0.15)$) -- ++(0,-\length - 0.15);

        \draw[blue_arrow] ($(3) + (-0.1, \height - \length)$) -- ++(0,\length + 0.15) -- ($(1) + (0, \height + 0.15)$) -- ++(0,-\length - 0.15);

    \end{tikzpicture}
  }}
        \caption{distance preserving}
        \label{fig:ring}
    \end{subfigure}
    \vspace{-1em}
    \caption{Different ring pattern implementations.}
    \label{fig:allreduce_patterns}
\end{figure}
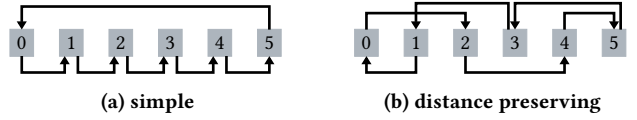

To address this problem, we consider the ring AllReduce \cite{DBLP:conf/sc/HoeflerBSGLHBGCS22, DBLP:journals/jpdc/BarnettLPG95} pattern. Because the network is a mesh and not a torus, we cannot have a ring in which a PE communicates with its nearest left and right neighbours. Instead, we propose two different mappings, which as we show result in the same predicted performance. The simplest way to map a ring is to to have each PE receive from its left neighbour and send to its right neighbour. Since the rightmost PE does not have a right neighbour, it sends a message to the leftmost PE in the row. See \Cref{fig:ring_simple} for an illustration. A problem with this design could be that the longest link is a bottleneck. We can also map a ring such that a PE will be communicating with PEs at a distance of at most two. See Figure~\ref{fig:ring} for an illustration. Notice that in both patterns we are utilizing bidirectional links.


The ring AllReduce first performs $P - 1$ rounds of reduce-scatter, after which each PE has a $\frac{B}{P}$ elements of the final vector. It then executes $P - 1$ rounds of allgather, after which each PE has the final vector. Let us assume that $B$ is divisible by $P$. In each round a PE sends a vector of length $\frac{B}{P}$ and receives a vector of length $\frac{B}{P}$.

\begin{lemma}
$$T_{\textsc{Ring}} = \frac{2(P - 1)B}{P} + 4P - 6 + 2(P - 1)(2T_R + 1)$$
\end{lemma}
\begin{proof}
We analyse the two mappings together. The depth is $2(P - 1)$, because each round depends on the previous one. In the first $P - 1$ rounds a wavelet traverses the whole ring minus one link. Since the ring maps $2(P - 1)$ links and we have two rounds, we get a distance of $2\cdot(2P - 3)$. In each round of the algorithm, we have $\frac{B}{P}$ wavelets travelling over each link. Since we have $2(P -1)$ links and $2(P - 1)$ rounds, the energy is $2(P - 1)\cdot \frac{2(P - 1)B}{P}$. The contention is $\frac{2(P - 1)}{P}$. Notice that in this case, the number of links is $2(P - 1)$ instead of $P - 1$, because we are using bidirectional links. 
\end{proof}

\subsection{Comparison}

\begin{figure}
\begin{center}
\includegraphics[width=0.5\textwidth, trim={1cm 1cm 11cm 8cm}, clip]{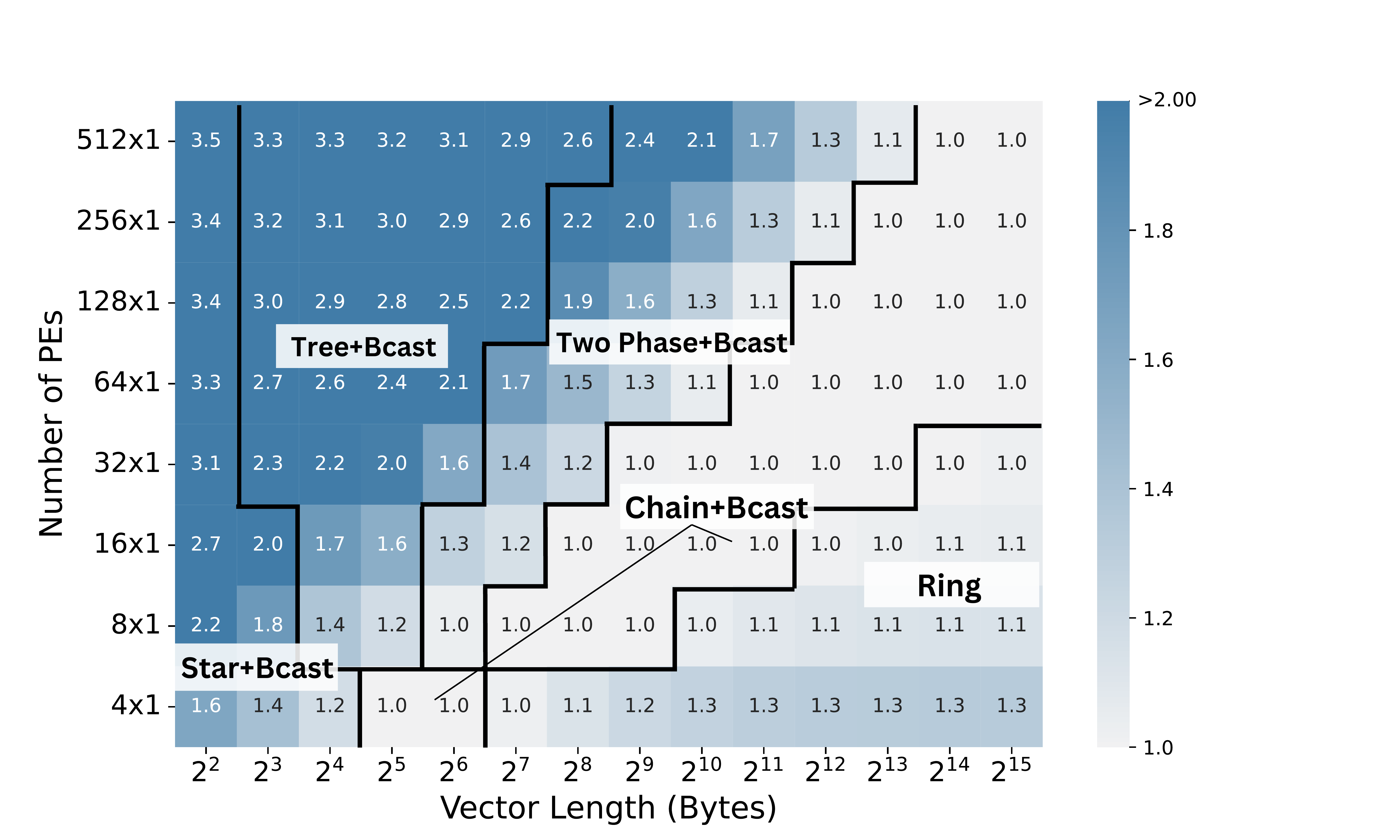}
\end{center}
\vspace{-1em}
\caption{Speedup of \emph{1D AllReduce} algorithm over Chain+Bcast, which is used by the vendor. The regions indicate which of the algorithms is the best fixed algorithm for the given combination of vector length and PE count.}
\label{fig:heatmap_1d_allreduce}
\end{figure}

Analytically, just like for reduce, we can determine the best choice of algorithm for a given $B$ and $P$. We plot the result in the heatmap \Cref{fig:heatmap_1d_allreduce}. It shows which algorithm the model predicts to perform best for a given combination of vector size $B$ and PE count $P$. 
As we could expect, different Reduce-then-Broadcast AllReduce patterns perform best for the same parameters as the underlying Reduce does. However, there is a part where the Chain+Broadcast is outperformed by the ring pattern. This is when the runtime is dominated by the contention due to a large vector length.

\section{2D Collectives}\label{sec:2d}
We discuss how to design collectives in a 2D setting. One problem we encounter is that the CS-2 has only one port from the processor to the router. This means we cannot e.g. send one packet on the y-axis and another on the x-axis each cycle. We would need to alternate between them each cycle. This means that in contrast to other works on AllReduce ~\cite{swing} where usually multiple ports per dimension are considered, we benefit less from the 2D setting. However, we still show how certain collectives, such as broadcast greatly benefit from this setting. In this section we assume that dimensions of the PE grid are $M \times N = P$.

\subsection{2D Broadcast}
Let us first analyse a broadcast where the root has position $(0, 0)$. A 2D broadcast can be implemented by performing 1D broadcast on the x-axis and multicasting to perform it simultaneously on the y-axis.
We get the following bound:
\begin{lemma}
$T_{\textsc{2D Broadcast}} = B + M + N - 2 + 2T_R + 1$
\end{lemma}
\begin{proof}
The depth is $1$. The distance is $M + N - 2$. The energy is $B(P - 1)$ with $P - 1$ links used. The contention is $B$.
\end{proof}
This means that if we had $P$ processes in a $\sqrt{P} \times \sqrt{P}$ grid, the broadcast would take $2\sqrt{P} + 2T_R - 1 + B$. This is much more efficient compared to the row broadcast on $P$ processes. Specifically, we see that there is a lot to benefit from the 2D setting.

\subsection{X-Y Reduce}
The simplest approach to performing 2D reduce, would be utilizing our existing 1D implementations. We can perform reduce on the x-axis followed by reduce on the y-axis. See Figure~\ref{fig:xyreduce} for an illustration. Then, the predicted runtime is going to be:

$$T_{\textsc{Reduce X}} + T_{\textsc{Reduce Y}}$$
\subsection{Snake Reduce}
\begin{figure}[t]
    \centering
    \begin{subfigure}[b]{0.3\textwidth}
        \centering
        \resizebox{0.7\linewidth}{!}{  \centering
  \scalebox{1}{
          \begin{tikzpicture}[]

    \def\nodeDist{0.4}    
    \foreach \x in {0,...,3}
    \foreach \y in {0,...,3} {
        \node[pe] (pe\x\y) at (\nodeDist*\x,\nodeDist*\y) {};
    }

    \draw[blue_arrow] (pe33.west) -- (pe03.east) node[midway, above, yshift=0.5mm] {\textbf{Reduce X}};;
                \node at (current bounding box.south) [below, yshift=-1mm] {Phase 1};
    \end{tikzpicture}
    \hspace{2mm}
    \begin{tikzpicture}[]

    \def\nodeDist{0.4}    
    \foreach \x in {0,...,3}
    \foreach \y in {0,...,3} {
        \node[pe] (pe\x\y) at (\nodeDist*\x,\nodeDist*\y) {};
    }

                    \node at (current bounding box.south) [below, yshift=-1mm] {Phase 2};
    \draw[blue_arrow] (pe00.north) -- (pe03.south) node[midway, sloped, allow upside down, above,yshift=0.5mm] {\textbf{Reduce Y}};
    \draw[blue_arrow] (pe10.north) -- (pe13.south);
    \draw[blue_arrow] (pe20.north) -- (pe23.south);
    \draw[blue_arrow] (pe30.north) -- (pe33.south);
    



    




    \end{tikzpicture}

  }}
        \caption{X-Y Reduce}
        \label{fig:xyreduce}
    \end{subfigure}
    \begin{subfigure}[b]{0.17\textwidth}
        \centering
        \resizebox{0.55\linewidth}{!}{  \centering
  \scalebox{1}{
    \begin{tikzpicture}[]

    \def\nodeDist{0.4}    
    \foreach \x in {0,...,3}
    \foreach \y in {0,...,3} {
        \node[pe] (pe\x\y) at (\nodeDist*\x,\nodeDist*\y) {};
    }

    \draw[blue_arrow] (pe00.east) -- (pe10.west);
    \draw[blue_arrow] (pe10.east) -- (pe20.west);
    \draw[blue_arrow] (pe20.east) -- (pe30.west);
    \draw[blue_arrow] (pe30.north) -- (pe31.south);
    
    \draw[blue_arrow] (pe32.north) -- (pe33.south);

    \draw[blue_arrow] (pe02.east) -- (pe12.west);
    \draw[blue_arrow] (pe12.east) -- (pe22.west);
    \draw[blue_arrow] (pe22.east) -- (pe32.west);
    \draw[blue_arrow] (pe01.north) -- (pe02.south);

    \draw[blue_arrow] (pe31.west) -- (pe21.east);
    \draw[blue_arrow] (pe21.west) -- (pe11.east);
    \draw[blue_arrow] (pe11.west) -- (pe01.east);

    \draw[blue_arrow] (pe33.west) -- (pe23.east);
    \draw[blue_arrow] (pe23.west) -- (pe13.east);
    \draw[blue_arrow] (pe13.west) -- (pe03.east);
        \node at (current bounding box.south) [below, yshift=-2mm] {};
    \end{tikzpicture}
  }}
        \caption{Snake Reduce}
        \label{fig:snake}
    \end{subfigure}
    \vspace{-2em}
    \caption{2D Reduce patterns.}
\end{figure}
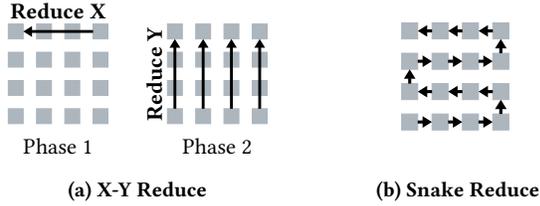

One problem with the previous approach is that the root PE will receive the vector at least twice, which is sub-optimal for $B \gg P$. We know that in the 1D setting, the chain pattern performs best in this case. We therefore propose to map the chain implementation in a snake-like pattern. See Figure~\Cref{fig:snake} for an illustration. Notice that this way the runtime is going to be the same as $T_{\textsc{chain}}$.

\subsection{2D Allreduce}
The simplest approach is executing AllReduce  on the x-axis followed by AllReduce on the y-axis. Then, the predicted runtime is:
$$T_{\textsc{allreduce x}} + T_{\textsc{allreduce y}}$$

By performing AllReduce  on the x-axis and y-axis, we are essentially going to be broadcasting twice, which is very bandwidth inefficient. Remember that we have a very efficient broadcast implementation. To improve the 2D AllReduce, we could perform first a 2D Reduce and then a 2D broadcast. The runtime is:
$$T_{\textsc{Allreduce}} = T_{\textsc{2D Reduce}} + T_{\textsc{2D Broadcast}}$$

\subsection{Lower Bound}

We provide a simple lower bound for the general 2D Reduce. Let $T^{\star}(M, N)$ be the time of an optimal Reduce on an $P= M \times N$ grid of PEs. 
\begin{lemma}
$T^{\star}(M, N) \geq \max\left(B, \frac{B}{8} + M + N -1\right) + 2T_R + 1$
\end{lemma}
\begin{proof}
    The contention is at least $B$ because if the root receives less than $B$ values there is no way to construct the result. Similarly, the energy is at least $PE$ because every PE has to send a value for each of the entries in the vector. There are at most $8P$ bidirectional links in a 2D grid of $P$ PEs. The distance is at least $M+N-1$ (from the bottom-right PE to the top-left PE). Finally, the depth is trivially at least $1$. The result follows by combining the terms.
\end{proof}
When comparing with the lower bound, we see that for $B \gg P$, the snake Reduce is optimal. When the vector length $B$ does not dominate the number of PEs $P$, there is room for improvement.

\subsection{Discussion}

\begin{figure}
\includegraphics[width=0.5\textwidth, trim={0cm 1cm 11cm 8cm}, clip]{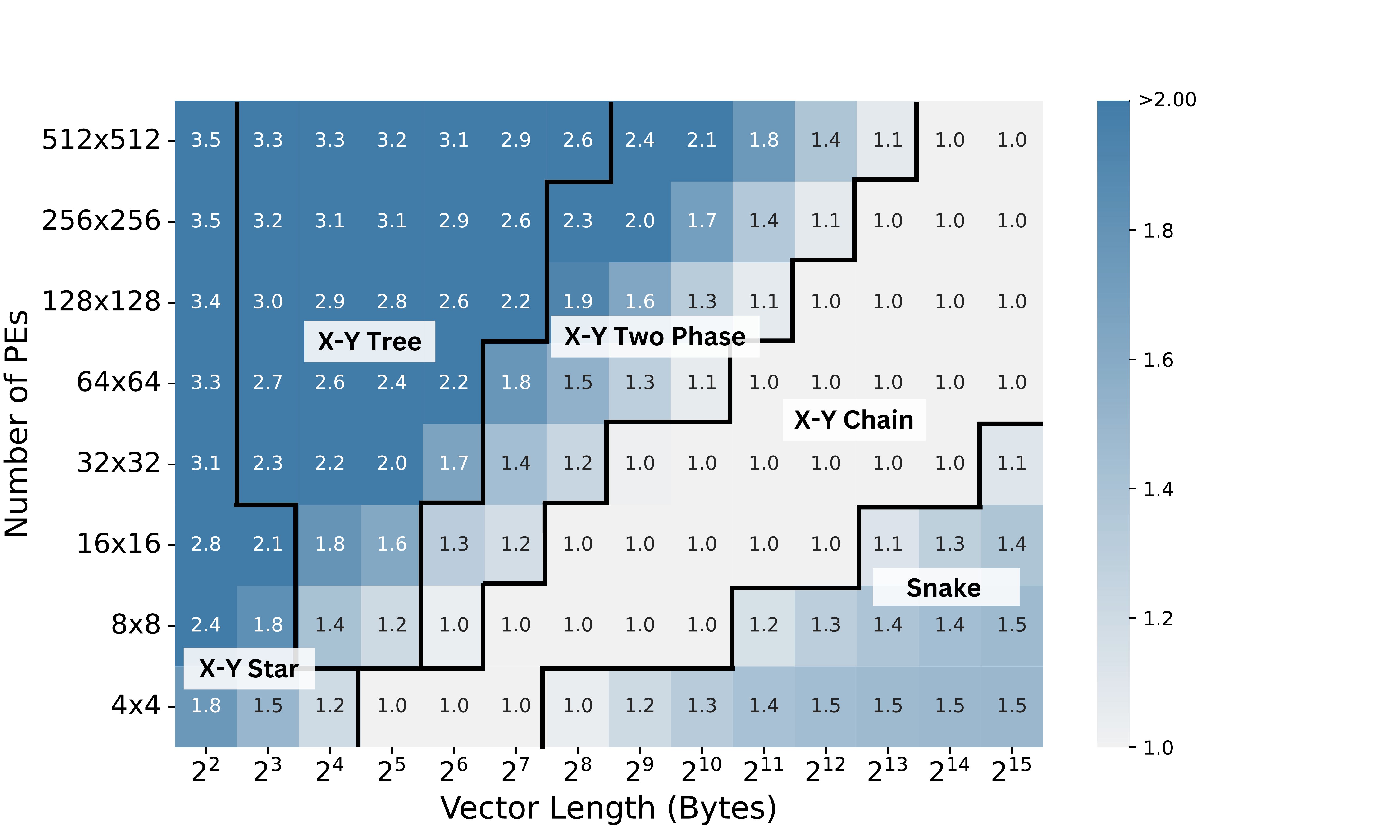}
\vspace{-1em}
\caption{Speedup of \emph{2D AllReduce} algorithm over the X-Y Chain, which is used by the vendor. The regions indicate which of the proposed algorithms is the best fixed algorithm for the given combination of vector length and PE count.}
\label{fig:2d_allreduce_heatmap}
\end{figure}

Analytically, we can determine the best choice of algorithm for a given $B$ and $P$. We plot the resulting heatmap in \Cref{fig:2d_allreduce_heatmap}, where we show which algorithm we predict to perform best for a given combination of vector size $B$ and PE count $P$. As expected, the results are very similar to what we observed in the 1D setting. However, here the bandwidth-limited area is occupied by the ring AllReduce in 1D is replaced by the snake pattern in 2D. 

\section{Experiments}\label{sec:experiments}
We evaluate the performance of different collectives for a row of PEs as well as a square grid of PEs. For each of the collective, we perform two types of experiments. All data types are 32-bit floating point numbers. In the first experiment type, we fix the number of PEs $P$ to be the maximum row length or grid size that is still a power of two and vary the vector length $B$. A large grid size caters to the fact that HPC applications need to utilize most of the PEs to achieve the best performance. The second experiment type fixes the vector length $B$ to 256 values (1 KB) and varies the number of PEs in a row or grid. This allows us to see how performance and our predictions are impacted by the number of PEs.


\subsection{Benchmarking} All benchmarks are on a CS-2 running at 850 MHz with $40$GB of on-chip SRAM. 
We run each benchmark $5$ times and plot the mean runtime. The observed standard deviation is negligible (<~4\%)~\cite{scbench}. Five measurements is enough because the CS-2 is mostly deterministic unlike traditional architectures. There is no caching and memory access time is deterministic. The time to travel between the routers is also always 1 cycle. One source of non-determinism is that PEs may insert no-ops to regulate thermal stress of the wafer. Additionally, although the cores all run at around 850 MHz, they are truly independent cores, with independent clocks. For these reasons we still see some deviation from the mean. The performance will also depend the specific CS-2 chip. If there are any defects, a proprietary process will route around them. 

\subsection{Implementation.} We implement all algorithms with the newest version of the Cerebras SDK 1.0~\cite{Cerebras-SDK} and runtime. For our Auto-Gen reduce, we compute the necessary parameters in Python. Based on that we generate the source code for each PE. Note that we provide our own (equivalent) implementation of Chain Reduce instead of the one provided in the SDK. A library call would cause reconfiguration of the routers which would yield artificially slow results. 

When implementing collectives for the CS-2 it is important to limit the number of colors as there are only 24 of them. Our 1D implementations utilize up to 3 colors, while the 2D implementations use up to 5. When using our collectives, the rest of the colors would be available to the application. 
The routing for the CS-2 is requires avoiding race conditions. Having two wavelets arrive at a router on the same color in the same cycle leads to undefined behaviour. To avoid this, we configure the routers such that at a given cycle they accept wavelets only from a single direction. We do this using control wavelets which alter the routing configuration at runtime.

%

\subsection{Time Measurements}
\begin{figure*}[ht]
    \centering
    \begin{subfigure}[b]{0.33\textwidth}
        \includegraphics[width=\textwidth]{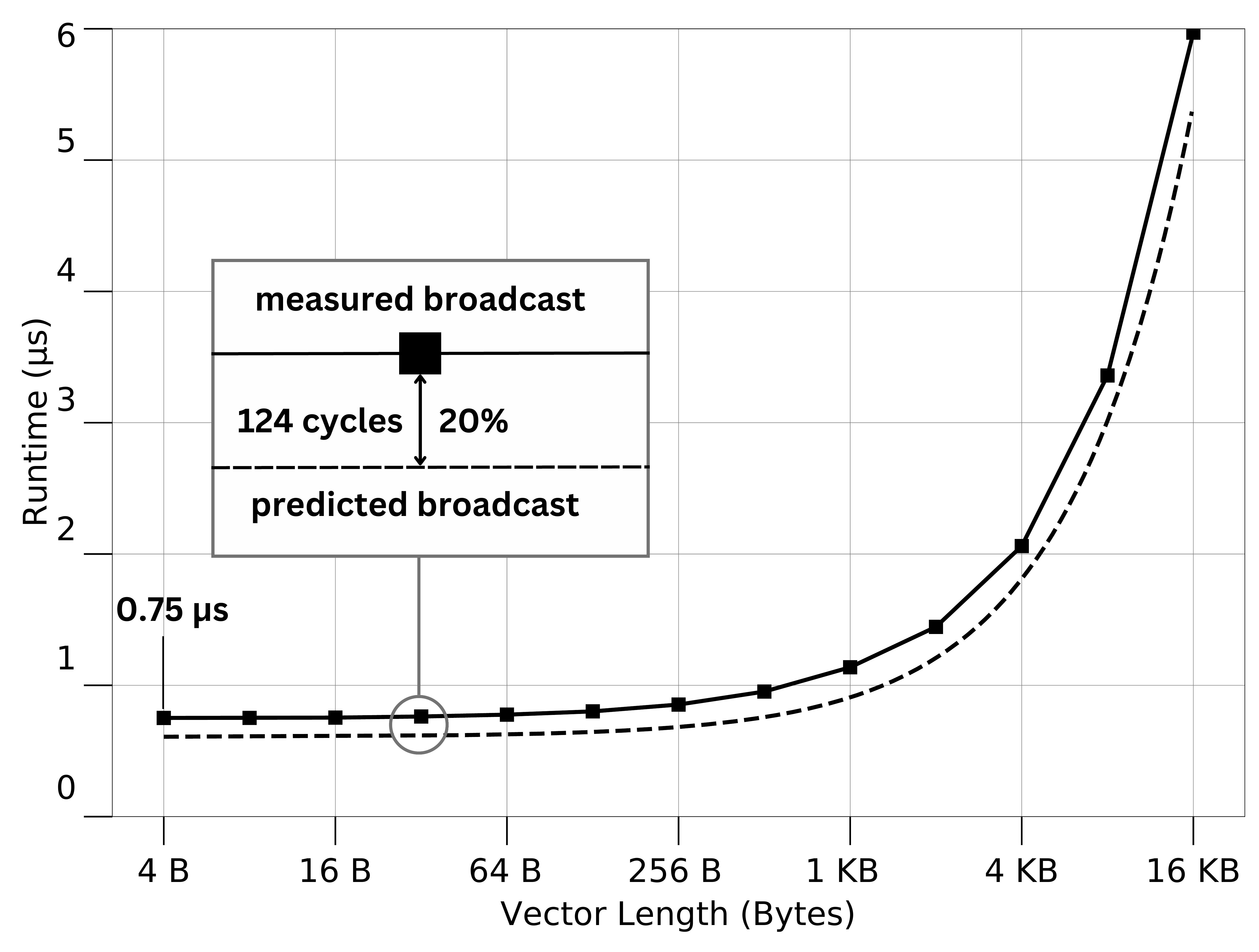}
        \caption{Broadcast}
        \label{fig:broadcast_fix}
    \end{subfigure}
    \begin{subfigure}[b]{0.33\textwidth}
        \includegraphics[width=\textwidth]{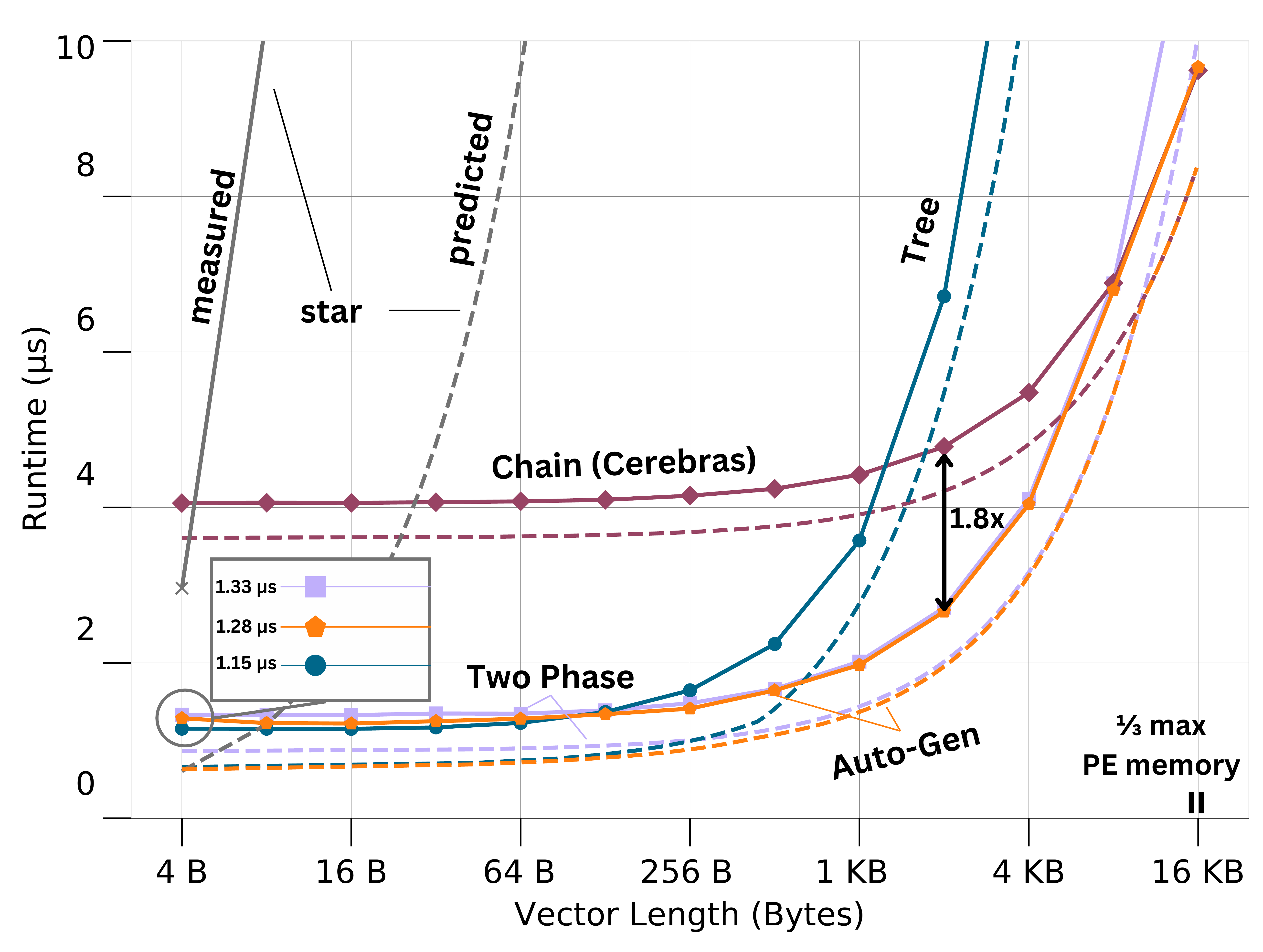}
        \caption{Reduce}
        \label{fig:reduce_fix}
    \end{subfigure}
    \begin{subfigure}[b]{0.33\textwidth}
        \includegraphics[width=\textwidth]{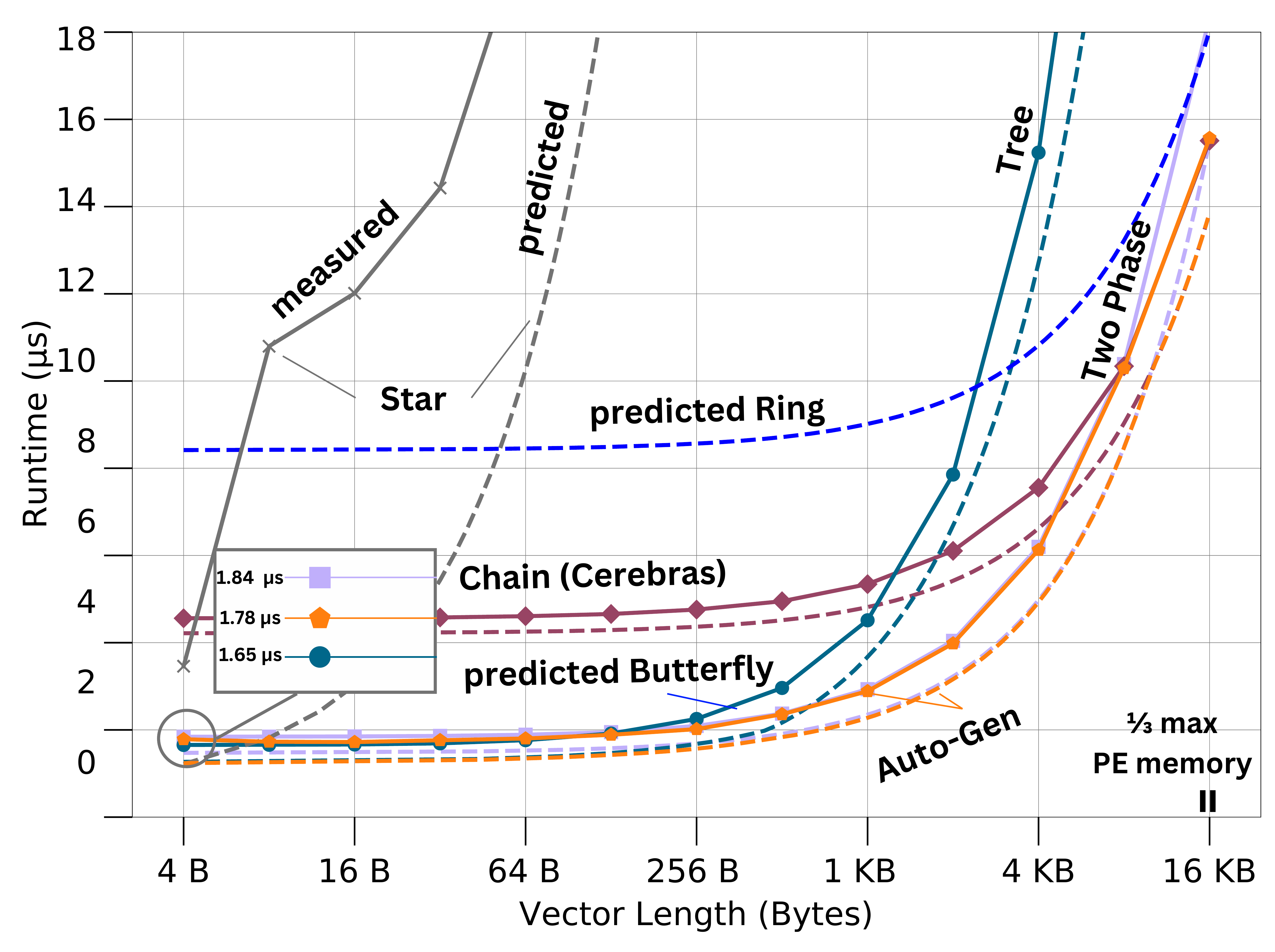}
        \caption{AllReduce}
        \label{fig:allreduce_fix}
    \end{subfigure}
    \vspace{-1.5em}
    \caption{Benchmarks for 1D row of 512x1 PEs and increasing vector length.}
    \vspace{0.5em}
    \label{fig:1d_fix}
\end{figure*}

Time measurements in distributed systems have been extensively studied~\cite{hoefler-collmea-sync, collective_benchmarking} and pose challenges, such as lack of a shared clock. A proper measurement methodology is particularly important for the sub-microsecond runtimes we observe. For broadcast, the methodology is easier because the computation starts from a single root, which synchronizes the start time. However, for Reduce we need to synchronize the clocks and ensure that each processor starts at the same time. For all measurements, we start by performing a Reduce to the PE at position $(0,0)$. This constitutes a barrier and ensures that all processes have no other ongoing computations.

To measure 1D broadcast performance, we use a ping-pong like approach. We execute a broadcast from the leftmost PE, then from the rightmost PE. We repeat this procedure $k$ times and report the end clock time - start clock time at the leftmost PE divided by $2k$. 

For Reduce and AllReduce we need a different approach, because the execution might start at multiple PEs. Hence, we need to ensure that all PEs start execution at the same time. We perform several calibration runs until a synchronized clock tells us that all PEs start at roughly the same time. The calibration adjusts a so-called \emph{wait parameter} $\alpha$ until the condition is satisfied. Initially, $\alpha=1$.

First, the PE at position $(0,0)$ performs a broadcast that triggers each PE at position $(i,j)$ to sample its local reference clock, called $T_R(i,j)$. Now, each PE $(i, j)$ executes $\alpha(M + N - i - j)$ writes to an empty memory location. After that, each PE samples the start clock $T_S(i,j)$, performs the collective and samples the end clock $T_E(i,j)$. For each PE we calibrate the start and end clock as follows:
\begin{align*}
T_S(i,j)' &= T_S(i,j) - (T_R(i,j) + (i + j + 2)) \\
T_E(i,j)' &= T_E(i,j) - (T_R(i,j) + (i + j + 2))
\end{align*}
This accounts for the difference in time when a PE samples the reference clock, which is $i+j+2$ for a PE at position $(i, j)$ as we use a broadcast to initiate the sampling. 
We adjust the wait parameter $\alpha$ and repeat the calibration until the difference in calibrated start times $\max_{i, j} T_S(i,j)' - \min_{i, j} T_S(i, j)'$ is small enough. We obtain a start difference below $57$ cycles for 1D and $129$ cycles for 2D Then, the final measurement is $\max_{i, j} T_E(i, j)' - \min_{i, j} T_S(i, j)'$. This methodology ensures accurate measurements in the wafer-scale setting. In an ideal system $\alpha = 1$ would make all PEs start at the same time since each write takes 1 cycle. However, in order to prevent the PE from overheating, the machine will start inserting no-ops, which we need to adjust for.

\subsection{1D Broadcast}

\textbf{Scaling Vector Length.}
Figure \ref{fig:broadcast_fix} shows the broadcast results for 512 PEs and increasing vector length. As expected, for small vector lengths the runtime is dominated by the distance. Hence, the runtime only grows slowly with the vector length. For vector lengths larger than $512$ bytes, the runtime grows roughly linearly with the vector length. The model matches the predictions closely, with a relative error of at most 21\%.

\begin{figure*}[t]
    \centering
    \begin{subfigure}[b]{0.33\textwidth}
        \includegraphics[width=\textwidth]{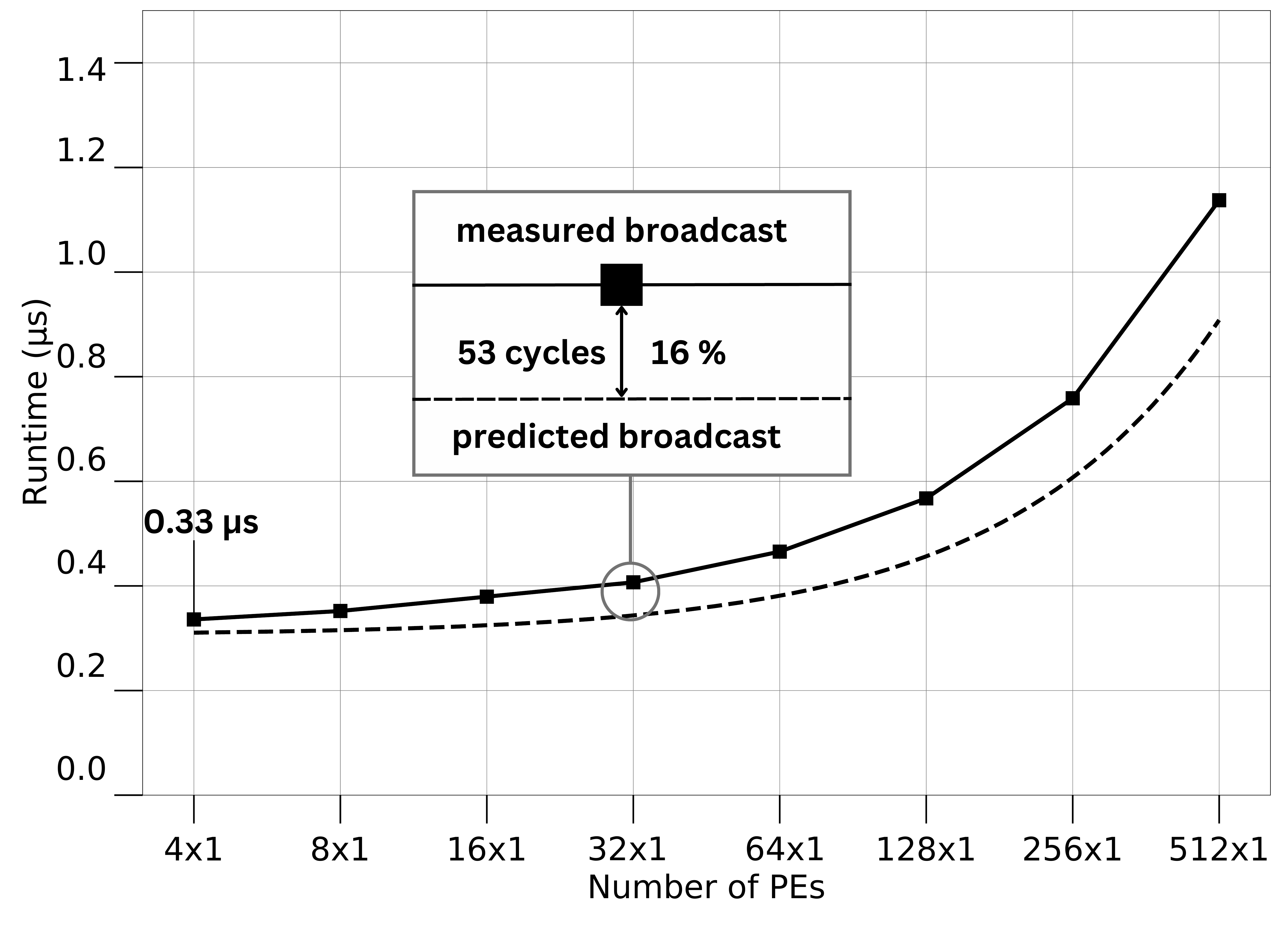}
        \caption{Broadcast}
        \label{fig:broadcast_inc}
    \end{subfigure}
    \begin{subfigure}[b]{0.33\textwidth}
        \includegraphics[width=\textwidth]{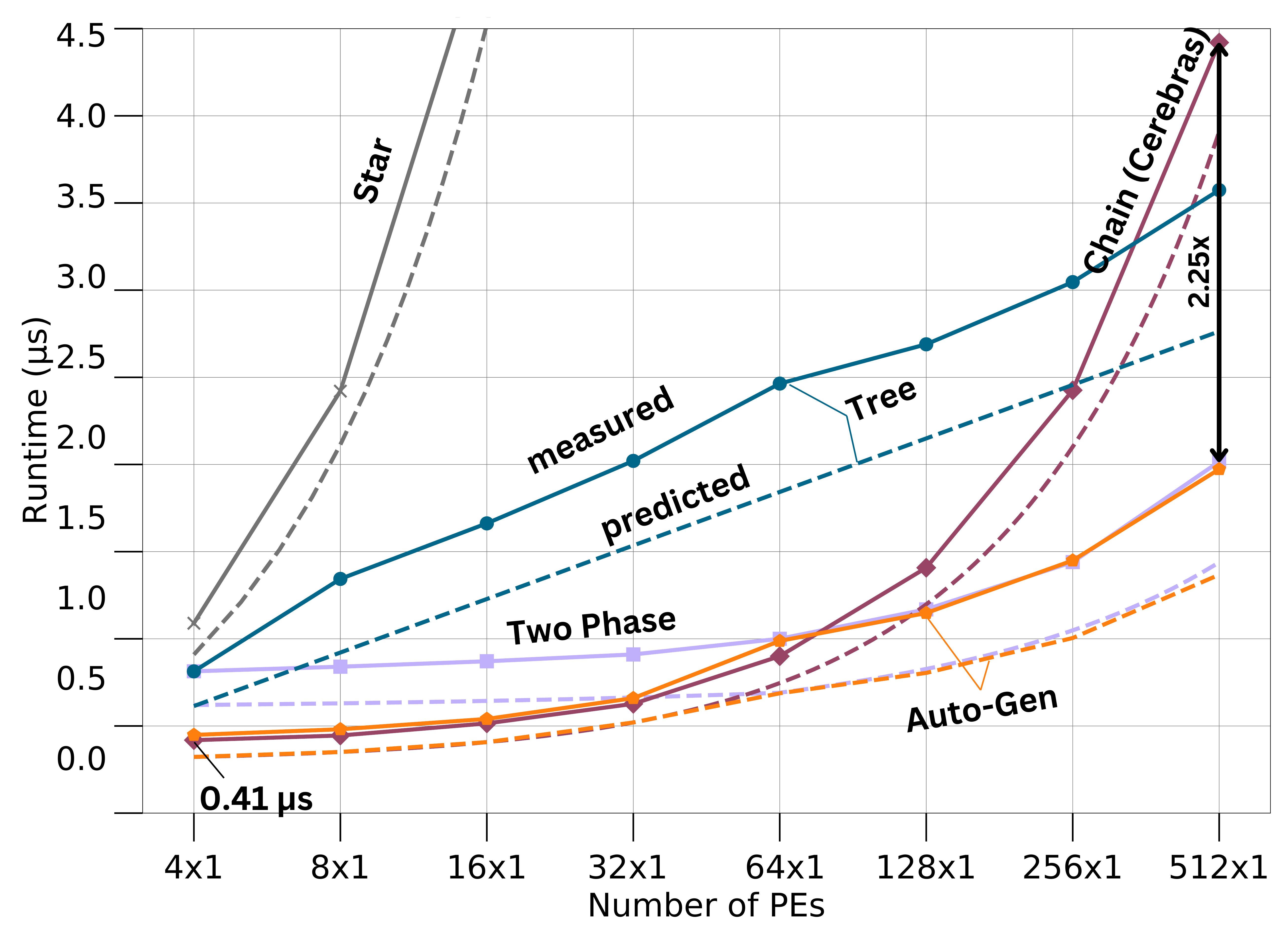}
        \caption{Reduce}
        \label{fig:reduce_inc}
    \end{subfigure}
    \begin{subfigure}[b]{0.32\textwidth}
        \includegraphics[width=\textwidth]{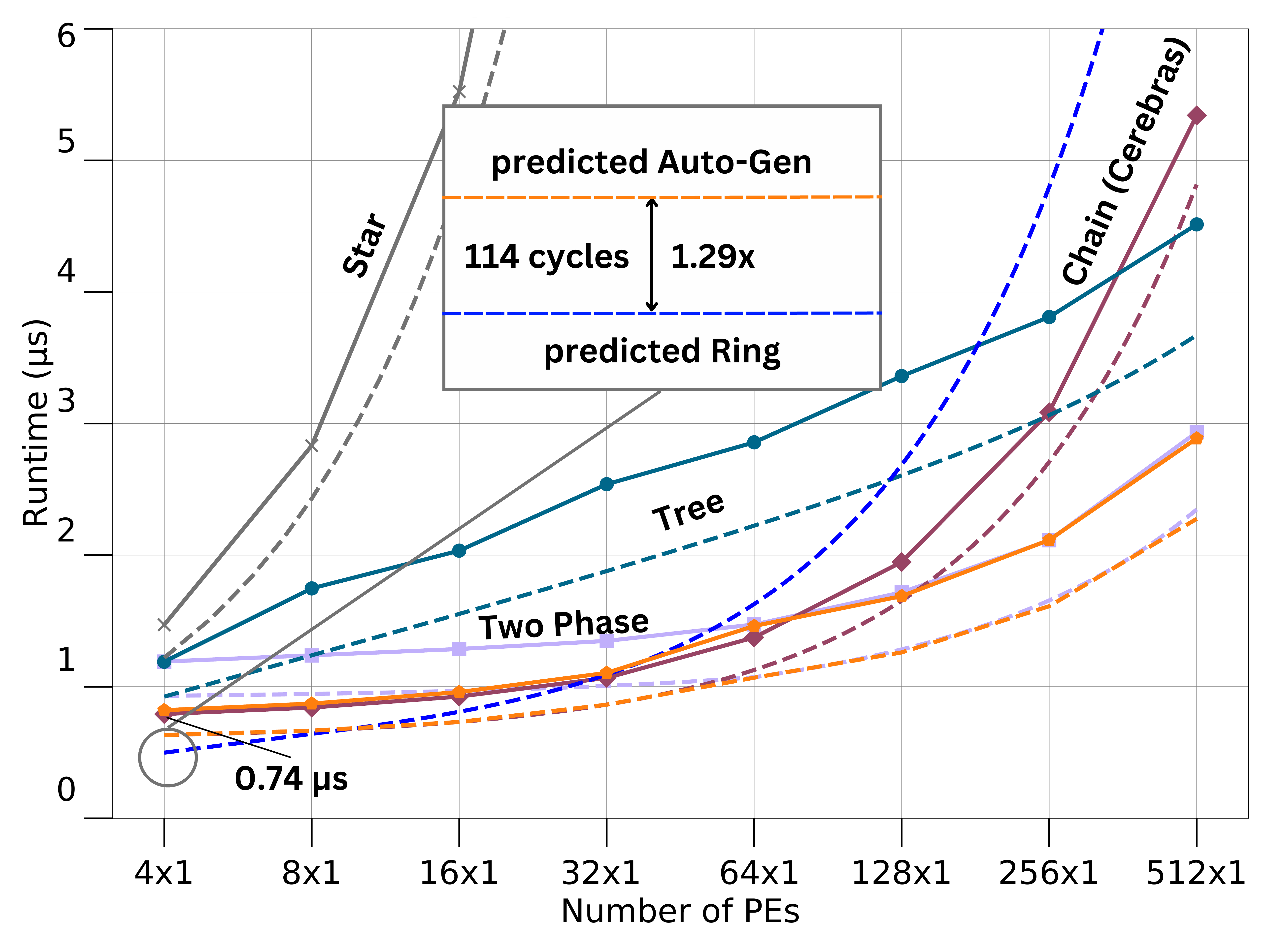}
        \caption{AllReduce}
        \label{fig:allreduce_inc}
    \end{subfigure}
    \vspace{-0.5em}
    \caption{Benchmarks for fixed vector length of 1 KB and increasing number of PEs.}
    \label{fig:1d_inc}
\end{figure*}

\textbf{Scaling PE count.}
\Cref{fig:broadcast_inc} shows the results for fixed vector length of 1 KB and increasing number of PEs. Our model predicts the correct trend, with a large initial runtime that accounts for sending the message (contention or energy term) and a gradually increasing contribution of the distance term. The relative error of the prediction is 8\%-21\%.

\subsection{1D Reduce}

\begin{figure*}[!t]
    \centering
    \begin{subfigure}[b]{0.33\textwidth}
        \includegraphics[width=\textwidth]{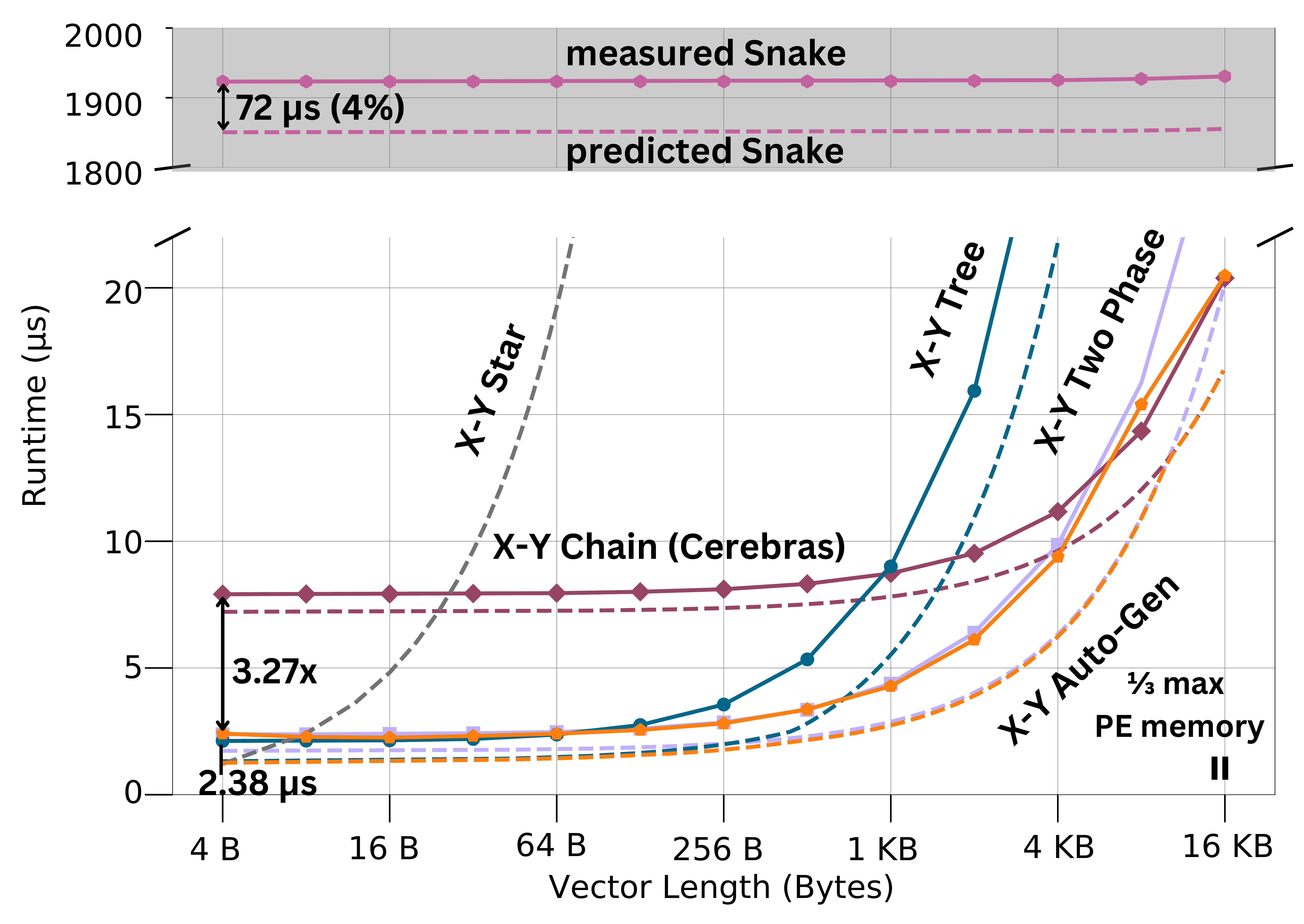}
        \caption{2D Reduce for $512\times 512$ PEs.}
        \label{fig:experiments-2d-reduce-fix}
    \end{subfigure}
    \begin{subfigure}[b]{0.33\textwidth}
        \includegraphics[width=\textwidth]{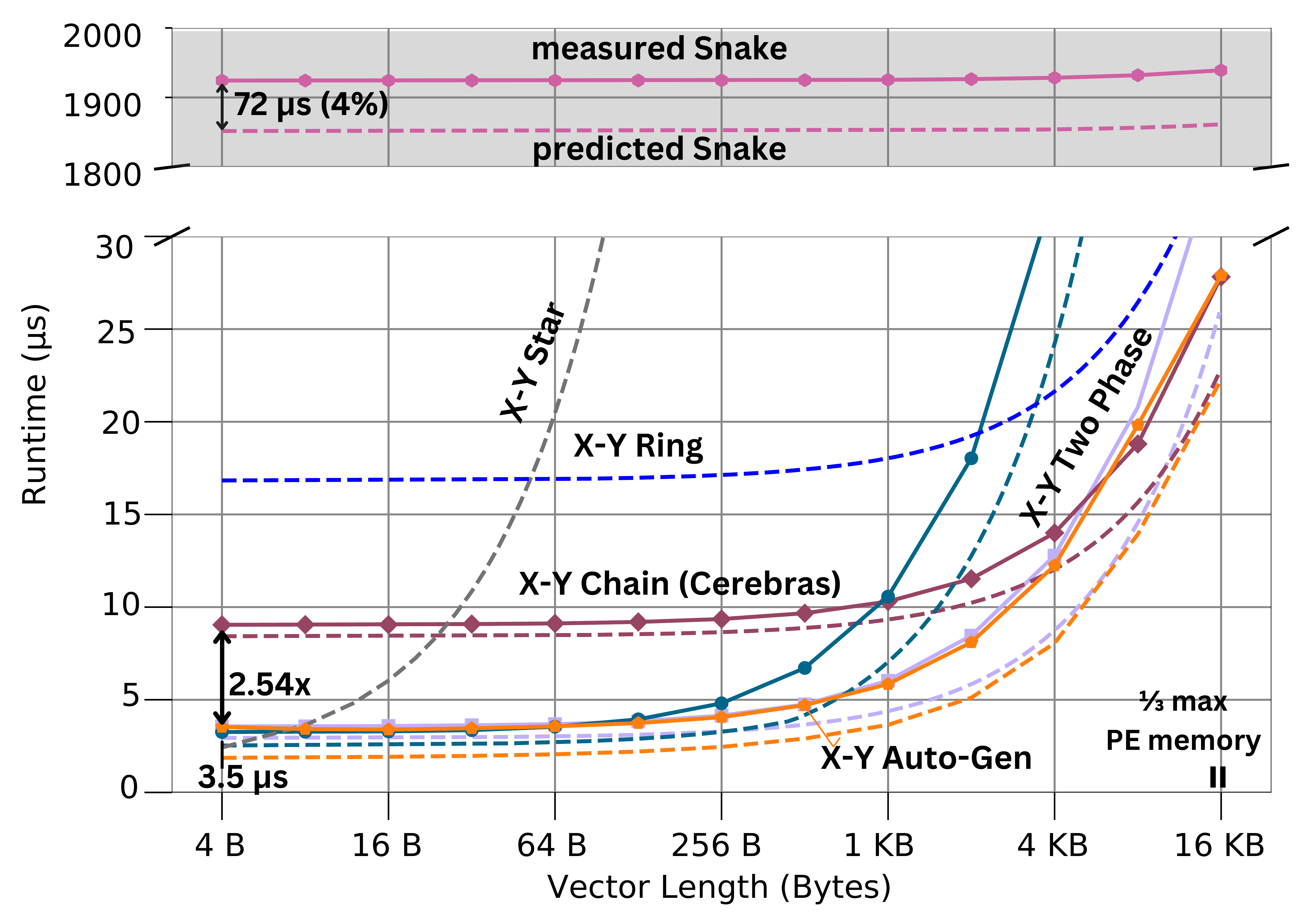}
        \caption{2D AllReduce for $512\times 512$ PEs.}
        \label{fig:experiments-2d-allreduce-fix}
    \end{subfigure}
    \begin{subfigure}[b]{0.33\textwidth}
        \includegraphics[width=\textwidth]{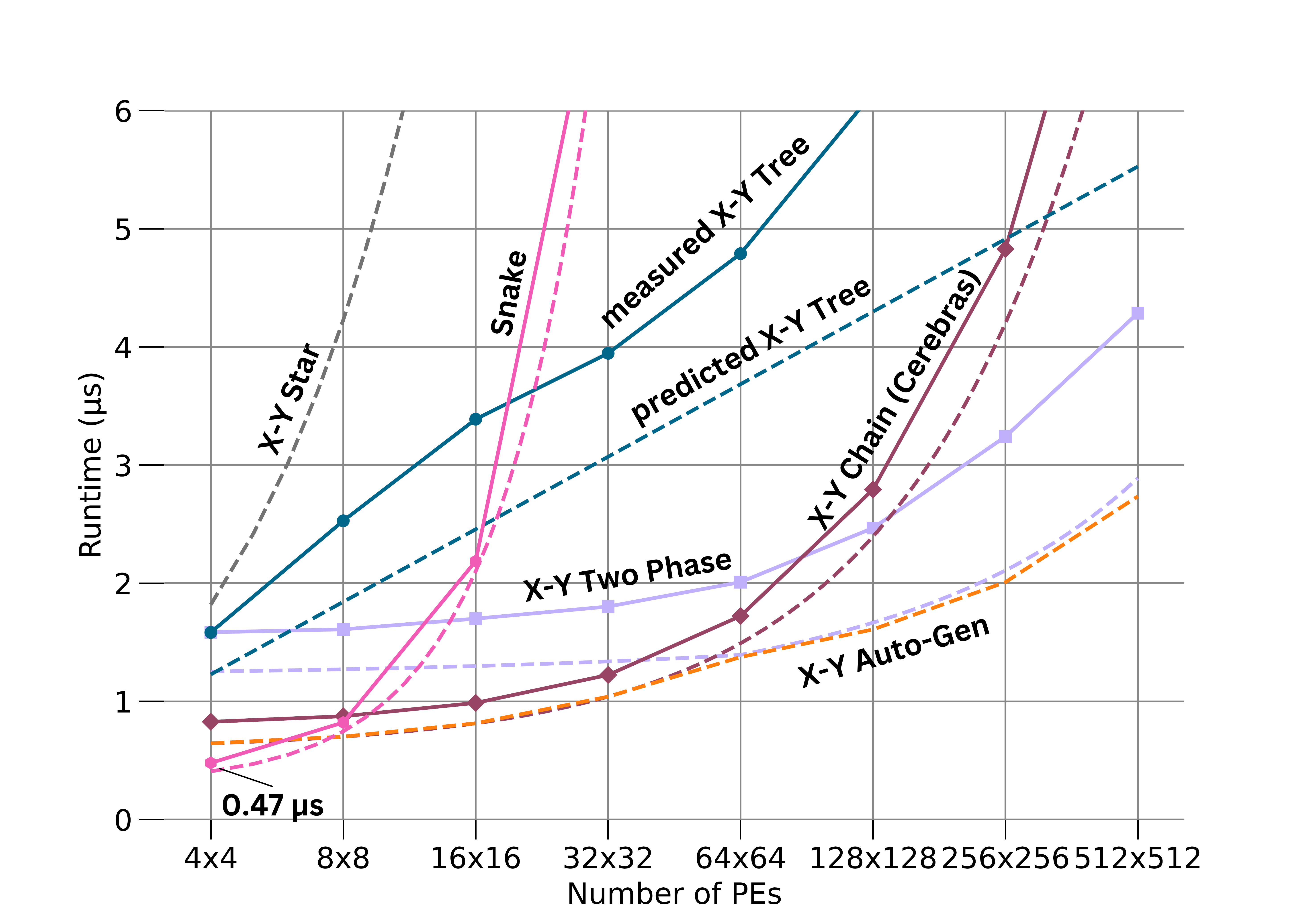}
        \caption{2D Reduce for vector length of 1KB.}
        \label{fig:experiments-2d-reduce-inc}
    \end{subfigure}
    \vspace{-1.5em}
    \caption{2D Reduce and AllReduce benchmarks.}
    \vspace{1em}
    \label{fig:experimental_plots_2d}
\end{figure*}

\textbf{Scaling Vector Length.}
We evaluate Reduce for 512 PEs with increasing vector length in~\Cref{fig:reduce_fix}. As predicted, among the fixed implementations, low depth patterns like the tree excel for small vector lengths, while those with greater depth, like the chain pattern, show inferior performance. 
With increasing vector lengths, energy and contention begin to outweigh depth in impact, resulting in the Two-Phase pattern outperforming others. Finally, for the largest vectors, the contention dominates the runtime and the chain pattern performs best.

The only exception to the above mentioned predictions is the star pattern. It performs worse than predicted in this scenario. This is likely because of the overhead associated when a PE starts receiving from another PE.  Because the star pattern receives elements from all PEs, this is much more pronounced than for other implementations, especially for large number of PEs. Except for scalars, our experiments and model both suggest that other low-depth patters, such as Tree or our Two-Phase algorithm, are faster.


Overall, our Auto-Gen Reduce is the fastest pattern except when reducing a scalar. There, it is slower by at most $110$ cycles. It outperforms the chain pattern, based on the current library implementation, by up to $3.16$x. This further confirms the choice of our model as a tool for automatic performance tuning.

With mean relative error per pattern ranging from $12\%$ to $35\%$, the general performance trends are well captured by our model. When choosing two different patterns, our model is able to very accurately predict which of the two performs best for the given vector length and number of PEs. In the cases where it mispredicts, the difference is at most $114$ cycles. This means that, even if the model's predictions are not perfect, the chosen algorithm remains highly competitive and close to the best possible performance.

\textbf{Scaling PE count.}
 Additionally, we evaluate the Reduce operation for a fixed vector length of 1 KB across an increasing number of PEs, as illustrated in \Cref{fig:reduce_inc}. Our model correctly predicts that initially the chain pattern performs best because with very few PEs, contention has a larger impact than the depth. With increasing number of PEs, the depth becomes more significant, and, as expected, the two phase pattern performs better. 

Just as for the results with the fixed number of PEs, we see that our Auto-Gen Reduce implementation is the fastest throughout. Two-Phase offers similar performance as Auto-Gen for $64$ or more PEs. Interestingly, predictions for the star pattern have high accuracy, with a $10\%$ minimum relative error. This is likely because the runtime is dominated by the vector length rather than the number of PEs in this case. Overall, our model predicts performance trends accurately, with a mean relative error between $13\%$ and $28\%$.

\subsection{1D AllReduce}

\textbf{Scaling Vector Length.} We evaluate AllReduce for 512 PEs and increasing vector length, see Figure~\ref{fig:allreduce_fix}. As expected, for the reduce-then-broadcast AllReduce implementations the runtime increases by the cost of performing a broadcast with respect to the corresponding Reduce variant. We can draw the same insights as from the Reduce results, e.g., how the optimal pattern changes based on the vector length. Our Auto-Gen AllReduce gains a $2.47$x improvement over the chain-then-broadcast approach, which the current library is based on. Additionally, just like for Reduce, our model accurately predicts performance trends and which pattern will perform best.

We also plotted predicted performance for ring pattern. We have already observed that our model accurately predicts which algorithm performs best. Even accounting for a $15\%$ prediction error, the largest observed overall, the ring algorithm is never be the best choice. Hence, we refrain from providing an implementation. This underscores the utility of our model is saving engineering effort on sub-optimal and unpromising approaches. Moreover, we see that algorithms designed for the traditional distributed memory setting do not translate well into the wafer-scale setting, where it is important to leverage multicasting and pipelining.


\textbf{Scaling PE count.} Moreover, we evaluate AllReduce for a fixed vector length of 1 KB and increasing number of PEs, see \Cref{fig:allreduce_inc}. Again, for the reduce-then-broadcast implementations, we observe similar results as for Reduce. We see that for 4 PEs, the predicted ring performance is a bit better than the chain AllReduce. However, the expected performance gain is not significant. For numbers of PEs larger than $8$, we see that the reduce-then-broadcast implementations would perform significantly better than the ring, outperforming it by possibly even $1.4$x. This further shows how powerful the multicast feature for the CS-2 is.

\subsection{2D Collectives}

\textbf{Scaling Vector Length.} We also evaluate our implementations on the full chip of $512\times512$ PEs and increasing vector length. For Reduce, the results can be found in Figure~\ref{fig:experiments-2d-reduce-fix}. The performance trends are as predicted. They are similar to to 1D setting. Our X-Y Auto-Gen Reduce outperforms the X-Y Chain by up to $3.27 \times$.

The predictions for the snake pattern are at most $10\%$ off. As expected, it performs very poorly. This is because of its linear depth in the number of PEs, which is over 200'000 for this experiment. Interestingly, these results indicate that $T_R = 2$ on average. Any other choice of $T_R$ would lead to significantly worse predictions.

Moreover, we have a relative error that is very similar to what we were seeing for 1D. The predictions are slightly worse because when we execute X-Y Reduce, after reducing on the X-axis, we need to load some values into registers which adds additional overhead. Again, our model is able to predict very well which pattern will perform best for a given vector length.

Figure~\ref{fig:experiments-2d-allreduce-fix} shows the  2D AllReduce results for fixed number of PEs and increasing vector length. 
The relative error of our predictions remains almost the same. Moreover, our X-Y Auto-Gen AllReduce implementation outperforms the X-Y Auto-Gen AllReduce implementation by up to $2.54 \times$.

\textbf{Scaling PE count.} Just like for 1D benchmarks, we measure performance for fixed vector length and increasing 2D grid of PEs from $4\times4$ to $512\times512$. The results can be found in Figure~\ref{fig:experiments-2d-reduce-inc}. As expected, when we have few PEs and we are bandwidth bound, the Snake pattern performs best. Then as the number of PEs grows, the X-Y Chain and finally the X-Y Two Phase pattern are best. 

Our X-Y Auto-Gen reduce once again achieves good overall performance across the board. The only exception is for $4\times 4$ PEs, where the Snake is better. This demonstrates that generating code based on our model works well both in the 1D and 2D setting.

\section{Related Work}



\subsection{Reduce on the WSE}
Orenes-Vera et al.~\cite{DBLP:conf/ics/cerebras_fft} introduced and implemented a wafer-scale \textbf{3D FFT algorithm}. The main communication bottleneck in their implementation is an all-to-all collective. They model the communication time by considering the most heavily congested link in the network. While the all-to-all is limited by congestion, our research demonstrates that, for the case of reduce and AllReduce, accurately modeling performance necessitates considering depth and distance, in addition to congestion factors.

Rocki et al.~\cite{DBLP:conf/sc/RockiESSMKPDS020} designed a wafer-scale \textbf{SpMV stencil}. It involves an AllReduce operation at its core. They use a variant of a 2D Star AllReduce with two PEs accumulating all the results and then broadcasting them. As our evaluation shows, such an approach is only efficient for small vector lengths because it creates large contention at the PEs that aggregate the vectors.


Hall et al.~\cite{Cerebras-Neural-Net-Training} designed a matrix-multiplication kernel as part of their \textbf{neural network training} stack. They utilize there the chain pattern to perform column-reduction, which as we have shown, is the best when the vector length is much larger than the number of PEs. They also perform row-reduce to specific PE by mapping the chain pattern onto a ring like in Figure~\ref{fig:ring}.

\subsection{Distributed Models of Computation}

The $\mathbf{\alpha-\beta}$ \textbf{model} of parallel computation~\cite{DBLP:journals/concurrency/ChanHPG07} considers the latency and bandwidth requirements of distributed programs. Specifically, sending a message of length $m$ costs $\alpha + m \beta$. Unlike our model, all processors are assumed to be at the same distance to each other -- the cost is independent of sender and receiver. More general models~\cite{DBLP:journals/corr/WickramasingheL16}, such as \textbf{LogGP}~\cite{DBLP:journals/ijhpca/ThakurRG05} share the same limitations. This means that those models are unable to capture the aspects of distance and energy. These are essential for accurately estimating performance in the spatial wafer-scale setting.

The \textbf{spatial computer model}~\cite{22SpatialGianinazzi, baumann2024lowdepth} introduced an asymptotic model that considers energy, depth, and distance. However, it assumes that the bandwidth of the PE is of the same order of magnitude as its local memory. This does not correspond to our practical setting, where the local SRAM memory is a several thousand times larger than the bandwidth from and to the PE. This difference in the model leads to a lack of pipelining in spatial computer algorithms. As we have seen, pipelining is necessary to obtain the best performance for very large vector lengths. To address issues arising from unequal bandwidth and memory, we introduced the contention term into the model. It reflects that PE-bandwidth is a scarce resource on the device. 

\subsection{Developments in  Accelerator Architectures}

The Versal ACAP~\cite{DBLP:conf/fpga/GaideGRB19} is a Course-Grained Reconfigurable Array (CGRA)~\cite{DBLP:conf/asap/ChinSRZKHA17} that contains both programmable FPGA fabric and software programmable accelerators. These accelerators consist of $8\times 50$ tiles with 48KB of local  memory connected via a mesh network~\cite{Roemer2023, Wierse2023}, similarly to the WSE. While the programming models differ, one can observe much of the same distance-dependent performance characteristics~\cite{Roemer2023}. Therefore, our model could provide a useful basis for algorithmic design on the Versal ACAP.

The SambaNova Reconfigurable Dataflow Unit (DFU)~\cite{DBLP:journals/cse/EmaniVAPSFJLNSK21} is a CGRA machine learning accelerator. In contrast to the WSE, it is not based on general-purpose cores. Instead, a large grid of tiny hardware units are reconfigured to perform the given operation in a pipelined dataflow. Note that the problem of mapping operations, such as communication collectives, onto a grid of compute elements is a shared problem. Hence, there could be insights and tangible benefits of our work that carry over to the DFU. 
\section{Conclusion}
We provided the first in-depth exploration of communication collectives on the Cerebras WSE. We introduced and analyzed different collective algorithms which outperform previous algorithms by up to $3.27\times$. Given the widespread use of these collectives in HPC applications, our improvements promise to significantly boost computational efficiency in various scientific fields.

To achieve those improvements, we introduced a streamlined model to design algorithms for the hardware. We demonstrated that this model accurately predicts performance on the WSE. While previous works focused on communication for specific problems on the Cerebras WSE, our approach is the first to enable systematic analysis of any program on this hardware.

Our findings demonstrate that achieving optimal performance on the WSE is contingent upon automatic code generation. Manual optimizations for the WSE, hindered by the complexity of hardware features like routing, are both tedious and challenging. Our newly introduced model significantly advances the creation of effective code generators by enabling accurate performance prediction. Utilizing our model-driven approach, we generate code that achieves near-optimal performance across a broad spectrum of input sizes. This marks the first time such an approach has been successfully adapted for a wafer-scale processor.

Overall, our work enhances the performance of communication collectives and provides valuable insights into optimizing wafer-scale programs for emerging architectures like the Cerebras WSE. The accurate theoretical framework for modeling algorithms significantly advances our understanding of the hardware and its limitations. Hence, our study represents a significant advancement in unlocking the full potential of this emerging architecture and boosting the efficiency of HPC applications.

\begin{acks}
This work received support from the the European Union's Horizon 2020 program (No. 955776, RED-SEA) and the ERC program (PSAP, No.101002047). This work was supported by Sapienza University under the SEED-2022 and "Progetti Grandi 2023" funding schemes.
\end{acks}

\bibliographystyle{ACM-Reference-Format}
\bibliography{citations}


\begin{thebibliography}{60}


\ifx \showCODEN    \undefined \def \showCODEN     #1{\unskip}     \fi
\ifx \showDOI      \undefined \def \showDOI       #1{#1}\fi
\ifx \showISBNx    \undefined \def \showISBNx     #1{\unskip}     \fi
\ifx \showISBNxiii \undefined \def \showISBNxiii  #1{\unskip}     \fi
\ifx \showISSN     \undefined \def \showISSN      #1{\unskip}     \fi
\ifx \showLCCN     \undefined \def \showLCCN      #1{\unskip}     \fi
\ifx \shownote     \undefined \def \shownote      #1{#1}          \fi
\ifx \showarticletitle \undefined \def \showarticletitle #1{#1}   \fi
\ifx \showURL      \undefined \def \showURL       {\relax}        \fi
\providecommand\bibfield[2]{#2}
\providecommand\bibinfo[2]{#2}
\providecommand\natexlab[1]{#1}
\providecommand\showeprint[2][]{arXiv:#2}

\bibitem[Barnett et~al\mbox{.}(1995)]%
        {DBLP:journals/jpdc/BarnettLPG95}
\bibfield{author}{\bibinfo{person}{Michael Barnett},
  \bibinfo{person}{Richard~J. Littlefield}, \bibinfo{person}{David~G. Payne},
  {and} \bibinfo{person}{Robert~A. van~de Geijn}.}
  \bibinfo{year}{1995}\natexlab{}.
\newblock \showarticletitle{Global Combine Algorithms for 2-D Meshes with
  Wormhole Routing}.
\newblock \bibinfo{journal}{\emph{J. Parallel Distributed Comput.}}
  \bibinfo{volume}{24}, \bibinfo{number}{2} (\bibinfo{year}{1995}),
  \bibinfo{pages}{191--201}.
\newblock
\urldef\tempurl%
\url{https://doi.org/10.1006/jpdc.1995.1018}
\showDOI{\tempurl}


\bibitem[Baumann et~al\mbox{.}(2024)]%
        {baumann2024lowdepth}
\bibfield{author}{\bibinfo{person}{Yves Baumann}, \bibinfo{person}{Tal
  Ben-Nun}, \bibinfo{person}{Maciej Besta}, \bibinfo{person}{Lukas Gianinazzi},
  \bibinfo{person}{Torsten Hoefler}, {and} \bibinfo{person}{Piotr Luczynski}.}
  \bibinfo{year}{2024}\natexlab{}.
\newblock \bibinfo{title}{Low-Depth Spatial Tree Algorithms}.
\newblock
\newblock
\showeprint[arxiv]{2404.12953}~[cs.DC]


\bibitem[Ben{-}Nun and Hoefler(2019)]%
        {DBLP:journals/csur/Ben-NunH19}
\bibfield{author}{\bibinfo{person}{Tal Ben{-}Nun} {and}
  \bibinfo{person}{Torsten Hoefler}.} \bibinfo{year}{2019}\natexlab{}.
\newblock \showarticletitle{Demystifying Parallel and Distributed Deep
  Learning: An In-depth Concurrency Analysis}.
\newblock \bibinfo{journal}{\emph{{ACM} Comput. Surv.}} \bibinfo{volume}{52},
  \bibinfo{number}{4} (\bibinfo{year}{2019}), \bibinfo{pages}{65:1--65:43}.
\newblock
\urldef\tempurl%
\url{https://doi.org/10.1145/3320060}
\showDOI{\tempurl}


\bibitem[Besta and Hoefler(2014)]%
        {DBLP:conf/sc/BestaH14}
\bibfield{author}{\bibinfo{person}{Maciej Besta} {and} \bibinfo{person}{Torsten
  Hoefler}.} \bibinfo{year}{2014}\natexlab{}.
\newblock \showarticletitle{Slim Fly: {A} Cost Effective Low-Diameter Network
  Topology}. In \bibinfo{booktitle}{\emph{International Conference for High
  Performance Computing, Networking, Storage and Analysis, {SC} 2014, New
  Orleans, LA, USA, November 16-21, 2014}},
  \bibfield{editor}{\bibinfo{person}{Trish Damkroger} {and}
  \bibinfo{person}{Jack~J. Dongarra}} (Eds.). \bibinfo{publisher}{{IEEE}
  Computer Society}, \bibinfo{pages}{348--359}.
\newblock
\urldef\tempurl%
\url{https://doi.org/10.1109/SC.2014.34}
\showDOI{\tempurl}


\bibitem[Besta and Hoefler(2022)]%
        {dist_gnn}
\bibfield{author}{\bibinfo{person}{Maciej Besta} {and} \bibinfo{person}{Torsten
  Hoefler}.} \bibinfo{year}{2022}\natexlab{}.
\newblock \showarticletitle{Parallel and Distributed Graph Neural Networks: An
  In-Depth Concurrency Analysis}.
\newblock \bibinfo{journal}{\emph{CoRR}}  \bibinfo{volume}{abs/2205.09702}
  (\bibinfo{year}{2022}).
\newblock
\urldef\tempurl%
\url{https://doi.org/10.48550/ARXIV.2205.09702}
\showDOI{\tempurl}
\showeprint[arXiv]{2205.09702}


\bibitem[Brandt and Lubrecht(1990)]%
        {brandt1990multilevel}
\bibfield{author}{\bibinfo{person}{Achi Brandt} {and} \bibinfo{person}{AA
  Lubrecht}.} \bibinfo{year}{1990}\natexlab{}.
\newblock \showarticletitle{Multilevel matrix multiplication and fast solution
  of integral equations}.
\newblock \bibinfo{journal}{\emph{J. Comput. Phys.}} \bibinfo{volume}{90},
  \bibinfo{number}{2} (\bibinfo{year}{1990}), \bibinfo{pages}{348--370}.
\newblock


\bibitem[Chan et~al\mbox{.}(2007)]%
        {DBLP:journals/concurrency/ChanHPG07}
\bibfield{author}{\bibinfo{person}{Ernie Chan}, \bibinfo{person}{Marcel
  Heimlich}, \bibinfo{person}{Avi Purkayastha}, {and}
  \bibinfo{person}{Robert~A. van~de Geijn}.} \bibinfo{year}{2007}\natexlab{}.
\newblock \showarticletitle{Collective communication: theory, practice, and
  experience}.
\newblock \bibinfo{journal}{\emph{Concurr. Comput. Pract. Exp.}}
  \bibinfo{volume}{19}, \bibinfo{number}{13} (\bibinfo{year}{2007}),
  \bibinfo{pages}{1749--1783}.
\newblock
\urldef\tempurl%
\url{https://doi.org/10.1002/cpe.1206}
\showDOI{\tempurl}


\bibitem[Chin et~al\mbox{.}(2017)]%
        {DBLP:conf/asap/ChinSRZKHA17}
\bibfield{author}{\bibinfo{person}{S.~Alexander Chin}, \bibinfo{person}{Noriaki
  Sakamoto}, \bibinfo{person}{Allan Rui}, \bibinfo{person}{Jim Zhao},
  \bibinfo{person}{Jin~Hee Kim}, \bibinfo{person}{Yuko Hara{-}Azumi}, {and}
  \bibinfo{person}{Jason~Helge Anderson}.} \bibinfo{year}{2017}\natexlab{}.
\newblock \showarticletitle{{CGRA-ME:} {A} unified framework for {CGRA}
  modelling and exploration}. In \bibinfo{booktitle}{\emph{28th {IEEE}
  International Conference on Application-specific Systems, Architectures and
  Processors, {ASAP} 2017, Seattle, WA, USA, July 10-12, 2017}}.
  \bibinfo{publisher}{{IEEE} Computer Society}, \bibinfo{pages}{184--189}.
\newblock
\urldef\tempurl%
\url{https://doi.org/10.1109/ASAP.2017.7995277}
\showDOI{\tempurl}


\bibitem[Cho et~al\mbox{.}(2021)]%
        {DBLP:conf/sc/ChoJE21}
\bibfield{author}{\bibinfo{person}{Benjamin~Y. Cho}, \bibinfo{person}{Jeageun
  Jung}, {and} \bibinfo{person}{Mattan Erez}.} \bibinfo{year}{2021}\natexlab{}.
\newblock \showarticletitle{Accelerating bandwidth-bound deep learning
  inference with main-memory accelerators}. In
  \bibinfo{booktitle}{\emph{International Conference for High Performance
  Computing, Networking, Storage and Analysis, {SC} 2021, St. Louis, Missouri,
  USA, November 14-19, 2021}}, \bibfield{editor}{\bibinfo{person}{Bronis~R.
  de~Supinski}, \bibinfo{person}{Mary~W. Hall}, {and} \bibinfo{person}{Todd
  Gamblin}} (Eds.). \bibinfo{publisher}{{ACM}}, \bibinfo{pages}{44}.
\newblock
\urldef\tempurl%
\url{https://doi.org/10.1145/3458817.3476146}
\showDOI{\tempurl}


\bibitem[Chunduri et~al\mbox{.}(2018)]%
        {DBLP:conf/sc/ChunduriPBHK18}
\bibfield{author}{\bibinfo{person}{Sudheer Chunduri}, \bibinfo{person}{Scott
  Parker}, \bibinfo{person}{Pavan Balaji}, \bibinfo{person}{Kevin Harms}, {and}
  \bibinfo{person}{Kalyan Kumaran}.} \bibinfo{year}{2018}\natexlab{}.
\newblock \showarticletitle{Characterization of {MPI} usage on a production
  supercomputer}. In \bibinfo{booktitle}{\emph{Proceedings of the International
  Conference for High Performance Computing, Networking, Storage, and Analysis,
  {SC} 2018, Dallas, TX, USA, November 11-16, 2018}}.
  \bibinfo{publisher}{{IEEE} / {ACM}}, \bibinfo{pages}{30:1--30:15}.
\newblock
\urldef\tempurl%
\url{http://dl.acm.org/citation.cfm?id=3291696}
\showURL{%
\tempurl}


\bibitem[De~Sensi et~al\mbox{.}(2024)]%
        {swing}
\bibfield{author}{\bibinfo{person}{Daniele De~Sensi}, \bibinfo{person}{Tommaso
  Bonato}, \bibinfo{person}{David Saam}, {and} \bibinfo{person}{Torsten
  Hoefler}.} \bibinfo{year}{2024}\natexlab{}.
\newblock \showarticletitle{Swing: Short-cutting Rings for Higher Bandwidth
  Allreduce}. In \bibinfo{booktitle}{\emph{21th USENIX Symposium on Networked
  Systems Design and Implementation (NSDI 24)}}. \bibinfo{publisher}{USENIX
  Association}, \bibinfo{address}{Santa Clara, CA}.
\newblock


\bibitem[De~Sensi et~al\mbox{.}(2021)]%
        {flare}
\bibfield{author}{\bibinfo{person}{Daniele De~Sensi},
  \bibinfo{person}{Salvatore Di~Girolamo}, \bibinfo{person}{Saleh Ashkboos},
  \bibinfo{person}{Shigang Li}, {and} \bibinfo{person}{Torsten Hoefler}.}
  \bibinfo{year}{2021}\natexlab{}.
\newblock \showarticletitle{Flare: Flexible in-Network Allreduce}. In
  \bibinfo{booktitle}{\emph{Proceedings of the International Conference for
  High Performance Computing, Networking, Storage and Analysis}} (St. Louis,
  Missouri) \emph{(\bibinfo{series}{SC '21})}. \bibinfo{publisher}{Association
  for Computing Machinery}, \bibinfo{address}{New York, NY, USA}, Article
  \bibinfo{articleno}{35}, \bibinfo{numpages}{16}~pages.
\newblock
\showISBNx{9781450384421}
\urldef\tempurl%
\url{https://doi.org/10.1145/3458817.3476178}
\showDOI{\tempurl}


\bibitem[De~Sensi et~al\mbox{.}(2020)]%
        {slingshot}
\bibfield{author}{\bibinfo{person}{Daniele De~Sensi},
  \bibinfo{person}{Salvatore Di~Girolamo}, \bibinfo{person}{Kim~H. McMahon},
  \bibinfo{person}{Duncan Roweth}, {and} \bibinfo{person}{Torsten Hoefler}.}
  \bibinfo{year}{2020}\natexlab{}.
\newblock \showarticletitle{An In-Depth Analysis of the Slingshot
  Interconnect}. In \bibinfo{booktitle}{\emph{Proceedings of the International
  Conference for High Performance Computing, Networking, Storage and Analysis}}
  (Atlanta, Georgia) \emph{(\bibinfo{series}{SC '20})}.
  \bibinfo{publisher}{IEEE Press}, Article \bibinfo{articleno}{35},
  \bibinfo{numpages}{14}~pages.
\newblock
\showISBNx{9781728199986}
\urldef\tempurl%
\url{https://doi.org/10.1109/sc41405.2020.00039}
\showDOI{\tempurl}


\bibitem[Dey et~al\mbox{.}(2023)]%
        {DBLP:journals/corr/abs-2304-03208}
\bibfield{author}{\bibinfo{person}{Nolan Dey}, \bibinfo{person}{Gurpreet
  Gosal}, \bibinfo{person}{Zhiming Chen}, \bibinfo{person}{Hemant Khachane},
  \bibinfo{person}{William Marshall}, \bibinfo{person}{Ribhu Pathria},
  \bibinfo{person}{Marvin Tom}, {and} \bibinfo{person}{Joel Hestness}.}
  \bibinfo{year}{2023}\natexlab{}.
\newblock \showarticletitle{Cerebras-GPT: Open Compute-Optimal Language Models
  Trained on the Cerebras Wafer-Scale Cluster}.
\newblock \bibinfo{journal}{\emph{CoRR}}  \bibinfo{volume}{abs/2304.03208}
  (\bibinfo{year}{2023}).
\newblock
\urldef\tempurl%
\url{https://doi.org/10.48550/ARXIV.2304.03208}
\showDOI{\tempurl}
\showeprint[arXiv]{2304.03208}


\bibitem[Emani et~al\mbox{.}(2021)]%
        {DBLP:journals/cse/EmaniVAPSFJLNSK21}
\bibfield{author}{\bibinfo{person}{Murali Emani}, \bibinfo{person}{Venkatram
  Vishwanath}, \bibinfo{person}{Corey Adams}, \bibinfo{person}{Michael~E.
  Papka}, \bibinfo{person}{Rick Stevens}, \bibinfo{person}{Laura Florescu},
  \bibinfo{person}{Sumti Jairath}, \bibinfo{person}{William Liu},
  \bibinfo{person}{Tejas Nama}, \bibinfo{person}{Arvind Sujeeth},
  \bibinfo{person}{Volodymyr~V. Kindratenko}, {and} \bibinfo{person}{Anne~C.
  Elster}.} \bibinfo{year}{2021}\natexlab{}.
\newblock \showarticletitle{Accelerating Scientific Applications With SambaNova
  Reconfigurable Dataflow Architecture}.
\newblock \bibinfo{journal}{\emph{Comput. Sci. Eng.}} \bibinfo{volume}{23},
  \bibinfo{number}{2} (\bibinfo{year}{2021}), \bibinfo{pages}{114--119}.
\newblock
\urldef\tempurl%
\url{https://doi.org/10.1109/MCSE.2021.3057203}
\showDOI{\tempurl}


\bibitem[Gaide et~al\mbox{.}(2019)]%
        {DBLP:conf/fpga/GaideGRB19}
\bibfield{author}{\bibinfo{person}{Brian Gaide}, \bibinfo{person}{Dinesh
  Gaitonde}, \bibinfo{person}{Chirag Ravishankar}, {and}
  \bibinfo{person}{Trevor Bauer}.} \bibinfo{year}{2019}\natexlab{}.
\newblock \showarticletitle{Xilinx Adaptive Compute Acceleration Platform:
  Versal\({}^{\mbox{TM}}\) Architecture}. In
  \bibinfo{booktitle}{\emph{Proceedings of the 2019 {ACM/SIGDA} International
  Symposium on Field-Programmable Gate Arrays, {FPGA} 2019, Seaside, CA, USA,
  February 24-26, 2019}}, \bibfield{editor}{\bibinfo{person}{Kia Bazargan}
  {and} \bibinfo{person}{Stephen Neuendorffer}} (Eds.).
  \bibinfo{publisher}{{ACM}}, \bibinfo{pages}{84--93}.
\newblock
\urldef\tempurl%
\url{https://doi.org/10.1145/3289602.3293906}
\showDOI{\tempurl}


\bibitem[Gianinazzi et~al\mbox{.}(2022)]%
        {22SpatialGianinazzi}
\bibfield{author}{\bibinfo{person}{Lukas Gianinazzi}, \bibinfo{person}{Tal
  Ben{-}Nun}, \bibinfo{person}{Saleh Ashkboos}, \bibinfo{person}{Yves Baumann},
  \bibinfo{person}{Piotr Luczynski}, {and} \bibinfo{person}{Torsten Hoefler}.}
  \bibinfo{year}{2022}\natexlab{}.
\newblock \showarticletitle{The spatial computer: {A} model for
  energy-efficient parallel computation}.
\newblock \bibinfo{journal}{\emph{CoRR}}  \bibinfo{volume}{abs/2205.04934}
  (\bibinfo{year}{2022}).
\newblock
\urldef\tempurl%
\url{https://doi.org/10.48550/arXiv.2205.04934}
\showDOI{\tempurl}
\showeprint[arXiv]{2205.04934}


\bibitem[Graham et~al\mbox{.}(2020)]%
        {Graham2020ScalableHA}
\bibfield{author}{\bibinfo{person}{Richard~L. Graham}, \bibinfo{person}{Lion
  Levi}, \bibinfo{person}{Devendar Bureddy}, \bibinfo{person}{Gil Bloch},
  \bibinfo{person}{Gilad Shainer}, \bibinfo{person}{David Cho},
  \bibinfo{person}{George Elias}, \bibinfo{person}{Daniel Klein},
  \bibinfo{person}{Joshua Ladd}, \bibinfo{person}{Ophir Maor},
  \bibinfo{person}{Ami Marelli}, \bibinfo{person}{Valentin Petrov},
  \bibinfo{person}{Evyatar Romlet}, \bibinfo{person}{Yong Qin}, {and}
  \bibinfo{person}{Ido Zemah}.} \bibinfo{year}{2020}\natexlab{}.
\newblock \showarticletitle{Scalable Hierarchical Aggregation and Reduction
  Protocol (SHARP)TM Streaming-Aggregation Hardware Design and Evaluation}.
\newblock \bibinfo{journal}{\emph{High Performance Computing}}
  \bibinfo{volume}{12151} (\bibinfo{year}{2020}), \bibinfo{pages}{41 -- 59}.
\newblock


\bibitem[Gropp et~al\mbox{.}(1996)]%
        {DBLP:journals/pc/GroppLDS96}
\bibfield{author}{\bibinfo{person}{William Gropp}, \bibinfo{person}{Ewing~L.
  Lusk}, \bibinfo{person}{Nathan~E. Doss}, {and} \bibinfo{person}{Anthony
  Skjellum}.} \bibinfo{year}{1996}\natexlab{}.
\newblock \showarticletitle{A High-Performance, Portable Implementation of the
  {MPI} Message Passing Interface Standard}.
\newblock \bibinfo{journal}{\emph{Parallel Comput.}} \bibinfo{volume}{22},
  \bibinfo{number}{6} (\bibinfo{year}{1996}), \bibinfo{pages}{789--828}.
\newblock
\urldef\tempurl%
\url{https://doi.org/10.1016/0167-8191(96)00024-5}
\showDOI{\tempurl}


\bibitem[Hoefler and Belli(2015)]%
        {scbench}
\bibfield{author}{\bibinfo{person}{Torsten Hoefler} {and}
  \bibinfo{person}{Roberto Belli}.} \bibinfo{year}{2015}\natexlab{}.
\newblock \showarticletitle{Scientific benchmarking of parallel computing
  systems: twelve ways to tell the masses when reporting performance results}.
  In \bibinfo{booktitle}{\emph{Proceedings of the International Conference for
  High Performance Computing, Networking, Storage and Analysis, {SC} 2015,
  Austin, TX, USA, November 15-20, 2015}},
  \bibfield{editor}{\bibinfo{person}{Jackie Kern} {and}
  \bibinfo{person}{Jeffrey~S. Vetter}} (Eds.). \bibinfo{publisher}{{ACM}},
  \bibinfo{pages}{73:1--73:12}.
\newblock
\urldef\tempurl%
\url{https://doi.org/10.1145/2807591.2807644}
\showDOI{\tempurl}


\bibitem[Hoefler et~al\mbox{.}(2022)]%
        {DBLP:conf/sc/HoeflerBSGLHBGCS22}
\bibfield{author}{\bibinfo{person}{Torsten Hoefler}, \bibinfo{person}{Tommaso
  Bonato}, \bibinfo{person}{Daniele~De Sensi}, \bibinfo{person}{Salvatore~Di
  Girolamo}, \bibinfo{person}{Shigang Li}, \bibinfo{person}{Marco Heddes},
  \bibinfo{person}{Jon Belk}, \bibinfo{person}{Deepak Goel},
  \bibinfo{person}{Miguel Castro}, {and} \bibinfo{person}{Steve Scott}.}
  \bibinfo{year}{2022}\natexlab{}.
\newblock \showarticletitle{HammingMesh: {A} Network Topology for Large-Scale
  Deep Learning}. In \bibinfo{booktitle}{\emph{{SC22:} International Conference
  for High Performance Computing, Networking, Storage and Analysis, Dallas, TX,
  USA, November 13-18, 2022}}. \bibinfo{publisher}{{IEEE}},
  \bibinfo{pages}{1--18}.
\newblock
\urldef\tempurl%
\url{https://doi.org/10.1109/SC41404.2022.00016}
\showDOI{\tempurl}


\bibitem[Hoefler and Moor(2014)]%
        {hoefler-moor-collectives}
\bibfield{author}{\bibinfo{person}{Torsten Hoefler} {and} \bibinfo{person}{D.
  Moor}.} \bibinfo{year}{2014}\natexlab{}.
\newblock \showarticletitle{{Energy, Memory, and Runtime Tradeoffs for
  Implementing Collective Communication Operations}}.
\newblock \bibinfo{journal}{\emph{Journal of Supercomputing Frontiers and
  Innovations}} \bibinfo{volume}{1}, \bibinfo{number}{2} (\bibinfo{date}{Oct.}
  \bibinfo{year}{2014}), \bibinfo{pages}{58--75}.
\newblock


\bibitem[Hoefler et~al\mbox{.}(2010)]%
        {hoefler-collmea-sync}
\bibfield{author}{\bibinfo{person}{Torsten Hoefler}, \bibinfo{person}{Timo
  Schneider}, {and} \bibinfo{person}{Andrew Lumsdaine}.}
  \bibinfo{year}{2010}\natexlab{}.
\newblock \showarticletitle{{Accurately Measuring Overhead, Communication Time
  and Progression of Blocking and Nonblocking Collective Operations at Massive
  Scale}}.
\newblock \bibinfo{journal}{\emph{International Journal of Parallel, Emergent
  and Distributed Systems}} \bibinfo{volume}{25}, \bibinfo{number}{4}
  (\bibinfo{date}{Jul.} \bibinfo{year}{2010}), \bibinfo{pages}{241--258}.
\newblock
\showISSN{1744-5779}


\bibitem[Inc.(2021)]%
        {Cerebras}
\bibfield{author}{\bibinfo{person}{Cerebras~Systems Inc.}}
  \bibinfo{year}{2021}\natexlab{}.
\newblock \showarticletitle{Cerebras Systems: Achieving Industry Best AI
  Performance Through A Systems Approach}.
\newblock  (\bibinfo{year}{2021}).
\newblock


\bibitem[Jacquelin et~al\mbox{.}(2022)]%
        {massively_scalable_stencil_cerebras}
\bibfield{author}{\bibinfo{person}{Mathias Jacquelin},
  \bibinfo{person}{Mauricio Araya{-}Polo}, {and} \bibinfo{person}{Jie Meng}.}
  \bibinfo{year}{2022}\natexlab{}.
\newblock \showarticletitle{Massively scalable stencil algorithm}.
\newblock \bibinfo{journal}{\emph{CoRR}}  \bibinfo{volume}{abs/2204.03775}
  (\bibinfo{year}{2022}).
\newblock
\urldef\tempurl%
\url{https://doi.org/10.48550/arXiv.2204.03775}
\showDOI{\tempurl}
\showeprint[arXiv]{2204.03775}


\bibitem[Jain and Sabharwal(2010)]%
        {DBLP:conf/ics/JainS10}
\bibfield{author}{\bibinfo{person}{Nikhil Jain} {and} \bibinfo{person}{Yogish
  Sabharwal}.} \bibinfo{year}{2010}\natexlab{}.
\newblock \showarticletitle{Optimal bucket algorithms for large {MPI}
  collectives on torus interconnects}. In \bibinfo{booktitle}{\emph{Proceedings
  of the 24th International Conference on Supercomputing, 2010, Tsukuba,
  Ibaraki, Japan, June 2-4, 2010}}, \bibfield{editor}{\bibinfo{person}{Taisuke
  Boku}, \bibinfo{person}{Hiroshi Nakashima}, {and} \bibinfo{person}{Avi
  Mendelson}} (Eds.). \bibinfo{publisher}{{ACM}}, \bibinfo{pages}{27--36}.
\newblock
\urldef\tempurl%
\url{https://doi.org/10.1145/1810085.1810093}
\showDOI{\tempurl}


\bibitem[Johnsson and Ho(1989)]%
        {DBLP:journals/tc/JohnssonH89}
\bibfield{author}{\bibinfo{person}{S.~Lennart Johnsson} {and}
  \bibinfo{person}{Ching{-}Tien Ho}.} \bibinfo{year}{1989}\natexlab{}.
\newblock \showarticletitle{Optimum Broadcasting and Personalized Communication
  in Hypercubes}.
\newblock \bibinfo{journal}{\emph{{IEEE} Trans. Computers}}
  \bibinfo{volume}{38}, \bibinfo{number}{9} (\bibinfo{year}{1989}),
  \bibinfo{pages}{1249--1268}.
\newblock
\urldef\tempurl%
\url{https://doi.org/10.1109/12.29465}
\showDOI{\tempurl}


\bibitem[Karonis et~al\mbox{.}(2000)]%
        {DBLP:conf/ipps/KaronisSFGLB00}
\bibfield{author}{\bibinfo{person}{Nicholas~T. Karonis},
  \bibinfo{person}{Bronis~R. de Supinski}, \bibinfo{person}{Ian~T. Foster},
  \bibinfo{person}{William Gropp}, \bibinfo{person}{Ewing~L. Lusk}, {and}
  \bibinfo{person}{John Bresnahan}.} \bibinfo{year}{2000}\natexlab{}.
\newblock \showarticletitle{Exploiting Hierarchy in Parallel Computer Networks
  to Optimize Collective Operation Performance}. In
  \bibinfo{booktitle}{\emph{Proceedings of the 14th International Parallel {\&}
  Distributed Processing Symposium (IPDPS'00), Cancun, Mexico, May 1-5, 2000}}.
  \bibinfo{publisher}{{IEEE} Computer Society}, \bibinfo{pages}{377--384}.
\newblock
\urldef\tempurl%
\url{https://doi.org/10.1109/IPDPS.2000.846009}
\showDOI{\tempurl}


\bibitem[Kim et~al\mbox{.}(2007)]%
        {DBLP:conf/isca/KimDA07}
\bibfield{author}{\bibinfo{person}{John Kim}, \bibinfo{person}{William~J.
  Dally}, {and} \bibinfo{person}{Dennis Abts}.}
  \bibinfo{year}{2007}\natexlab{}.
\newblock \showarticletitle{Flattened butterfly: a cost-efficient topology for
  high-radix networks}. In \bibinfo{booktitle}{\emph{34th International
  Symposium on Computer Architecture {(ISCA} 2007), June 9-13, 2007, San Diego,
  California, {USA}}}, \bibfield{editor}{\bibinfo{person}{Dean~M. Tullsen}
  {and} \bibinfo{person}{Brad Calder}} (Eds.). \bibinfo{publisher}{{ACM}},
  \bibinfo{pages}{126--137}.
\newblock
\urldef\tempurl%
\url{https://doi.org/10.1145/1250662.1250679}
\showDOI{\tempurl}


\bibitem[Kim et~al\mbox{.}(2008)]%
        {DBLP:conf/isca/KimDSA08}
\bibfield{author}{\bibinfo{person}{John Kim}, \bibinfo{person}{William~J.
  Dally}, \bibinfo{person}{Steve Scott}, {and} \bibinfo{person}{Dennis Abts}.}
  \bibinfo{year}{2008}\natexlab{}.
\newblock \showarticletitle{Technology-Driven, Highly-Scalable Dragonfly
  Topology}. In \bibinfo{booktitle}{\emph{35th International Symposium on
  Computer Architecture {(ISCA} 2008), June 21-25, 2008, Beijing, China}}.
  \bibinfo{publisher}{{IEEE} Computer Society}, \bibinfo{pages}{77--88}.
\newblock
\urldef\tempurl%
\url{https://doi.org/10.1109/ISCA.2008.19}
\showDOI{\tempurl}


\bibitem[Kumar and Faraj(2013)]%
        {allreduce_multicast}
\bibfield{author}{\bibinfo{person}{Sameer Kumar} {and} \bibinfo{person}{Daniel
  Faraj}.} \bibinfo{year}{2013}\natexlab{}.
\newblock \showarticletitle{Optimization of MPI\_Allreduce on the Blue Gene\/Q
  Supercomputer}. In \bibinfo{booktitle}{\emph{Proceedings of the 20th European
  MPI Users' Group Meeting}} (Madrid, Spain) \emph{(\bibinfo{series}{EuroMPI
  '13})}. \bibinfo{publisher}{Association for Computing Machinery},
  \bibinfo{address}{New York, NY, USA}, \bibinfo{pages}{97–103}.
\newblock
\showISBNx{9781450319034}
\urldef\tempurl%
\url{https://doi.org/10.1145/2488551.2488557}
\showDOI{\tempurl}


\bibitem[Kumar and Jouppi(2020)]%
        {DBLP:journals/corr/abs-2011-03605}
\bibfield{author}{\bibinfo{person}{Sameer Kumar} {and} \bibinfo{person}{Norm
  Jouppi}.} \bibinfo{year}{2020}\natexlab{}.
\newblock \showarticletitle{Highly Available Data Parallel {ML} training on
  Mesh Networks}.
\newblock \bibinfo{journal}{\emph{CoRR}}  \bibinfo{volume}{abs/2011.03605}
  (\bibinfo{year}{2020}).
\newblock
\showeprint[arXiv]{2011.03605}
\urldef\tempurl%
\url{https://arxiv.org/abs/2011.03605}
\showURL{%
\tempurl}


\bibitem[Laguna et~al\mbox{.}(2019)]%
        {DBLP:conf/sc/LagunaMMRSS19}
\bibfield{author}{\bibinfo{person}{Ignacio Laguna}, \bibinfo{person}{Ryan~J.
  Marshall}, \bibinfo{person}{Kathryn Mohror}, \bibinfo{person}{Martin
  Ruefenacht}, \bibinfo{person}{Anthony Skjellum}, {and}
  \bibinfo{person}{Nawrin Sultana}.} \bibinfo{year}{2019}\natexlab{}.
\newblock \showarticletitle{A large-scale study of {MPI} usage in open-source
  {HPC} applications}. In \bibinfo{booktitle}{\emph{Proceedings of the
  International Conference for High Performance Computing, Networking, Storage
  and Analysis, {SC} 2019, Denver, Colorado, USA, November 17-19, 2019}},
  \bibfield{editor}{\bibinfo{person}{Michela Taufer}, \bibinfo{person}{Pavan
  Balaji}, {and} \bibinfo{person}{Antonio~J. Pe{\~{n}}a}} (Eds.).
  \bibinfo{publisher}{{ACM}}, \bibinfo{pages}{31:1--31:14}.
\newblock
\urldef\tempurl%
\url{https://doi.org/10.1145/3295500.3356176}
\showDOI{\tempurl}


\bibitem[Lie(2021)]%
        {DBLP:conf/hotchips/Lie21}
\bibfield{author}{\bibinfo{person}{Sean Lie}.} \bibinfo{year}{2021}\natexlab{}.
\newblock \showarticletitle{Multi-Million Core, Multi-Wafer {AI} Cluster}. In
  \bibinfo{booktitle}{\emph{{IEEE} Hot Chips 33 Symposium, {HCS} 2021, Palo
  Alto, CA, USA, August 22-24, 2021}}. \bibinfo{publisher}{{IEEE}},
  \bibinfo{pages}{1--41}.
\newblock
\urldef\tempurl%
\url{https://doi.org/10.1109/HCS52781.2021.9567153}
\showDOI{\tempurl}


\bibitem[Lie(2023)]%
        {DBLP:journals/micro/Lie23}
\bibfield{author}{\bibinfo{person}{Sean Lie}.} \bibinfo{year}{2023}\natexlab{}.
\newblock \showarticletitle{Cerebras Architecture Deep Dive: First Look Inside
  the Hardware/Software Co-Design for Deep Learning}.
\newblock \bibinfo{journal}{\emph{{IEEE} Micro}} \bibinfo{volume}{43},
  \bibinfo{number}{3} (\bibinfo{year}{2023}), \bibinfo{pages}{18--30}.
\newblock
\urldef\tempurl%
\url{https://doi.org/10.1109/MM.2023.3256384}
\showDOI{\tempurl}


\bibitem[Ltaief et~al\mbox{.}(2023)]%
        {DBLP:conf/sc/LtaiefHWJRK23}
\bibfield{author}{\bibinfo{person}{Hatem Ltaief}, \bibinfo{person}{Yuxi Hong},
  \bibinfo{person}{Leighton Wilson}, \bibinfo{person}{Mathias Jacquelin},
  \bibinfo{person}{Matteo Ravasi}, {and} \bibinfo{person}{David~Elliot Keyes}.}
  \bibinfo{year}{2023}\natexlab{}.
\newblock \showarticletitle{Scaling the "Memory Wall" for Multi-Dimensional
  Seismic Processing with Algebraic Compression on Cerebras {CS-2} Systems}. In
  \bibinfo{booktitle}{\emph{Proceedings of the International Conference for
  High Performance Computing, Networking, Storage and Analysis, {SC} 2023,
  Denver, CO, USA, November 12-17, 2023}},
  \bibfield{editor}{\bibinfo{person}{Dorian Arnold}, \bibinfo{person}{Rosa~M.
  Badia}, {and} \bibinfo{person}{Kathryn~M. Mohror}} (Eds.).
  \bibinfo{publisher}{{ACM}}, \bibinfo{pages}{6:1--6:12}.
\newblock
\urldef\tempurl%
\url{https://doi.org/10.1145/3581784.3627042}
\showDOI{\tempurl}


\bibitem[{Message Passing Interface Forum}(2021)]%
        {mpi40}
\bibfield{author}{\bibinfo{person}{{Message Passing Interface Forum}}.}
  \bibinfo{year}{2021}\natexlab{}.
\newblock \bibinfo{booktitle}{\emph{{MPI}: A Message-Passing Interface Standard
  Version 4.0}}.
\newblock
\urldef\tempurl%
\url{https://www.mpi-forum.org/docs/mpi-4.0/mpi40-report.pdf}
\showURL{%
\tempurl}


\bibitem[Oliveira et~al\mbox{.}(2022)]%
        {gemv_nn}
\bibfield{author}{\bibinfo{person}{G.~F. Oliveira}, \bibinfo{person}{J.
  Gomez-Luna}, \bibinfo{person}{S. Ghose}, \bibinfo{person}{A. Boroumand},
  {and} \bibinfo{person}{O. Mutlu}.} \bibinfo{year}{2022}\natexlab{}.
\newblock \showarticletitle{Accelerating Neural Network Inference With
  Processing-in-DRAM: From the Edge to the Cloud}.
\newblock \bibinfo{journal}{\emph{IEEE Micro}} \bibinfo{volume}{42},
  \bibinfo{number}{06} (\bibinfo{date}{nov} \bibinfo{year}{2022}),
  \bibinfo{pages}{25--38}.
\newblock
\showISSN{1937-4143}
\urldef\tempurl%
\url{https://doi.org/10.1109/MM.2022.3202350}
\showDOI{\tempurl}


\bibitem[Orenes{-}Vera et~al\mbox{.}(2023)]%
        {DBLP:conf/ics/cerebras_fft}
\bibfield{author}{\bibinfo{person}{Marcelo Orenes{-}Vera},
  \bibinfo{person}{Ilya Sharapov}, \bibinfo{person}{Robert Schreiber},
  \bibinfo{person}{Mathias Jacquelin}, \bibinfo{person}{Philippe Vandermersch},
  {and} \bibinfo{person}{Sharan Chetlur}.} \bibinfo{year}{2023}\natexlab{}.
\newblock \showarticletitle{Wafer-Scale Fast Fourier Transforms}. In
  \bibinfo{booktitle}{\emph{Proceedings of the 37th International Conference on
  Supercomputing, {ICS} 2023, Orlando, FL, USA, June 21-23, 2023}},
  \bibfield{editor}{\bibinfo{person}{Kyle~A. Gallivan},
  \bibinfo{person}{Efstratios Gallopoulos}, \bibinfo{person}{Dimitrios~S.
  Nikolopoulos}, {and} \bibinfo{person}{Ram{\'{o}}n Beivide}} (Eds.).
  \bibinfo{publisher}{{ACM}}, \bibinfo{pages}{180--191}.
\newblock
\urldef\tempurl%
\url{https://doi.org/10.1145/3577193.3593708}
\showDOI{\tempurl}


\bibitem[Patarasuk and Yuan(2009)]%
        {DBLP:journals/jpdc/PatarasukY09}
\bibfield{author}{\bibinfo{person}{Pitch Patarasuk} {and} \bibinfo{person}{Xin
  Yuan}.} \bibinfo{year}{2009}\natexlab{}.
\newblock \showarticletitle{Bandwidth optimal all-reduce algorithms for
  clusters of workstations}.
\newblock \bibinfo{journal}{\emph{J. Parallel Distributed Comput.}}
  \bibinfo{volume}{69}, \bibinfo{number}{2} (\bibinfo{year}{2009}),
  \bibinfo{pages}{117--124}.
\newblock
\urldef\tempurl%
\url{https://doi.org/10.1016/j.jpdc.2008.09.002}
\showDOI{\tempurl}


\bibitem[Qi et~al\mbox{.}(2014)]%
        {qi2014krux}
\bibfield{author}{\bibinfo{person}{Jianlong Qi}, \bibinfo{person}{Hassan
  Foroughi~Asl}, \bibinfo{person}{Johan Bj{\"o}rkegren}, {and}
  \bibinfo{person}{Tom Michoel}.} \bibinfo{year}{2014}\natexlab{}.
\newblock \showarticletitle{kruX: matrix-based non-parametric eQTL discovery}.
\newblock \bibinfo{journal}{\emph{BMC bioinformatics}}  \bibinfo{volume}{15}
  (\bibinfo{year}{2014}), \bibinfo{pages}{1--7}.
\newblock


\bibitem[Rabenseifner(2004)]%
        {DBLP:conf/iccS/Rabenseifner04}
\bibfield{author}{\bibinfo{person}{Rolf Rabenseifner}.}
  \bibinfo{year}{2004}\natexlab{}.
\newblock \showarticletitle{Optimization of Collective Reduction Operations}.
  In \bibinfo{booktitle}{\emph{Computational Science - {ICCS} 2004, 4th
  International Conference, Krak{\'{o}}w, Poland, June 6-9, 2004, Proceedings,
  Part {I}}} \emph{(\bibinfo{series}{Lecture Notes in Computer Science},
  Vol.~\bibinfo{volume}{3036})}, \bibfield{editor}{\bibinfo{person}{Marian
  Bubak}, \bibinfo{person}{G.~Dick van Albada}, \bibinfo{person}{Peter M.~A.
  Sloot}, {and} \bibinfo{person}{Jack~J. Dongarra}} (Eds.).
  \bibinfo{publisher}{Springer}, \bibinfo{pages}{1--9}.
\newblock
\urldef\tempurl%
\url{https://doi.org/10.1007/978-3-540-24685-5\_1}
\showDOI{\tempurl}


\bibitem[Rabenseifner and Tr{\"{a}}ff(2004)]%
        {DBLP:conf/pvm/RabenseifnerT04}
\bibfield{author}{\bibinfo{person}{Rolf Rabenseifner} {and}
  \bibinfo{person}{Jesper~Larsson Tr{\"{a}}ff}.}
  \bibinfo{year}{2004}\natexlab{}.
\newblock \showarticletitle{More Efficient Reduction Algorithms for
  Non-Power-of-Two Number of Processors in Message-Passing Parallel Systems}.
  In \bibinfo{booktitle}{\emph{Recent Advances in Parallel Virtual Machine and
  Message Passing Interface, 11th European {PVM/MPI} Users' Group Meeting,
  Budapest, Hungary, September 19-22, 2004, Proceedings}}
  \emph{(\bibinfo{series}{Lecture Notes in Computer Science},
  Vol.~\bibinfo{volume}{3241})}, \bibfield{editor}{\bibinfo{person}{Dieter
  Kranzlm{\"{u}}ller}, \bibinfo{person}{P{\'{e}}ter Kacsuk}, {and}
  \bibinfo{person}{Jack~J. Dongarra}} (Eds.). \bibinfo{publisher}{Springer},
  \bibinfo{pages}{36--46}.
\newblock
\urldef\tempurl%
\url{https://doi.org/10.1007/978-3-540-30218-6\_13}
\showDOI{\tempurl}


\bibitem[Rocki et~al\mbox{.}(2020)]%
        {DBLP:conf/sc/RockiESSMKPDS020}
\bibfield{author}{\bibinfo{person}{Kamil Rocki}, \bibinfo{person}{Dirk~Van
  Essendelft}, \bibinfo{person}{Ilya Sharapov}, \bibinfo{person}{Robert
  Schreiber}, \bibinfo{person}{Michael Morrison}, \bibinfo{person}{Vladimir
  Kibardin}, \bibinfo{person}{Andrey Portnoy}, \bibinfo{person}{Jean{-}Francois
  Dietiker}, \bibinfo{person}{Madhava Syamlal}, {and} \bibinfo{person}{Michael
  James}.} \bibinfo{year}{2020}\natexlab{}.
\newblock \showarticletitle{Fast stencil-code computation on a wafer-scale
  processor}. In \bibinfo{booktitle}{\emph{Proceedings of the International
  Conference for High Performance Computing, Networking, Storage and Analysis,
  {SC} 2020, Virtual Event / Atlanta, Georgia, USA, November 9-19, 2020}},
  \bibfield{editor}{\bibinfo{person}{Christine Cuicchi}, \bibinfo{person}{Irene
  Qualters}, {and} \bibinfo{person}{William~T. Kramer}} (Eds.).
  \bibinfo{publisher}{{IEEE/ACM}}, \bibinfo{pages}{58}.
\newblock
\urldef\tempurl%
\url{https://doi.org/10.1109/SC41405.2020.00062}
\showDOI{\tempurl}


\bibitem[Roemer(8 06)]%
        {Roemer2023}
\bibfield{author}{\bibinfo{person}{Niklas Roemer}.}
  \bibinfo{year}{2023-08-06}\natexlab{}.
\newblock \emph{\bibinfo{title}{Designing of a communication library for Versal
  devices using Stream-Based API}}.
\newblock Bachelor Thesis. \bibinfo{school}{ETH Zurich},
  \bibinfo{address}{Zurich}.
\newblock
\urldef\tempurl%
\url{https://doi.org/10.3929/ethz-b-000635928}
\showDOI{\tempurl}


\bibitem[Saad and Schultz(1989)]%
        {DBLP:journals/pc/SaadS89}
\bibfield{author}{\bibinfo{person}{Yousef Saad} {and}
  \bibinfo{person}{Martin~H. Schultz}.} \bibinfo{year}{1989}\natexlab{}.
\newblock \showarticletitle{Data communication in parallel architectures}.
\newblock \bibinfo{journal}{\emph{Parallel Comput.}} \bibinfo{volume}{11},
  \bibinfo{number}{2} (\bibinfo{year}{1989}), \bibinfo{pages}{131--150}.
\newblock
\urldef\tempurl%
\url{https://doi.org/10.1016/0167-8191(89)90024-0}
\showDOI{\tempurl}


\bibitem[Sack and Gropp(2015)]%
        {DBLP:journals/topc/SackG15}
\bibfield{author}{\bibinfo{person}{Paul Sack} {and} \bibinfo{person}{William
  Gropp}.} \bibinfo{year}{2015}\natexlab{}.
\newblock \showarticletitle{Collective Algorithms for Multiported Torus
  Networks}.
\newblock \bibinfo{journal}{\emph{{ACM} Trans. Parallel Comput.}}
  \bibinfo{volume}{1}, \bibinfo{number}{2} (\bibinfo{year}{2015}),
  \bibinfo{pages}{12:1--12:33}.
\newblock
\urldef\tempurl%
\url{https://doi.org/10.1145/2686882}
\showDOI{\tempurl}


\bibitem[Selig(2023)]%
        {Cerebras-SDK}
\bibfield{author}{\bibinfo{person}{Justin Selig}.}
  \bibinfo{year}{2023}\natexlab{}.
\newblock \bibinfo{booktitle}{\emph{The Cerebras Software Development Kit: A
  Technical Overview}}.
\newblock \bibinfo{type}{{T}echnical {R}eport}. \bibinfo{institution}{Cerebras
  Systems, Inc.} \bibinfo{pages}{8} pages.
\newblock
\urldef\tempurl%
\url{https://f.hubspotusercontent30.net/hubfs/8968533/Cerebras%20SDK%20Technical%20Overview%20White%20Paper.pdf}
\showURL{%
\tempurl}


\bibitem[Shabalin(2012)]%
        {shabalin2012matrix}
\bibfield{author}{\bibinfo{person}{Andrey~A Shabalin}.}
  \bibinfo{year}{2012}\natexlab{}.
\newblock \showarticletitle{Matrix eQTL: ultra fast eQTL analysis via large
  matrix operations}.
\newblock \bibinfo{journal}{\emph{Bioinformatics}} \bibinfo{volume}{28},
  \bibinfo{number}{10} (\bibinfo{year}{2012}), \bibinfo{pages}{1353--1358}.
\newblock


\bibitem[Stewart~Hall(2023)]%
        {Cerebras-Neural-Net-Training}
\bibfield{author}{\bibinfo{person}{Sean~Lie Stewart~Hall, Rob~Schreiber}.}
  \bibinfo{year}{2023}\natexlab{}.
\newblock \bibinfo{booktitle}{\emph{Training Giant Neural Networks Using Weight
  Streaming on Cerebras Wafer-Scale Clusters}}.
\newblock \bibinfo{type}{{T}echnical {R}eport}. \bibinfo{institution}{Cerebras
  Systems, Inc.} \bibinfo{pages}{34} pages.
\newblock
\urldef\tempurl%
\url{https://f.hubspotusercontent30.net/hubfs/8968533/Virtual\%20Booth\%20Docs/CS\%20Weight\%20Streaming\%20White\%20Paper\%20111521.pdf}
\showURL{%
\tempurl}


\bibitem[Thakur et~al\mbox{.}(2005)]%
        {DBLP:journals/ijhpca/ThakurRG05}
\bibfield{author}{\bibinfo{person}{Rajeev Thakur}, \bibinfo{person}{Rolf
  Rabenseifner}, {and} \bibinfo{person}{William Gropp}.}
  \bibinfo{year}{2005}\natexlab{}.
\newblock \showarticletitle{Optimization of Collective Communication Operations
  in {MPICH}}.
\newblock \bibinfo{journal}{\emph{Int. J. High Perform. Comput. Appl.}}
  \bibinfo{volume}{19}, \bibinfo{number}{1} (\bibinfo{year}{2005}),
  \bibinfo{pages}{49--66}.
\newblock
\urldef\tempurl%
\url{https://doi.org/10.1177/1094342005051521}
\showDOI{\tempurl}


\bibitem[Tramm et~al\mbox{.}(2024)]%
        {DBLP:journals/corr/cerebras_monte_carlo}
\bibfield{author}{\bibinfo{person}{John Tramm}, \bibinfo{person}{Bryce Allen},
  \bibinfo{person}{Kazutomo Yoshii}, \bibinfo{person}{Andrew Siegel}, {and}
  \bibinfo{person}{Leighton Wilson}.} \bibinfo{year}{2024}\natexlab{}.
\newblock \showarticletitle{Efficient algorithms for Monte Carlo particle
  transport on AI accelerator hardware}.
\newblock \bibinfo{journal}{\emph{Computer Physics Communications}}
  \bibinfo{volume}{298} (\bibinfo{year}{2024}), \bibinfo{pages}{109072}.
\newblock
\showISSN{0010-4655}
\urldef\tempurl%
\url{https://doi.org/10.1016/j.cpc.2023.109072}
\showDOI{\tempurl}


\bibitem[Vadhiyar et~al\mbox{.}(2000)]%
        {DBLP:conf/sc/VadhiyarFD00}
\bibfield{author}{\bibinfo{person}{Sathish~S. Vadhiyar},
  \bibinfo{person}{Graham~E. Fagg}, {and} \bibinfo{person}{Jack~J. Dongarra}.}
  \bibinfo{year}{2000}\natexlab{}.
\newblock \showarticletitle{Automatically Tuned Collective Communications}. In
  \bibinfo{booktitle}{\emph{Proceedings Supercomputing 2000, November 4-10,
  2000, Dallas, Texas, {USA.} {IEEE} Computer Society, {CD-ROM}}},
  \bibfield{editor}{\bibinfo{person}{Jed Donnelley}} (Ed.).
  \bibinfo{publisher}{{IEEE} Computer Society}, \bibinfo{pages}{3}.
\newblock
\urldef\tempurl%
\url{https://doi.org/10.1109/SC.2000.10024}
\showDOI{\tempurl}


\bibitem[Vaswani et~al\mbox{.}(2017)]%
        {DBLP:conf/nips/VaswaniSPUJGKP17}
\bibfield{author}{\bibinfo{person}{Ashish Vaswani}, \bibinfo{person}{Noam
  Shazeer}, \bibinfo{person}{Niki Parmar}, \bibinfo{person}{Jakob Uszkoreit},
  \bibinfo{person}{Llion Jones}, \bibinfo{person}{Aidan~N. Gomez},
  \bibinfo{person}{Lukasz Kaiser}, {and} \bibinfo{person}{Illia Polosukhin}.}
  \bibinfo{year}{2017}\natexlab{}.
\newblock \showarticletitle{Attention is All you Need}. In
  \bibinfo{booktitle}{\emph{Advances in Neural Information Processing Systems
  30: Annual Conference on Neural Information Processing Systems 2017, December
  4-9, 2017, Long Beach, CA, {USA}}},
  \bibfield{editor}{\bibinfo{person}{Isabelle Guyon}, \bibinfo{person}{Ulrike
  von Luxburg}, \bibinfo{person}{Samy Bengio}, \bibinfo{person}{Hanna~M.
  Wallach}, \bibinfo{person}{Rob Fergus}, \bibinfo{person}{S.~V.~N.
  Vishwanathan}, {and} \bibinfo{person}{Roman Garnett}} (Eds.).
  \bibinfo{pages}{5998--6008}.
\newblock
\urldef\tempurl%
\url{https://proceedings.neurips.cc/paper/2017/hash/3f5ee243547dee91fbd053c1c4a845aa-Abstract.html}
\showURL{%
\tempurl}


\bibitem[Veen(1986)]%
        {dataflow_architectures}
\bibfield{author}{\bibinfo{person}{Arthur~H. Veen}.}
  \bibinfo{year}{1986}\natexlab{}.
\newblock \showarticletitle{Dataflow Machine Architecture}.
\newblock \bibinfo{journal}{\emph{ACM Comput. Surv.}} \bibinfo{volume}{18},
  \bibinfo{number}{4} (\bibinfo{date}{dec} \bibinfo{year}{1986}),
  \bibinfo{pages}{365–396}.
\newblock
\showISSN{0360-0300}
\urldef\tempurl%
\url{https://doi.org/10.1145/27633.28055}
\showDOI{\tempurl}


\bibitem[Wickramasinghe and Lumsdaine(2016)]%
        {DBLP:journals/corr/WickramasingheL16}
\bibfield{author}{\bibinfo{person}{Udayanga Wickramasinghe} {and}
  \bibinfo{person}{Andrew Lumsdaine}.} \bibinfo{year}{2016}\natexlab{}.
\newblock \showarticletitle{A Survey of Methods for Collective Communication
  Optimization and Tuning}.
\newblock \bibinfo{journal}{\emph{CoRR}}  \bibinfo{volume}{abs/1611.06334}
  (\bibinfo{year}{2016}).
\newblock
\showeprint[arXiv]{1611.06334}
\urldef\tempurl%
\url{http://arxiv.org/abs/1611.06334}
\showURL{%
\tempurl}


\bibitem[Wierse(3 02)]%
        {Wierse2023}
\bibfield{author}{\bibinfo{person}{Max Wierse}.}
  \bibinfo{year}{2023-02}\natexlab{}.
\newblock \emph{\bibinfo{title}{Evaluation of Xilinx Versal Device}}.
\newblock Bachelor Thesis. \bibinfo{school}{ETH Zurich},
  \bibinfo{address}{Zurich}.
\newblock
\urldef\tempurl%
\url{https://doi.org/10.3929/ethz-b-000600880}
\showDOI{\tempurl}


\bibitem[Wilson(2023)]%
        {Cerebras_Collectives}
\bibfield{author}{\bibinfo{person}{Leighton Wilson}.}
  \bibinfo{year}{2023}\natexlab{}.
\newblock \bibinfo{title}{What’s New in R0.6 of the Cerebras SDK}.
\newblock
  \bibinfo{howpublished}{\url{https://www.cerebras.net/blog/whats-new-in-r0.6-of-the-cerebras-sdk}}.
\newblock
\newblock
\shownote{Accessed: 2023-08-09}.


\bibitem[Woo et~al\mbox{.}(2022)]%
        {field_equation_modeling_cerebras}
\bibfield{author}{\bibinfo{person}{Mino Woo}, \bibinfo{person}{Terry Jordan},
  \bibinfo{person}{Robert Schreiber}, \bibinfo{person}{Ilya Sharapov},
  \bibinfo{person}{Shaheer Muhammad}, \bibinfo{person}{Abhishek Koneru},
  \bibinfo{person}{Michael James}, {and} \bibinfo{person}{Dirk~Van
  Essendelft}.} \bibinfo{year}{2022}\natexlab{}.
\newblock \showarticletitle{Disruptive Changes in Field Equation Modeling: {A}
  Simple Interface for Wafer Scale Engines}.
\newblock \bibinfo{journal}{\emph{CoRR}}  \bibinfo{volume}{abs/2209.13768}
  (\bibinfo{year}{2022}).
\newblock
\urldef\tempurl%
\url{https://doi.org/10.48550/arXiv.2209.13768}
\showDOI{\tempurl}
\showeprint[arXiv]{2209.13768}


\bibitem[Worsch et~al\mbox{.}(2002)]%
        {collective_benchmarking}
\bibfield{author}{\bibinfo{person}{Thomas Worsch}, \bibinfo{person}{Ralf~H.
  Reussner}, {and} \bibinfo{person}{Werner Augustin}.}
  \bibinfo{year}{2002}\natexlab{}.
\newblock \showarticletitle{On Benchmarking Collective {MPI} Operations}. In
  \bibinfo{booktitle}{\emph{Recent Advances in Parallel Virtual Machine and
  Message Passing Interface, 9th European {PVM/MPI} Users' Group Meeting, Linz,
  Austria, September 29 - October 2, 2002, Proceedings}}
  \emph{(\bibinfo{series}{Lecture Notes in Computer Science},
  Vol.~\bibinfo{volume}{2474})}, \bibfield{editor}{\bibinfo{person}{Dieter
  Kranzlm{\"{u}}ller}, \bibinfo{person}{P{\'{e}}ter Kacsuk},
  \bibinfo{person}{Jack~J. Dongarra}, {and} \bibinfo{person}{Jens Volkert}}
  (Eds.). \bibinfo{publisher}{Springer}, \bibinfo{pages}{271--279}.
\newblock
\urldef\tempurl%
\url{https://doi.org/10.1007/3-540-45825-5\_43}
\showDOI{\tempurl}


\end{thebibliography}

\clearpage


\end{document}